\documentclass[11pt]{article}%

\usepackage[english]{babel}
\RequirePackage[T1]{fontenc}

\usepackage{comment}
\usepackage{color}
\usepackage{parskip}
\usepackage{appendix}
\usepackage{amssymb}
\usepackage{booktabs}
\usepackage{comment}
\usepackage{graphicx}
\usepackage{caption}
\usepackage{subcaption}
\usepackage{mathrsfs}
\usepackage{tabularx}
\usepackage{arydshln}
\usepackage{amsmath}
\usepackage{multirow}
\usepackage{multicol}
\usepackage{adjustbox}
\usepackage{authblk}
\usepackage{relsize}
\usepackage{wrapfig}
\usepackage{float}
\RequirePackage{mathptmx}      
\RequirePackage{flushend}
\RequirePackage[numbers,sort&compress]{natbib}
\RequirePackage[colorlinks,citecolor=blue,urlcolor=blue,linkcolor=blue]{hyperref}

\usepackage[utf8]{inputenc}
\usepackage[T1]{fontenc}
\usepackage{amsmath}
\usepackage{graphicx}
\usepackage[scale=0.8]{geometry}
\usepackage{lineno}
\usepackage{xcolor}

\def\babar{\mbox{\slshape B\kern-0.1em{\smaller A}\kern-0.1em B\kern-0.1em{\smaller A\kern-0.2em R~}}}

\usepackage{sectsty}

\sectionfont{\fontsize{16}{18}\selectfont}
\subsectionfont{\fontsize{14}{16}\selectfont}
\subsubsectionfont{\fontsize{11}{13}\selectfont}

\makeatletter
\renewcommand\@author{
    \AB@authlist\\[\affilsep]
    \begin{multicols}{2}
          \begin{flushleft}
        \AB@affillist
           \end{flushleft}
    \end{multicols}
    }

\def\@maketitle{%
  \begin{center}
    {\large\bfseries\@title\par}
    \vspace{1em}

    \begin{minipage}{0.95\textwidth}
      \centering
      \@author
    \end{minipage}

    \vspace{0.5cm}
    {\large\@date}
  \end{center}
}
\makeatother

\usepackage{amsmath}
\usepackage{comment}
\usepackage{graphicx}

\title{
{\raggedleft\small DPHEP-2026-01\par}
\vspace{0.5em}
\centering
\huge Data Preservation in High Energy Physics:\\
\huge Global Report 2026\\
\vspace{0.3cm}
\normalsize DPHEP Collaboration
\vspace{1.3cm}
}

\author[6]{Matthew~Bellis}
\author[4]{Jamie~Boyd}
\author[16]{Daniel~Britzger}
\author[21]{Andy~Buckley}
\author[4]{Chris~Burr}
\author[4]{Micheal~Buchar}
\author[13]{Gang~Chen}
\author[19]{Andrew~Chisholm}
\author[7]{Jiri~Chudoba}
\author[4]{Michael~Davis}
\author[5]{Cristinel~Diaconu}
\author[1]{David~Dobrigkeit~Chinellato}
\author[28]{Marcus~Ebert}
\author[4,9,22,27]{Apranik~Fatehi}
\author[24]{Dillon~S.~Fitzgerald}
\author[4]{Jose~Benito~Gonzalez~Lopez}
\author[25]{Richard~William~Gran}
\author[4]{Giovanni~Guerrieri}
\author[21]{Martin~Habedank}
\author[13]{Hao~Hu}
\author[8]{David~Horvat}
\author[3]{Luka~Lambrecht}
\author[17]{Clemens~Lange}
\author[11]{Kati~Lassila-Perini}
\author[2]{Jerome~Lauret}
\author[4]{Jean-Yves~Le~Meur}
\author[1]{Dietrich~Liko}
\author[4]{Panna~Liptak}
\author[15]{Zach~Marshall}
\author[26]{Thomas~McCauley}
\author[4]{Cameron~Duncan~McClymont}
\author[4]{Wesley~Middelbos}
\author[4]{Micha~Moskovic}
\author[20]{Piet~Nogga}
\author[4]{Ryunosuke~O'Neil}
\author[12]{Stefano~Piano}
\author[14]{Alan~Price}
\author[14]{Tomasz~Procter}
\author[13]{Fazhi~Qi}
\author[4]{Diana~Rand}
\author[29]{Hossein~Rashidi}
\author[6]{Emily~Rensch}
\author[4]{Pablo~Saiz}
\author[4]{Ulrich~Schwickerath}
\author[4]{Tibor \v{S}imko}
\author[16]{Andrii~Verbytskyi}
\author[10]{Dirk~Zerwas}
\author[13]{Zhengde~Zhang}
\author[23]{Graham~Wilson}

\affil[1]{\footnotesize Austrian Academy of Sciences (AT)}
\affil[2]{\footnotesize Brookhaven National Laboratory (US)}
\affil[3]{\footnotesize Brown University (US)}
\affil[4]{\footnotesize CERN (CH)}
\affil[5]{\footnotesize CPPM, CNRS/IN2P3 and Aix-Marseille University (FR)}
\affil[6]{\footnotesize Siena University (US)}
\affil[7]{\footnotesize Czech Academy of Sciences, FZU (CZ)}
\affil[8]{\footnotesize Deggendorf Institute of Technology (DE)}
\affil[9]{\footnotesize DESY (DE)}
\affil[10]{\footnotesize DMLab, CNRS/IN2P3 and DESY (DE)}
\affil[11]{\footnotesize Helsinki Institute of Physics (FI)}
\affil[12]{\footnotesize INFN Trieste (IT)}
\affil[13]{\footnotesize Institute of High Energy Physics(IHEP), CAS (CN)}
\affil[14]{\footnotesize Jagiellonian University (PL)}
\affil[15]{\footnotesize Lawrence Berkeley National Lab. (US)}
\affil[16]{\footnotesize Max-Planck-Institut f\"{u}r Physik (DE)}
\affil[17]{\footnotesize Paul Scherrer Institute PSI (CH)}
\affil[18]{\footnotesize UCL (UK)}
\affil[19]{\footnotesize University of Birmingham (UK)}
\affil[20]{\footnotesize University of Bonn (DS)}
\affil[21]{\footnotesize University of Glasgow (UK)}
\affil[22]{\footnotesize University of Hamburg (DE)}
\affil[23]{\footnotesize University of Kansas (US)}
\affil[24]{\footnotesize University of Michigan (US)}
\affil[25]{\footnotesize University of Minnesota Duluth (US)}
\affil[26]{\footnotesize University of Notre Dame (US)}
\affil[27]{\footnotesize University of Tehran (IR)}
\affil[28]{\footnotesize University of Victoria (CA)}
\affil[29]{\footnotesize Bu-Ali Sina University (IR)}

\begin{document}
\maketitle 

\begin{abstract}
This document summarizes the contributions to the 5th DPHEP workshop March 5-6, 2026, CERN, and reflects the advancements since 2024, as well as future milestones and tendencies. Impressive progress in HEP data preservation is observed.  Legacy data revival was showcased through successful reanalysis of archived data using contemporary methods, demonstrating the long-term scientific value of preservation. Sustainability challenges were noted, emphasizing the need for long-term funding and institutional support to maintain data preservation infrastructure, particularly for legacy experiments transitioning to archival modes. Innovative transverse projects display constant progress towards common technologies for a robust and transferrable DP. In particular, there is a clear shift toward automation, with increasing use of AI and machine learning for data curation, metadata extraction, and workflow optimization. Open science momentum is growing, with wider adoption of FAIR principles and open data policies, and experiments committing to public releases.  
\end{abstract}
\newpage
\tableofcontents
\newpage
\section{Executive summary}
The 5th DPHEP Collaboration Workshop, held at CERN on March 5--6, 2026,  emphasized the increasing importance of long-term data preservation and open science.\footnote{\url{https://indico.cern.ch/e/dphep5}} Several aspects were addressed: the larger landscape of open data policies and their concrete implementation in the running experiments at CERN, the status of data preservation at experiments and sites, and the transverse projects proposing either new strategies or technologies for addressing data preservation beyond experiment-specific constrains.

\begin{figure}[hhh]
    \centering
    \includegraphics[width=0.48\textwidth]{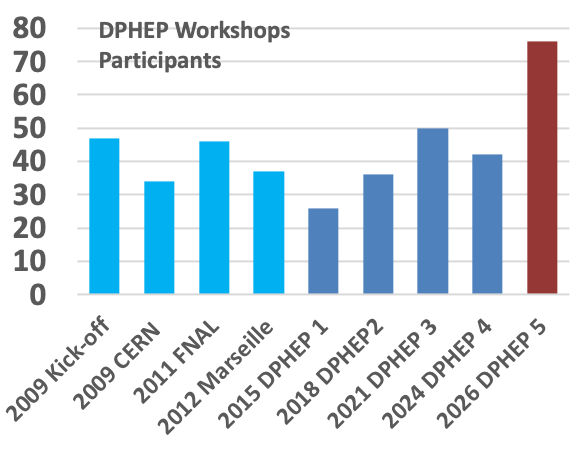}
    \includegraphics[width=0.48\textwidth]{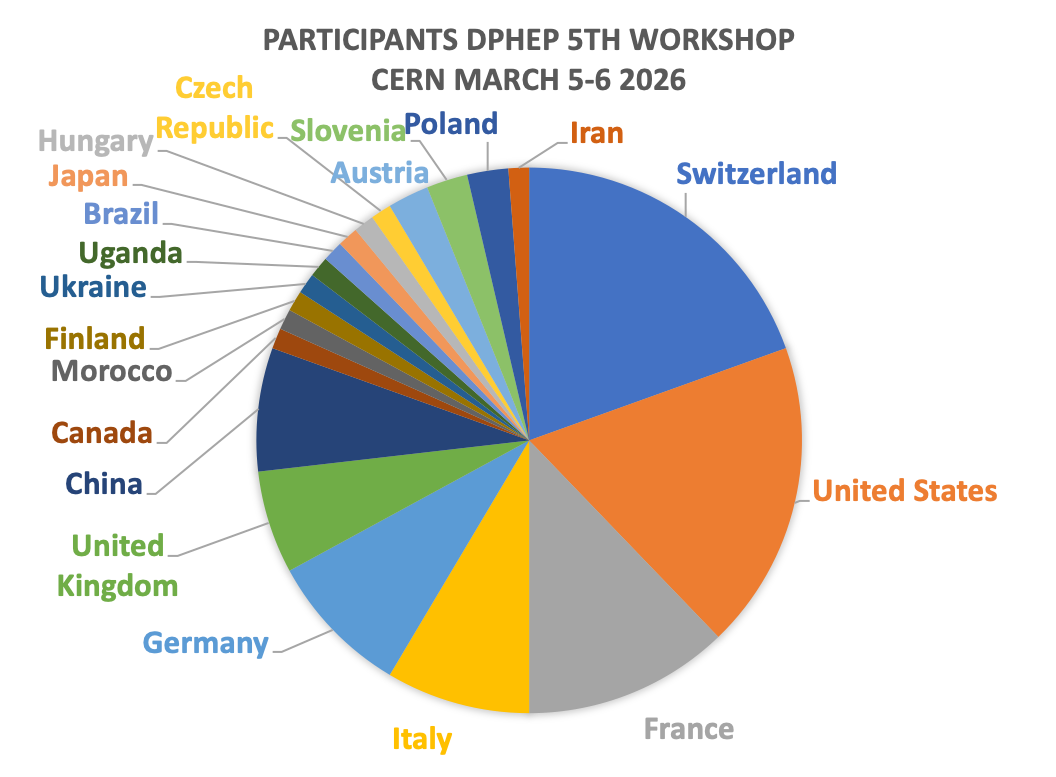}
    \caption{
Participation at the DPHEP workshops (left) and countries of participants at the 5th workshop (right).}
\label{dphep_participation}
\end{figure}
The implementation of CERN’s Open Data policy aims to balance data ownership concerns with the need for openness. The policy mandates the release of substantial datasets within five years of collection, with growing adoption across LHC experiments. Updates were provided on the long-term preservation of data from various experiments, including ALICE, ATLAS, CMS, LHCb, \babar, H1, OPAL, and DELPHI. These updates included reanalysis efforts, such as the reprocessing of OPAL’s 26-year-old data with modern tools. Progress was also reported on preserving data from BESIII, JUNO, and LHAASO, emphasizing comprehensive frameworks for raw and processed data, as well as analysis software.

Transverse projects were a key focus, with discussions on cross-disciplinary preservation standards like OAIS, FAIR, and TRUST through the EOSC EDEN project. The REANA platform was showcased for its reproducible analysis workflows and AI-assisted tools for open data reuse, including federated workflows across the European Open Science Cloud. The CERN Preserve Platform was highlighted for its OAIS-compliant pipelines for long-term preservation of research content such as theses, publications, and documentation.

Emerging technologies and AI integration were significant themes. The use of generative AI and large language models to extract dataset metadata from journal articles was explored, aiming to improve data discoverability. The Data Orchestration Agent (DORA) was introduced as an AI-driven framework for automating data lifecycle management and optimizing datasets for AI training. Prototypes for AI-assisted workflow authoring were discussed to reduce errors and enhance reproducibility. Modern tools for legacy data, such as the transition from TCL to Awkward for reanalyzing 20-year-old \babar data, were demonstrated, bridging legacy and contemporary analysis frameworks. Advanced b-tagging with ALEPH data using deep-learning techniques was also presented, showing improvements in heavy-flavor tagging performance.

The challenges and future directions were addressed, including cold storage solutions such as tape-based archival systems within the CERN Open Data portal. These systems balance cost and accessibility for large datasets exceeding 5 PB. The discussion also covered metadata and standards, including the introduction of a Python package for benchmarking Monte Carlo generators to ensure reproducibility in future collider simulations. The conversion of DELPHI and ALEPH legacy data to modern formats such as EDM4HEP was also discussed, facilitating integration with current tools. Most notably, it was demonstrated that modern analysis techniques can drastically improve b-tagging performance at ALEPH, paving the way for significantly greater precision in certain key LEP measurements.

Collaboration and policy were key topics, with an outline of assessment frameworks for ICFA’s data preservation guidelines, emphasizing FAIR principles and community engagement. A holistic approach from Brookhaven was described, combining AI assistants and document management for sustainable knowledge retention.


In conclusion, the workshop underscored that data preservation is a dynamic process enabling new discoveries, education, and interdisciplinary innovation. 
Next steps involve policy refinement, technological adoption, and global coordination to ensure the enduring accessibility and utility of unique datasets in high energy physics. Through its long term and pluri-experimental/multi-site approach, DPHEP complements and supports the dynamics of multi-annual projects and ensures a long term vision of a fully efficient data life-cycle  in HEP.

\section{Experiments and sites}

\subsection{DPHEP greetings from the past: news around LEP}
{\small
\it Author: Ulrich Schwickerath (CERN)
}

\begin{figure}[hhh]
\includegraphics[width=\textwidth]{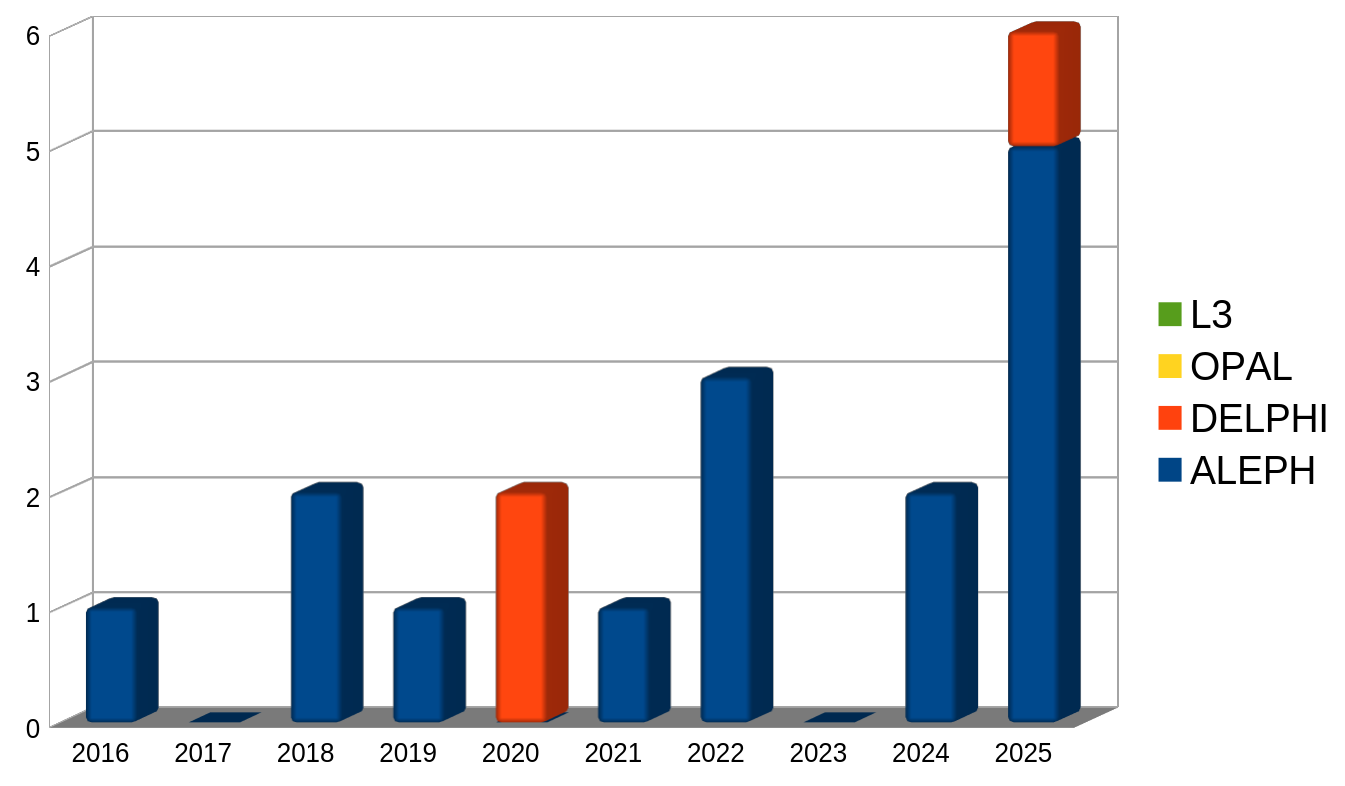}
\caption{\label{fig_3rdparty}3rd party publications mentioning LEP data since 2016.}
\end{figure}

Since the last workshop in 2024, the number of publications mentioning LEP archived or open data has significantly increased. This can be seen in fig.~\ref{fig_3rdparty}. The increasing number of publications reflects an increased interest in this kind of data. Publications have been dominated by ALEPH data, which may well be related to their fairly open data access policy, allowing 3rd party people to access and analyse their archived data. Since the DELPHI experiment changed their data access policy, a first publication has been released in 2025. The OPAL experiment announced a new access policy recently as well, and their data will be published during 2026 on the CERN open data portal. Successful contact has been made to people from L3, and during the workshop the wish was expressed to get hands on this data as well, which would complete the LEP data recovery.

Binaries for the common tools CERNLIB and OpenPHIGS are now available on /cvmfs/dphep.cern.ch, including simple setup scripts to identify the correct version and setup paths and environment variables.

\subsubsection{CERNLIB}
CERNLIB is a pre-requisite for many old experiment stacks, including the LEP experiments. 
Since the last workshop, several new releases have been published. Among bug fixes and addition of support for new operating system versions, the ancient support for the insecure and obsolete RFIO protocol has been dropped and replaced by the possibility to link with XrootD. This allows reading data directly over the network, including access to protected data with strong authentication. Maintenance cost have been reduced significantly by making extensive use of CIs for building, testing and releasing. Also, it can be expected that AIs will be useful to sort out potentially upcoming challenges in the future.

\subsubsection{OpenPHIGS}
OpenPHIGS is a free implementation of the PHIGS standard, which is used by both DELPHI and OPAL. The implementation was presented already during the last workshop. In collaboration with the CERN OSPO, the original author has been contacted and permission has been given by him to re-publish the code, including the extensions needed for data preservation, under the CERN umbrella on \hyperlink{https://github.com/cern/openphigs}{https://github.com/cern/openphigs}.

\subsubsection{LEP experiments}
Apart from the data access policy change for OPAL, not much has been changed with respect to the last workshop status.
Both OPAL and DELPHI have extended their portfolio of supported operating systems. While the software stack for these two works natively on recent Linux distributions, ALEPH software is preserved on an SLC6 container. Binaries are available on /cvmfs/<experiment>.cern.ch, in the case of ALEPH only for an older release for SLC4.

Software sources for ALEPH, DELPHI and OPAL are available on gitlab.cern.ch. The data for all experiments is available on EOS, including L3 data.

\begin{figure}
\centering
\includegraphics[width=0.49\linewidth]{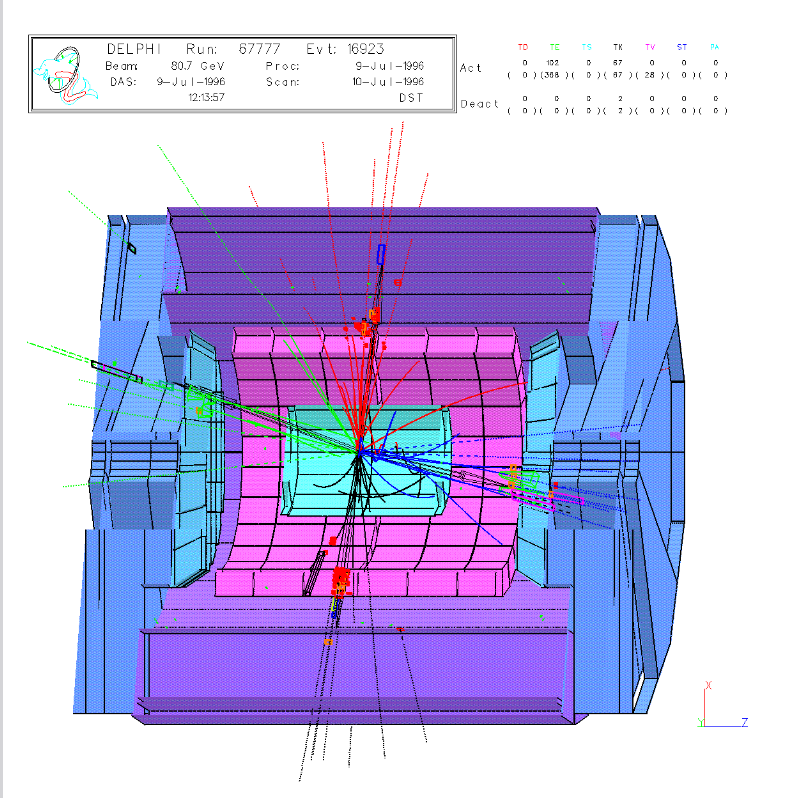}
\includegraphics[width=0.49\linewidth]{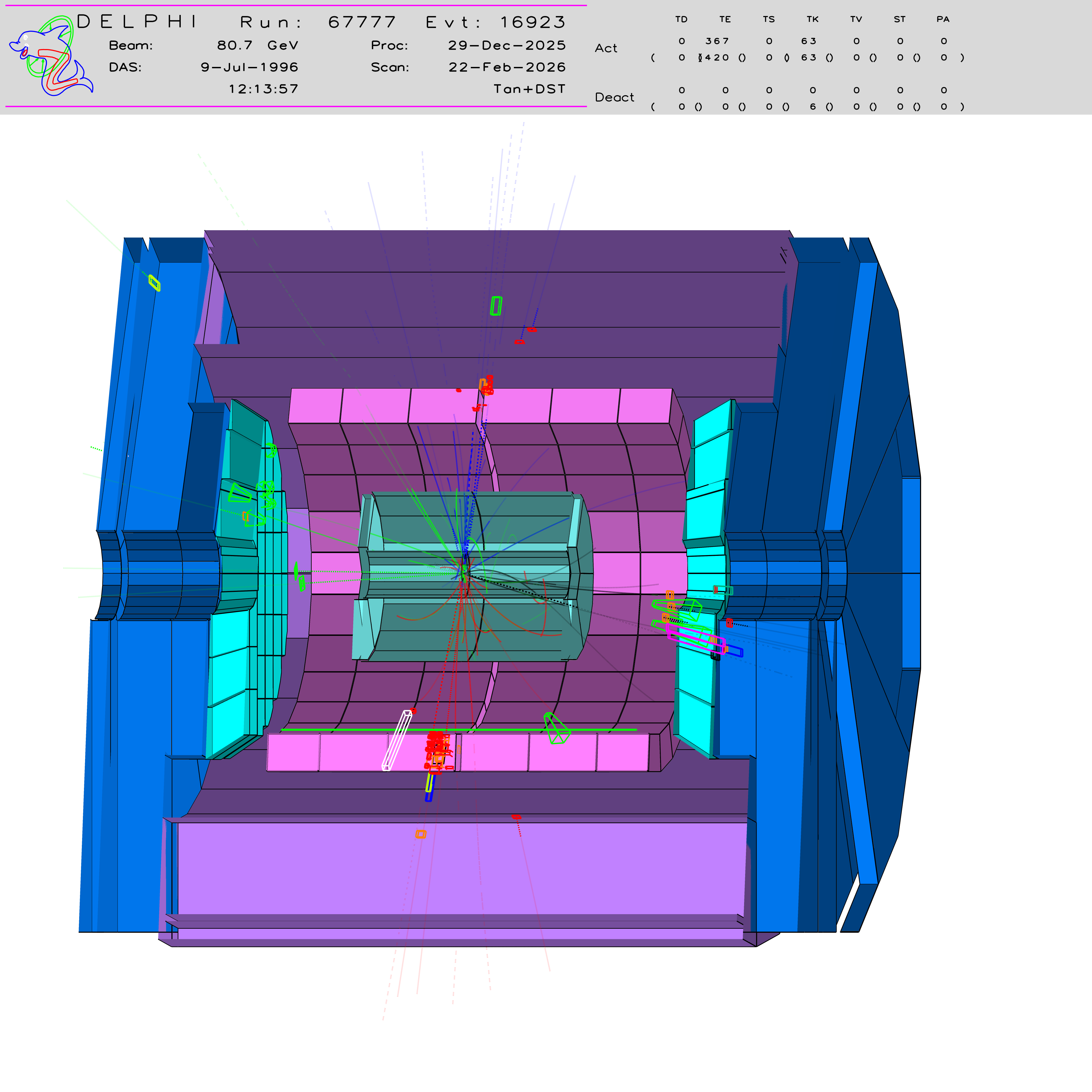}
\caption{\label{fig_ww}First $W^+W^-$ event recorded at LEP. Original figure (left), recovered, reprocessed and re-displayed in high resolution (right).}
\end{figure}

DELPHI is currently performing a data curation exercise which is unfinished but already unveiled some surprises, like the unexpected corruption of some files. As an example, the original file containing the raw data of the first $W^+W^-$ event which has ever been recorded at an $e^+e^-$ collider has been lost. Fortunately, a copy of the raw data has been located within the full DST of this run, from which it is possible to reprocess the event, shown in fig.~\ref{fig_ww}. The global impact of the data loss is expected to be limited, but needs to be kept in mind when re-analysing the data.

\subsubsection{The JADE experiment}
JADE is another experiment which depends on CERNLIB.  While only briefly mentioned during the workshop, in the mean time the JADE collaboration finished the publication of their data via the Open Data Portal at CERN shortly after the workshop. Contributions to the recovery of CERNLIB enabled the recovery of the LEP experiment software stacks, show casing the significance of even old data for the field.

\subsubsection{Risks}
DELPHI and OPAL data run on RHEL10 and compatible clones of this Linux flavor. This ensures continued access until the early 30th of the century. Nevertheless, the ever changing IT infrastructure continues to bear some risks which have to be detected and addressed in time:
\begin{itemize}
\item CERNLIB relies on static linking on 64bit operating system. The reason for this is that it assumes 32bit address space. The ZEBRA system allows to use Fortran integers as pointers, which are by definition only 32bit wide. There are indications that support for static linking is vanishing. This is happening already on MacOS.
\item X11 replacement by Wayland: OpenPHIGS and CERNLIB have been tested on systems with Wayland only. Things work but rely on X11 compatibility libraries, including Xt, GLX and similar. It may become necessary to modernise the whole graphics parts, which is expected to require some effort.
\item Both CERNLIB and OpenPHIGS have a dependency on Motif. This was already missing in the first release of RHEL10, and only added later for dependency reasons with other tools. In case Motif is removed, the graphics parts will have to be rewritten.
\item Computer security requirements are another risk, for example the retirement and removal of ciphers and encryption types which are now considered insecure. This is in particular a threat for frozen binaries.
\item Long term ownership of data and software, including their maintenance, has to be clarified. 
\item In case of LEP, there is a strong dependency on CERN services which may change. This include gitlab, EOS, CVMFS and others. Many of these require CERN logins as well, which makes it more difficult to external people to contribute.
\item While there are several copies of LEP data on tape and at least one on disk, all of them are at CERN, and actually in the same building. At the same time, the tape capacity is so large nowadays, that the data of a single LEP experiment easily fits on a single cartridge, and these can break. Experiments should therefore try to find a way to have an external copy of their data. For DELPHI, this is already happening, and there was a report during this workshop on the status.
\end{itemize}

\subsection{Advanced jet flavour tagging with archived ALEPH data}

{\small \it 
Authors: Apranik Fatehi (CERN \& University of Tehran (IR)); Birgit Stapf (CERN); Gerardo Ganis (CERN); Jacopo Fanini (CERN \& Université Paris-Saclay); Loukas Gouskos (Brown University (US)); Luka Lambrecht (Brown University (US)); Matteo Defranchis (CERN); Michele Selvaggi (CERN); Taj Sebastian Gillin (Brown University (US)); Zihan Ma (Brown University (US))

 The authors would like to note that this work has recently been submitted to JHEP~\cite{defranchis2026modernjetflavourtagging}, and refer there for all details. The text below provides a quick summary of the results. When citing this work specifically, please cite the dedicated publication.}

The 1994 dataset from the ALEPH experiment at LEP was recently converted into the inter-experiment and general-purpose EDM4HEP data format~\cite{fanini2026lepdataedm4hepmitigatingdata}. Using this archived and converted ALEPH data, and the corresponding simulation, we apply modern software intended for FCC studies to process the data and train and employ state-of-the-art, deep-learning based jet flavour tagging techniques.

\begin{figure}
    \centering
    \includegraphics[width=0.49\linewidth]{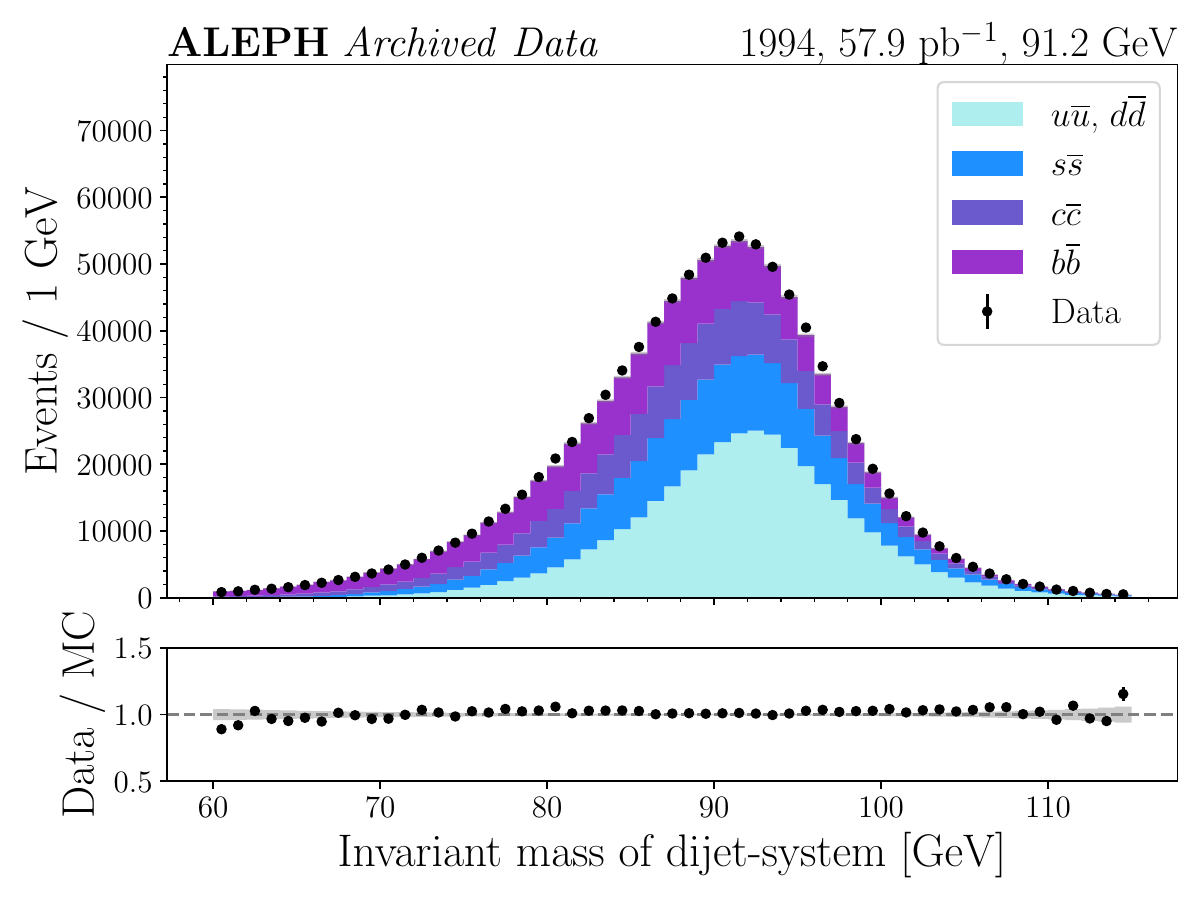}
    \includegraphics[width=0.49\linewidth]{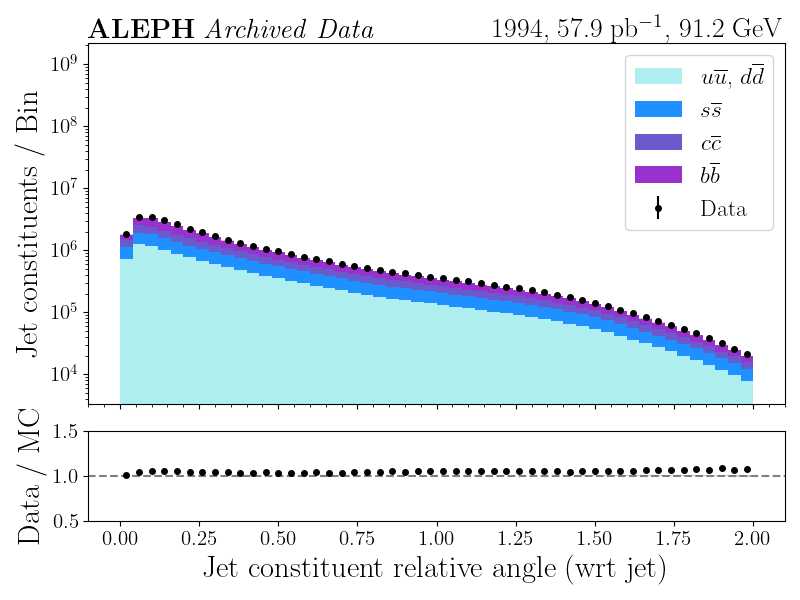}
    \caption{Comparison between data and simulation for the dijet invariant mass (left) and the relative angle of each particle with respect to the axis of the jet it is clustered in (right). The coloured histograms are simulated $Z \rightarrow q\bar{q}$ events, split by quark flavour. The black markers are the data. The grey band represents the statistical uncertainty in the simulation.}
    \label{fig:section_2.2:mcvsdata}
\end{figure}

The dataset consists of electron-positron collisions at a center-of-mass energy corresponding to the $Z$ pole of 91.2\,GeV. We select $Z \rightarrow q\bar{q}$ events based on the presence of two hadronic jets and a sufficient number of tracks with some basic quality criteria. The simulation-to-data agreement for some kinematic variables are shown in Fig.~\ref{fig:section_2.2:mcvsdata}, showing good modeling for both jet- and particle-level features.

We fit the primary vertex and calculate the impact parameters for each track, reconstruct secondary vertices with a special emphasis on $K^0_S$ and $\Lambda^0$ candidates, and derive particle identification probabilities from the energy loss measurements of ALEPH's time projection chamber. All this information, together with basic kinematic properties of each particle within the jet, is combined in a transformer-based jet flavour classifier, using the \textsc{ParticleTransformer} architecture~\cite{qu2024particletransformerjettagging} developed at CMS.

We achieve up to one order of magnitude improvement in background rejection for $b$-jet tagging compared to the legacy algorithms used for the most recent ALEPH measurements~\cite{Barate:321135, ALEPH:2001mdb}, for the same identification efficiency. Alternatively, for the same background rate, the signal efficiency is approximately doubled. To the best of our knowledge, we also present the first implementation and performance studies of strange jet tagging at LEP, allowing the selection of an event sample enriched in $Z \rightarrow s\bar{s}$, albeit with significantly lower purity than $Z \rightarrow b\bar{b}$ or $Z \rightarrow c\bar{c}$. We observe good agreement between simulation and data as a function of the flavour tagger output scores, and residual differences can be accounted for with a tag-and-probe method.

These studies pave the way for improving the precision of measurements of electroweak observables in $Z$ boson hadronic decays, by employing modern analysis techniques to archived data. They can also serve as guidance for the development of detectors and algorithms for future electron-positron colliders, by providing a testbed for software and algorithms with real data rather than simulation only.

In more general terms, these results demonstrate the value of long term data preservation, where archived datasets remain a valuable resource for future analyses whenever new techniques or ideas become available.

\subsection{Converting DELPHI Legacy Open Data to the EDM4HEP Format}
{\small \it Author: Dietrich Liko (Marietta Blau Institute, Austrian Academy of Sciences)}

The DELPHI experiment, one of the four major detectors at the Large Electron-Positron Collider (LEP) at CERN, has achieved a significant milestone in Open Science by releasing its complete legacy dataset. This release prioritizes the "Short DST" (SDST) format for physics analysis while including RAW data primarily for visualization purposes. While the full suite of DELPHI software has been ported to modern platforms, the recommended legacy analysis tools, such as SKELANA, rely on a fading ecosystem of FORTRAN77, PATCHY, ZEBRA, and HBOOK.The reliance on these legacy tools presents several critical risks:
\begin{itemize}
\item \textbf{Sustainability:} Expertise in Fortran and ZEBRA is rapidly disappearing; without intervention, the data risks becoming inaccessible within a single generation.
\item \textbf{Software Fragility:} ZEBRA depends on legacy software that is increasingly difficult to maintain on contemporary systems.
\item \textbf{Isolation:} The legacy format prevents the integration of modern analysis techniques, such as current Machine Learning (ML) frameworks.
\end{itemize}To address these challenges, Project Amphitrite was established to transform legacy DELPHI data into the modern EDM4HEP format. Named after the Greek goddess who brings order to chaotic seas, the project "tames" the data for use in the modern Key4hep and ROOT ecosystems. This conversion ensures long-term preservation and enables direct benchmarking between LEP data and future-collider simulations for projects like the FCC, ILC, and CLIC.The technical architecture utilizes a C++ wrapper over original DELPHI tools to provide C++20 idiomatic access to the hierarchical ZEBRA banks. This wrapper exposes the legacy structures as typesafe, STL-like containers. The conversion is then executed through a pipeline of \texttt{CollectionWriters}. The current pipeline focuses on several key stages:
\begin{itemize}
\item \textbf{Event Metadata:} Capturing run information and event headers.
\item \textbf{Tracking and Vertexing:} Converting primary and secondary vertices, charged tracks, and track states.
\item \textbf{Monte Carlo (MC) Truth:} Linking reconstructed particles to their MC truth counterparts.
\end{itemize}

Ongoing physics validation is essential to ensure the converted data accurately reproduces the original detector's performance. Study areas include Z lineshape, tracking resolution for the Microvertex (VD) and Very Forward Tracker (VFT) silicon detectors, and combined Particle Identification (PID) from the RICH and TPC.Project Amphitrite aims for a first release focusing on vertices, particles, and MC information around Easter of this year. Future updates will extend the conversion to include calorimetry and advanced PID. 

By moving to EDM4HEP, this irreplaceable LEP record remains a living dataset accessible through modern tools like ROOT \texttt{RDataFrame} and Python-based frameworks.
\newpage
\subsection{A New Repository for DELPHI Data}
{\small \it Author: Jiri Chudoba (Czech Academy of Sciences (CZ)); Michal Lukes (FZU)}

EOCS-CZ activities include the creation of several data repositories for various scientific fields. The repository for High Energy Physics and Astroparticle Physics is currently under construction. Data from the LEP experiment DELPHI, already preserved in the CERN Open Data Portal, serve as the first use case for the new repository. The motivation is to maintain an additional copy of these valuable data in a geographically distinct location, to test the performance and limits of data access at the current site, and to compare them with those of the new repository.
 
The new prototype data repository is based on the Invenio framework. This prototype explores an experiment-agnostic approach to publishing preserved HEP (and other) data. The repository technology is conceptually aligned with but independent from the CERN Open Data portal. 

We have built the tools to automatically download all 12727 datasets plus supplementaries, 32TB of data, ~450k files, to our servers. From there we uploaded them to the prototype instance of our repository~\footnote{\url{https://doi.org/10.83100/8wej-ry35} \url{https://test1.physics.du.cesnet.cz}}. We kept the records as similar to the original as practical with the two exceptions: 
\begin{itemize}
    \item Due to technical limitations we had to zip larger datasets with more than 20 files
    \item Because we were able to extract the exact energy used in simulated experiments from the metadata we chose to use this exact value instead of range for better filtering. But we also provide the original metadata as a backup solution.
\end{itemize}

 \begin{figure}[hhh]
    \centering
    \includegraphics[width=0.7\linewidth]{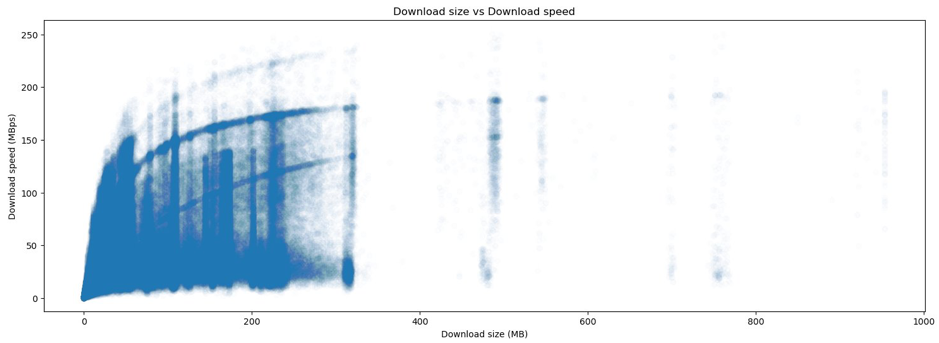}
    \includegraphics[width=0.7\linewidth]{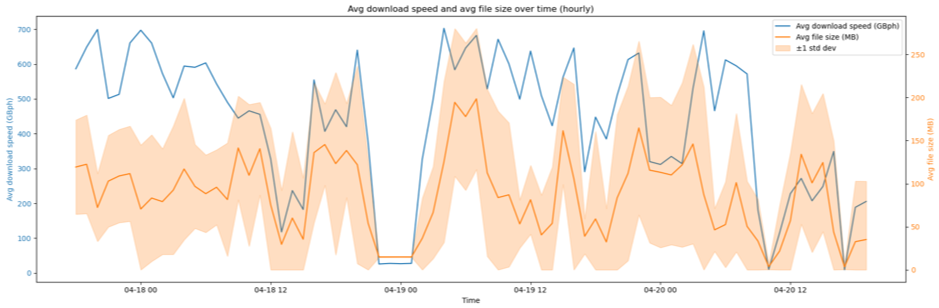}
    \caption{Performance parameters of the data repository prototype.}
    \label{fig:section_2.4:repo}
\end{figure}

The original download took almost three weeks due to technical issues while the implementation was still in progress, but we were able to get it down to a couple of days, depending on the usage of the network. We encountered a couple of bugs in the implementation of cernopendata-client but they were promptly fixed thanks to the work from Tibor Šimko. The only lingering issue is caused by OpenSearch which limits search results to 10000. This limit prevented us from listing all records using the API, but it is not an issue for regular users. It is worth noting that our repository presents the same problem.

Our repository is based on InvenioRDM and many features are still in development by another project~\footnote{\url{https://github.com/FZU-AV-CR/invenio-fzu}}. Its performance is illustrated in Figure~\ref{fig:section_2.4:repo}. We were able to get our first DOI and a record in NMA~\footnote{\url{https://nma.eosc.cz/datasets/records/doi/10.83100/8wej-ry35}}.

In future we plan to perform performance experiments on our repository to compare it to the open data portal and add many more communities some of which are already under development like Astro particle physics, Crystallography, Telescopes and Other communities from HEP. Other future improvements include any new features provided by the developers of our infrastructure like hierarchical records, usage monitoring or alternative PIDs.

\subsection{26 years on: Experience with OPAL Data Re-analysis}
{\small \it Author: Graham Wilson (University of Kansas (US))}

Recently, I have been exploring the feasibility of doing analysis 
with the OPAL data again. I have been amazed and excited by the possibilities at hand. After quickly reproducing the results on radiative neutrino counting ($\mathrm{e}^{+} \mathrm{e}^{-} \to \nu \bar{\nu} \gamma (\gamma)$)
from a partial dataset published in 2000~\cite{OPAL:2000puu}, it is planned to 
complete, publish, and document an improved analysis for 
the complete LEP2 dataset of OPAL with a fully reproducible workflow; 
the target is an experimental systematic uncertainty of 1\% necessitating 
several analysis upgrades. 
Several other topics in physics analysis, detector response modelling, and measurements for future accelerators 
also suggest themselves.

It is not just data availability, but that the whole software stack can run locally on a laptop (via CVMFS), including event simulation, event reconstruction, and event display, 
enabling development/fixes to the software.

I had previously been interested in further exploiting OPAL data, but was deterred by several issues. Recent developments have made this much easier: the software stack is supported across multiple architectures, the data are in EOS, CERNLIB lives again, and with the Open-PHIGS solution for the event display, there are no show-stoppers. Leveraging 26 years of Moore's law, together with well-documented software and access to the source code, enables 
data analysis methods that go far beyond what was computationally feasible last century. 
While the code base is in ANSI-compliant F77, Fortran-callable C++ functions have been used for self-contained new developments. In the context of the radiative neutrino counting analysis, it has been straightforward to generate large simulated data samples of instrumental backgrounds arising from cosmic-rays and beam-halo muons, and to 
make analysis-relevant improvements to the event display 
and GEANT3-based full detector simulation.

\begin{figure}
    \centering
    \includegraphics[width=0.6\linewidth]{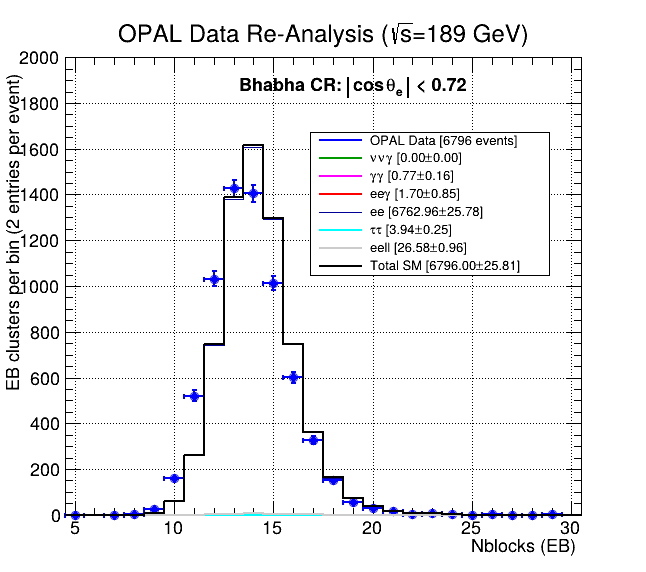}
    \caption{Comparison between data (blue markers) and simulation (black histogram) 
    for the number of blocks per cluster in the Electromagnetic Barrel (EB) calorimeter 
    in a Bhabha control region for cluster polar angles satisfying $|\cos{\theta_{\mathrm{e}}}| < 0.72$ using the OPAL 1998 data at $\sqrt{s}=189$~GeV. The simulation model normalization has been adjusted to exactly match the data event count.}
    \label{fig:OPAL-Nblocks}
\end{figure}

Progress has been made on a number of issues for the radiative neutrino counting analysis:
\begin{itemize}
\item
  Resurrecting and improving the shower fit algorithm that adjusts 
the calorimeter impact point to best describe the energy depositions per lead glass block 
in barrel calorimeter clusters. This improves photon 4-vector reconstruction and allows for 
powerful rejection of spurious photon candidates from non-collision background events.
\item 
  The sensitivity of the analysis to detector occupancy has been reduced, for example by tightening the 
muon veto requirement in the muon barrel detector system from at least two to at least three muon chamber hits.
\item 
  The rejection of cosmic-ray muons in wide-angle photon conversion candidates has been 
  improved by a factor of ten by adding a track transverse 
impact parameter criterion.
\item 
  Systematic studies with a high statistics Bhabha scattering control sample.
Fig.~\ref{fig:OPAL-Nblocks} shows a comparison between data and simulation for these 
beam energy electrons in the barrel region for the 1998 data-set. 
The calorimeter clusters observed in data are approximately 3\% smaller than predicted by the simulation.
A possible explanation is an overestimate of the upstream material (pressure vessel + coil) in 
the legacy detector simulation.
\item
The multi-photon vetoes used 
to reject events with two or more photons such as $\mathrm{e}^{+} \mathrm{e}^{-} \to \gamma \gamma (\gamma)$ 
with no genuine missing transverse momentum have been improved. This increases the signal efficiency 
and also reduces sensitivity to the modelling of multi-photon signal 
contributions ($\mathrm{e}^{+} \mathrm{e}^{-} \to \nu \bar{\nu} \gamma \gamma (\gamma)$) 
while retaining negligible background acceptance.
\end{itemize}

This progress is encouraging for the ongoing effort to bring this revamped analysis to completion, 
with the aim of producing competitive results that exploit the full LEP2 data set.

\subsection{Status of \babar's Data and Analysis Preservation Efforts}
{\small \it 
 \it Author: Marcus Ebert (University of Victoria, Canada)
}

\begin{figure}
    \centering
    \includegraphics[width=0.8\linewidth]{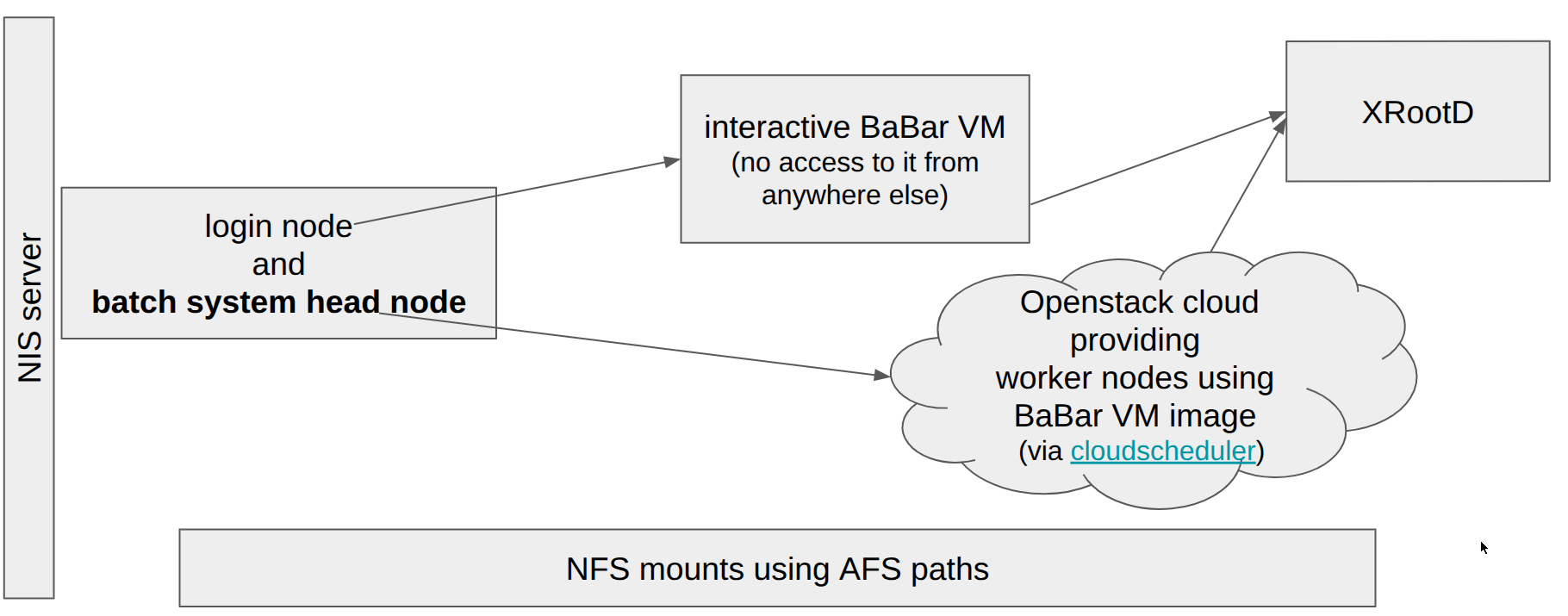}\\
    \vspace{0.5cm}
    \includegraphics[width=0.7\linewidth]{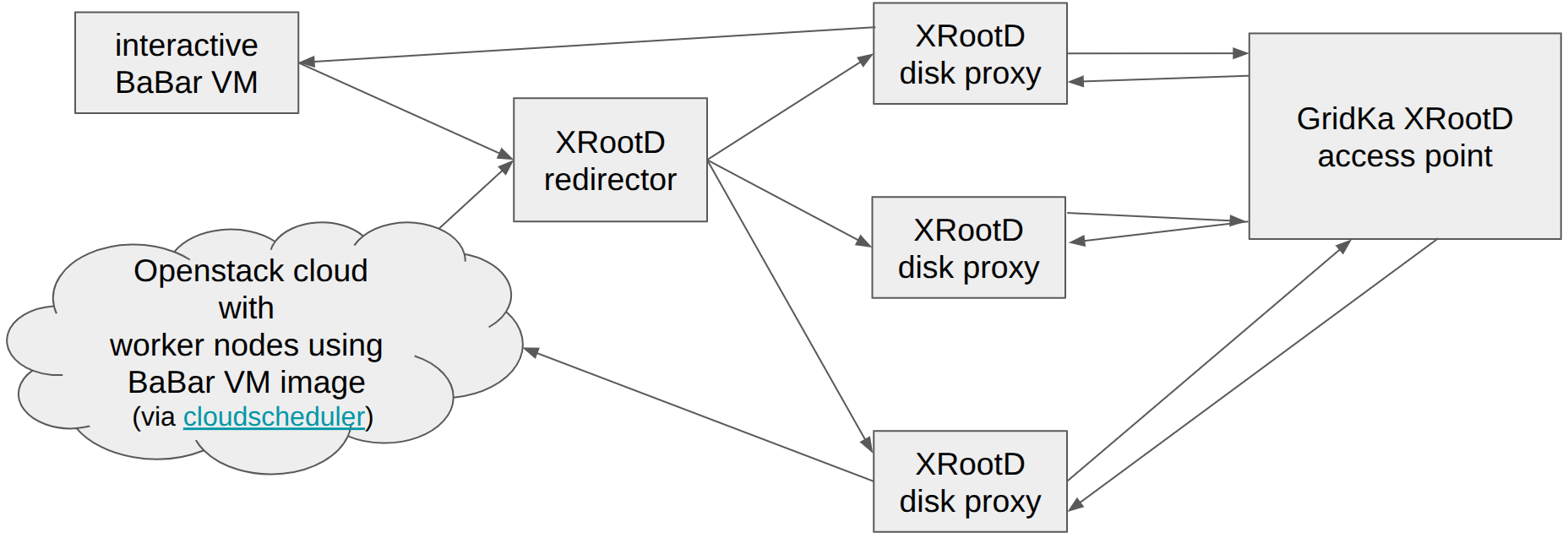}
    \caption{Overview of the \babar LTDA system; on top a general overview is shown, on the bottom the XRootD data access infrastructure is shown.}
    \label{fig:babar_overview}
\end{figure}

\babar's current Long Term Data and Analysis (LTDA) system is located at the University of Victoria, Canada, and the data for analysis access is located at GridKa, Germany. A copy of the data is also located at CC-IN2P3 and at CERN. This system is successfully in production since 2022. An overview of the LTDA is shown in Fig.~\ref{fig:babar_overview}.
12 analyses have been published since 2022, one in 2026 so far. In addition, \babar is still presenting at many conferences each year. In 2026, two new analyses started so far. In addition, via the "\babar Associates Open-access Program"~\cite{babar:openaccess}, five new associate membership requests were received in the first half of 2026, leading to five new analyses that will start soon. While the \babar $\Upsilon(4S)$ data sample may become less important with the increase in data collected by the Belle II experiment, \babar also has the largest $\Upsilon(3S)$ data sample collected which will still be important in the future. As far as \babar is aware, there is currently no plan to collect data at the $\Upsilon(3S)$ resonance by Belle II.

\begin{figure} [hhh]
    \centering
    \includegraphics[width=\linewidth]{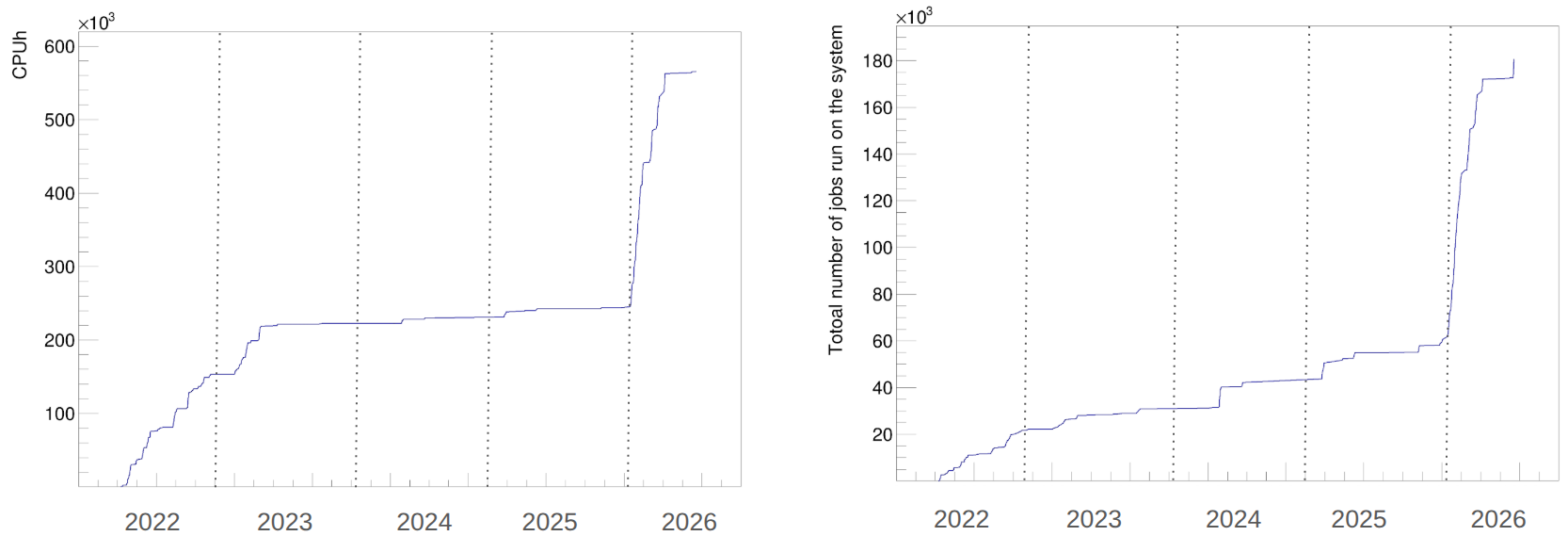}
    \caption{Usage of the new \babar LTDA system since mid 2022; on the left the total CPUh used over time is shown and on the right the total number of jobs that run on the system.}
    \label{fig:babar_LTDA}
\end{figure}

It is expected that the use of the LTDA system is not consistent over time. Analysts usually generate output data within the \babar framework at the beginning of their analysis, which is then further analyzed at their home institutions with ROOT~\cite{ROOT_NIMA_1997}. From time to time, the LTDA system may be needed again for small productions required for additional cross-checks.
The usage of the LTDA system since 2022 is shown in Fig.~\ref{fig:babar_LTDA}. The shown {\it CPUh} are not scaled by {\it HEPScore23} values.

While the LTDA system is in good use, the underlying hardware is old and starts to fail. Except for one new NFS server, all hardware is between 10 and 20 years old.  To take the age and probability of failure into account, raid systems on the machines are used as well as redundant server set up to account for whole machine failures. So far, multiple XRootD proxy machines as well as infrastructure machines have been lost due to hardware failures. Failed infrastructure machines include multiple NFS server as well as the hypervisor that runs the Virtual Machines (VM) that provide the login system, the NIS server, the interactive \babar VM, as well as the XRootD redirector. Due to redundancy on the machine level, so far the LTDA is still operational. However, without funding it is just a matter of time until the system will fully fail. In addition, the collaboration becomes also less active as a whole. Especially older members retire or are not reachable anymore for various reasons, while younger \babar members move on to other experiments and do not necessarily stay in contact.

For the above reasons, a decision for the future of the \babar data access needs to be made. In general, there are only two possibilities available: 
\begin{enumerate}
    \item No data preservation - when the LTDA system becomes inoperable, all of \babar, including collected collider data, analysis framework, and documentation, will be lost,
    \item Real data preservation - find a new archival home for the whole \babar infrastructure with fully open access.
\end{enumerate}

\babar decided to find a way of doing real data preservation which will provide full open access to the data, the analysis framework as well as to the documentation. Already in 2025, the \babar Collaboration Council voted to make the \babar software public, on demand while \babar is looking for a permanent public home for the software. In May 2026, the \babar Collaboration Council also voted unanimously in favor of finding a solution that makes the whole \babar framework public and preserves it.

For a real data preservation, a long term home for the different parts of the framework need to be found. This includes
\begin{itemize}
    \item Data availability,
    \item Documentation,
    \item Software, and
    \item the analysis environment in which the software can run.
\end{itemize}

The CERN Open Data portal already provides a system that gives open access to experiment's data together with some documentation on how to make use of the data \cite{OpenDataPortal}. While the \babar data is already at CERN, it is not available through the Open Data portal at the moment. \babar needs to approach CERN to see if it would be possible to make the already stored data available through the Open Data portal. Similar, the software itself may be hosted in the same system as well, or in a gitlab/github repository.
All of \babar's documentation is based on web servers - some are in plain html pages and some newer documentation lives within a wiki. It should also not be too complicated to find a permanent home for such pages since the infrastructure may already exist at sites.
However, while documentation and software are important to make use of the data, one also needs the environment to analyze all of the data. 

\babar's analysis environment is preserved within a VM image and currently available for \babar members to download and use at home. Such image contains the latest software release and has full access to the data currently stored at GridKa. This image is also used as worker node image for the LTDA batch system. 
While the analysis environment is preserved and the image could be made public, one would not be able to run over all the data in a reasonable time when using such image alone. To be able to really make use of the data, some kind of batch system is needed to analyze the data. In \babar's case, such batch system would need to utilize the preserved VM image as base for the worker nodes. However, being able to analyze the data will be a general problem of all experiments which want to do a real data preservation.

While data preservation was initially understood to have the collected physics data files preserved, it was very early on clear that this is not enough and it also needs to include the software, documentation and in general {\it "comprises all digitally encoded
information which has been created as a result of the planning, running and exploiting of an experiment."}~\cite{DPHEP2025}.
Now that \babar is at a stage where all digital encoded information is somehow preserved and there are possibilities to preserve it in the future as well, it turns out that such preservation alone is not enough. To have a useful preservation, the data also needs to be usable for which an analysis framework within a batch system with open access or similar needs to exists. Without that, the data may be accessible but is not processable. As far as \babar is aware, no such open access system exists currently which will sooner or later also become an issue for other experiments. Some effort and requests for it, across experiments and institutions, would be needed to make such a system a workable reality and data preservation useful.

\subsection{Data preservation in the H1 experiment at HERA}
{\small
 \it Authors: Daniel Britzger (Max-Planck-Institut für Physik München); Stefan Schmitt (Deutsches Elektronen-Synchrotron (DE)); Zhiqing Philippe Zhang (IJCLab, Orsay (FR))
}


HERA at DESY remains the only high-energy electron-- and positron--proton collider operated so far. Between 1992 and 2007, the H1 experiment recorded a unique data set of lepton--proton collisions at center-of-mass energies of up to 319\,GeV~\cite{H1:1996jzy,H1:1996prr}. The preservation
and continued analysis of these data are therefore of lasting scientific interest, in particular in view of renewed activity in lepton--hadron physics and future facilities such as the EIC.
In addition, the H1 data provide a valuable testbed for the development and validation of modern machine-learning techniques, such as ML-based unfolding and full-event unfolding.

The H1 Collaboration remains open to new members interested in analysing the preserved data set. Scientific exchange and coordination of ongoing work are facilitated through regular monthly analysis meetings.

The H1 data preservation model follows the DPHEP level-4 strategy, aiming to retain the full capability to analyze the data, including access to the collision data and simulated samples, but also to maintain the simulation, reconstruction and analysis software, all databases, calibration and detector information, and the corresponding documentation.
The long-term availability of documentation is therefore a central element
of the preservation effort. 
The H1 web server at DESY provides access to the collaboration-internal and public material, including more than 12\,000 digital notes and documents, about 4000 internal presentations, and archived web pages. 
In addition, non-digital documentation is kept in the DESY library archive, and efforts are still ongoing to scan these documents and incorporate them into the H1 digital archive.

The actual data volume is modest compared to that of present LHC experiments: the complete RAW data set amounts to about 75~TB, the compressed DST data to about 20~TB, and the analysis-level data to about 4~TB. The H1 software stack, developed over several decades, consists of core reconstruction and simulation packages, largely written in
\texttt{Fortran}, and the object-oriented analysis framework
\texttt{H1oo}~\cite{Steder:2011zz}, written in \texttt{C++} and based on
\texttt{ROOT}. An Oracle database is used to store run-dependent details, detector geometry, monitoring, calibration information, etc.

Since the end of data taking, H1 has operated in a preserved analysis mode, and published the first paper in the developed data preservation mode in 2016~\cite{H1:2016goa}. 
DESY continues to act as host laboratory, while the collaboration is organized under a renewed agreement based on individual membership. 
Access to data, software and internal documentation is granted through DESY
computing accounts. The resources required by H1 are comparatively small, and the experiment is well integrated into the common DESY analysis infrastructure, in particular the NAF cluster. This allows H1 to profit from centrally maintained storage, batch and interactive computing, while imposing only a limited additional support load.

To ensure the continued usability of the data in this environment, the preserved analysis environment has undergone several migrations since 2012. 
In particular, the complete software stack was moved to 64-bit Linux systems around 2012 and subsequently to newer DESY-supported platforms. 
A further major modernization of the software environment was carried out in recent years, as described in Ref.~\cite{Britzger:2021xcx}.
This included the transition to CentOS~7 and more recently also to AlmaLinux~9. The \texttt{H1oo} analysis framework has been migrated to \texttt{ROOT}~6 and supports modern \texttt{C++} standards, including \texttt{C++20}.
The migration also integrated selected external dependencies from the LCG software stack~\cite{Roiser_2010}. All software packages are now maintained in \texttt{git}. 
%
The preserved framework retains the ability to perform complete Monte Carlo production, including full detector simulation and run-dependent conditions.



Overall, the status of H1 data preservation is stable. While there have
been no major structural changes since the start of the preservation
programme, continuous maintenance and periodic migrations remain essential
to its central goal: ensuring that the H1 data can continue to be analysed
with validated software on supported computing platforms.
The maintained environment provides the basis for new physics measurements
with the preserved HERA data, as demonstrated by recent H1 results
\cite{H1:2024nde,H1:2024aze,H1:2024pvu}. It also enables the use of modern
analysis methods, including machine-learning-based reconstruction and
unfolding techniques, as explored for example in
Refs.~\cite{H1:2021wkz,Arratia:2021tsq,H1:2023fzk,H1:2024mox}.

Recent developments and ongoing tasks include ensuring compliance with modern data-protection requirements, in particular the EU General Data Protection Regulation (GDPR), where personal data are involved, as well as with DESY IT security policies. 
In addition, applications of automated analysis workflows using coding agents are being explored. To support access to the preserved code and documentation, a retrieval-augmented generation system based on large language models, referred to as \texttt{ChatH1}, is currently under development.

\subsection{MINERvA neutrino experiment open data}

{\small
 \it Author: Richard William Gran (University of Minnesota Duluth (US))
}

The MINERvA experiment specializes in studying neutrino interactions and producing neutrino cross section measurements.  We took data from 2009 to 2019 with a scintillator tracker detector where CH is also the target material for the neutrino interactions.  The experiment also had a tracking calorimeter system, a downstream muon spectrometer, and embedded passive targets He, C, H2O, Fe, Pb that span the entire range of nucleus size.  The detectors were at Fermilab in the NuMI neutrino beam, the same beam as the MINOS and NOvA neutrino oscillation experiments.  As of this writing, MINERvA is the neutrino experiment with the first or second most (tied?) published physics measurements.

In contrast to oscillation experiments, we are a dedicated neutrino interaction experiment.   Our goal is to solve systematic shortcomings in the models for neutrino interaction rates and final state spectra.   Without the progress from these efforts, interaction systematic uncertainties would become the limiting factor in future, high precision neutrino oscillation experiments.  The scientific scope includes both particle and nuclear physics, GeV scale cross sections, A-dependence, and MeV scale effects.   Our dataset reflects a generational advance in statistical power and systematic (especially flux) precision for a program in the few-GeV neutrino energy range.  Some aspects of the MINERvA experiment’s data have no equivalent in the international experimental neutrino program future as currently prioritized.

Because of the importance to current and future experiments, and with encouragement of the U.S. funding agencies Department of Energy (and Fermilab) and the National Science Foundation, we have opened our data set for community use.  We announced the availability to the most engaged community at the NuInt conference in Mainz, Germany in October 2025.  The rest of this summary describes the parameters of the open data product and some ways our decisions are unique\footnote{\url{https://minerva.fnal.gov/opendata}}.

The landing page is at minerva.fnal.gov/opendata and we now have a DOI: 10.15484/3022562  \cite{MINERvA:2025cfj}.   The data are available as flat ROOT trees containing branches of calibrated and reconstructed event details that support all current, a few recently past, and a few prospective analyses.  In rough numbers, the statistics of the sample are highlighted by 20 million events with a muon in the final state.  The data also include events with an electron in the final state, neutral current events with electromagnetic showers, and events where a neutrino scatters from atomic electrons,  Fully simulated events are available with statistics of at least four times the data itself and include generator and GEANT4 truth matching information.  

These primary files are currently being hosted by Fermilab and can be accessed through xrootd methods, including xrdcp for downloads.  This solution works in our case because we are the first neutrino experiment with a data release of this type and scale, because our primary files consume only 20 TB on disk, and because we are no longer accumulating or processing additional samples.  Notably, this disk footprint is small enough to download and fits on modest desktop workstations at someone’s home institution.  Fermilab is designing a platform to host ours and future data, so we anticipate an additional (or replacement) interface may become available in a few years. 

We have also released our software framework.   Technically this is not needed because the files are flat ROOT trees.  On the other hand, the software framework encodes a number of adjustments that are applied to the data or simulated data at analysis time that would be required to reproduce or extend one of our analyses.  The software framework is currently hosted on github and a 2 GB set of parameter input files is available using the same xrootd copy service.  Most of the software was public already, along with code for many specific analyses, and provide examples of workflow and techniques described in our publications.   It runs on current versions of Linux (e.g. AlmaLinux 9) and MacOS, and by extension it runs on many modern systems.

The MINERvA experiment will co-exist with its open data for several years.   As of the DPHEP 5th workshop, we have fifteen publications in our pipeline with twelve active and supported graduate students and post docs.  We have an additional list of analyses we think have scientific merit but we do not anticipate taking on ourselves.  That said, the effort on MINERvA will rapidly decline in the next three years. 

Because we are already seven years past the end of operations, we decided it was efficient to release all the data at once, rather than release it piecewise while we finish our own publications.  Perhaps the overlap of the open data with the final few years of the experimental collaboration will make it easier for us to assign collaborators to give technical support and encouragement to the most interesting groups seeking to use our data.

In addition to the primary data and simulated data files, we have additional files that may or may not be necessary for some use cases.   Some are special simulated samples that provide higher statistics for otherwise rare types of events, or are samples that were not part of the original production but were needed for a later analysis.  An additional set of files contains hit level (single energy deposit time, location) information and supports our html5 compliant event display called Arachne \cite{MINERvA:2011bff}.  Currently the Arachne server is behind authentication at Fermilab, but we hope to find another host who can make it generally accessible.  We have other files in hdf5 format that contain similar information that we use to support a machine learning event vertex (interaction point) locator.   These last two may have additional uses beyond analysis with the primary files with high-level reconstructed objects.  When these additional files are include, the total size is about 100 TB.   Thanks to the very small neutrino interaction cross section, this is still tiny compared to LHC data.

We plan that our open data will be used primarily for scientific analysis, furthering our mission to produce cross sections and drive improvements in modeling neutrino interactions.  Main use-cases include cross-checking something that is seen in the analysis of current and future oscillation experiments, doing forward-folding rather than unfolding, and fitting model variations directly to reconstructed data.  Some uses could simply be expanding our analyses to additional channels, special regions of kinematic space, or to higher dimensionality.  In this era of expansive machine learning and AI/LLM techniques, there may be exceptional opportunities to harvest measurements from the data or test the robustness of new techniques using a well characterized and high statistics data set.   

One thing that is subdominant in our thinking of applications is traditional outreach.   We already have a robust Particle Physics Master Class with activities originally created through a U.S. NSF grant in an early phase of the experiment.  It is more than adequate for one-day set of activities, both at the high school level but also for an advanced particle physics course.  Our open data could be used for longer-format projects in the context of a course or an undergraduate senior thesis.  This could especially support groups who would not otherwise be able to engage directly (e.g. due to geography or limited funding) with an active experiment at laboratories like Fermilab or CERN.

As of this writing, we have run two tutorials, one at Fermilab the same week of the DPHEP 5th workshop, and one in June 2026 at CERN, both with remote video connections.   They are based on a long-standing MINERvA 101 tutorial taken by nearly two decades of MINERvA students.  The tutorial has been updated to conform to the new audience and data files, then updated again with lessons learned from interacting with users through the tutorials.  These materials are also preserved as part of the github repository that hosts our software framework.

We have not chosen to preserve our simulation and processing workflow.   Currently it is done using ScientificLinux 7 containers.   Porting that layer of processing out to the current and next supported versions of Linux is more effort than MINERvA has.  At best, we will save a snapshot that could potentially be reverse engineered, though missing the muon spectrometer component that was run by the MINOS experiment.  One consequence of this is that new simulated data can not be obtained, not with an updated or alternate neutrino event generator nor with an updated version of Geant4.  We have extensive experience using reweighting techniques, most illustrated in our software framework release, enough to satisfy many of the physics possibilities someone may want for an analysis.

Related to reweighting, our version 2.12 of the GENIE neutrino event generator is needed for comparison to any new model developments.   The predictions of this generator configuration are of course already in the simulated event files.   The original generator release is no longer available for download from the original GENIE collaboration authors.  However, we have a patched version that includes both MINERvA specific modifications described in our publications and modifications that allow building it on current and near future Linux platforms such as AlmaLinux 9.  Upon request, we can supply that to persons interested in generating high statistics generator-level samples to compare with other generators, perhaps to make a direct comparison or ML-assisted translation like \cite{MINERvA:2025lya}.  We have started discussions with the GENIE generator authors about their support for a public, static release, then will inquire about external dependencies that also need patches.

Finally, the collaboration has extensively discussed the terms of use of our open data products.  We have purposely chosen to have as few barriers to use as possible.  We have a suggested acknowledgment that can be cut and pasted into a publication and citations that acknowledge MINERvA and also allow us to track the use of the data.   We do not require MINERvA review of non-MINERvA analyses using these data.  We suggest that limited but generous support remains available during the time we have active graduate students and postdocs.  Another asset for the community is a huge base of alumni still active in particle physics.   One model we think could be productive is to recruit a current or former MINERvA member to collaborate and be a co-author on a paper using the open data.  The collaboration decided such a step does not automatically require the inclusion of the entire MINERvA collaboration as co-authors as we do for our own publications.  Our official policy at this time is that the current co-spokespersons take questions of support, acknowledgement, or authorship, and will investigate who should best address the situation and how.

In conclusion, the MINERvA experiment is pleased to be leading neutrino experiments with a full data and software release.   This is the culmination of an effort started about four years earlier that involved a large fraction of the collaboration.  We are looking forward to seeing it gain use in expected and novel ways as the next generation of precision neutrino experiments come online and most need it.

\subsection{Status of Open and Preserved Data in ATLAS}
{\small
 \it Author: Zach Marschall (LBNL)
}

Since the last report from the DPHEP Collaboration~\cite{DPHEP:2023blx}, the ATLAS Collaboration has made considerable strides in its Open Data programme~\cite{ATLASOpenData}. The first Open Data for research were released in July 2024, comprising 36~fb$^{-1}$ of $\sqrt{s}=13$~TeV proton--proton collision data in the PHYSLITE format, a light-weight, pre-calibrated format used for data analysis inside the collaboration. These data were simplified and used to create a new beta release of the Open Data for education and outreach. Late in 2024, the first heavy ion collision Open Data were released, including 486~$\mu$b$^{-1}$ of minimum bias data at 5~TeV per nucleon centre-of-mass energy in a HION14 data format that was specially designed to support track-based data analyses. Most recently, more than 12B events of event generator output over more than 6000 datasets have been released, covering both $\sqrt{s}=13$~TeV and 13.6~TeV, in the HEPMC text format. Several other bespoke datasets have been released as well, primarily focusing on AI/ML applications. All these data are released under a CC0 license~\cite{CC0} on the CERN Open Data Portal~\cite{OpenDataPortal}, allowing free use, with a request for citation.

Because of the complexity of these data releases, a new Python package, \texttt{atlasopenmagic}~\cite{atlasopenmagic}, has been developed. This package provides convenient access to all the metadata for all samples across all releases, including cross sections, generator version descriptions, and process information. The package provides a variety of search functions to help identify samples of interest, which is more extensive than the functionality available directly in the Open Data Portal. Users can also get lists of file locations, with support for a variety of access methods and back-ends depending on the tools and accounts available to them.

Alongside these many new data releases, the documentation has been significantly expanded and improved. The goal of the Open Data effort is to provide a broad spectrum of documented learning materials that serve a wide variety of audiences, from high school students and interested members of the public who don’t know how to write software, to HEP experimentalists and phenomenologists hoping to develop high-quality research publications. A ``classroom application'' was developed~\cite{ATLASClassroomApp} that is entirely web-based and targets introductory audiences. The spectrum of ``histogram analyzer'' applications, which are also no-code, but are somewhat more complex than the classroom application, was also expanded to include machine learning and dark matter applications. 

With the new release of Open Data for education and outreach, the available analysis examples have grown enormously, particularly via Jupyter notebook offerings. For each new release that includes a new data format, an introductory notebook has been constructed to demonstrate simple data access and explain the features of the format. A new category of notebook tutorials called ``Concepts’’ has been introduced, with advanced-undergraduate-level introductions to a variety of concepts in particle physics like statistics, systematic uncertainties, acceptance and efficiency, and non-collision backgrounds. The number of Standard Model analyses has also grown to cover several different Higgs boson analyses and new final states. These are complemented by a refresh of the C++-based analysis framework and a new interactive in-browser event display based on Phoenix~\cite{Phoenix}. The Python notebooks have been improved to support Binder, Colab, and SWAN, as well as local running options, to ensure maximum accessibility for audiences worldwide.

\begin{figure}
    \centering
    \includegraphics[width=0.9\linewidth]{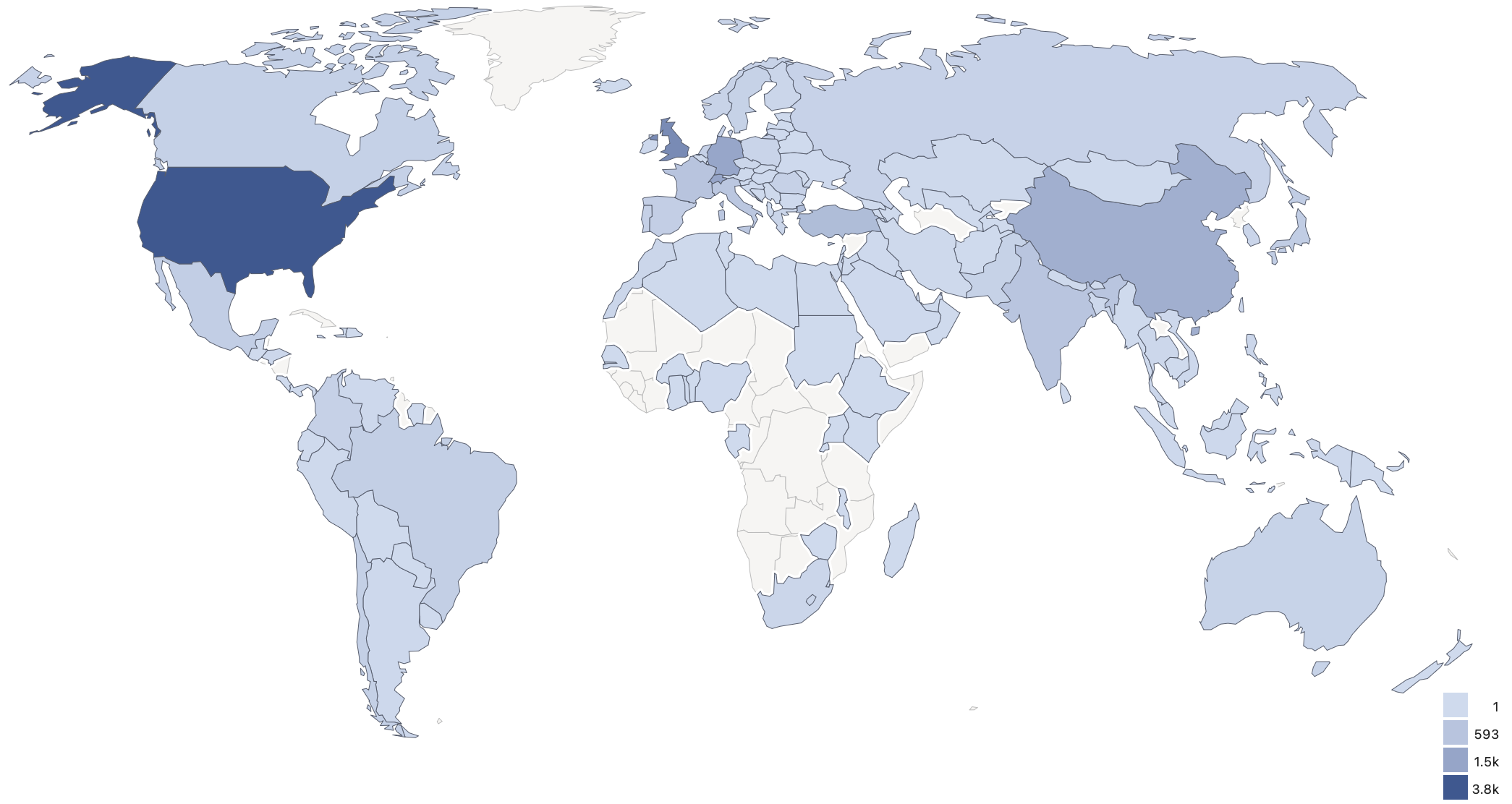}
    \caption{Geolocation of visitors to the ATLAS Open Data website between June 2, 2025 and January 23, 2026. These 20,000 visits included 10,000 unique visitors and a 5-minute average visit time, with only 12\% of visitors leaving without taking some action.}
    \label{fig:section_2.9:ATLASWebVisits}
\end{figure}

\begin{figure}
    \centering
    \includegraphics[width=0.8\linewidth]{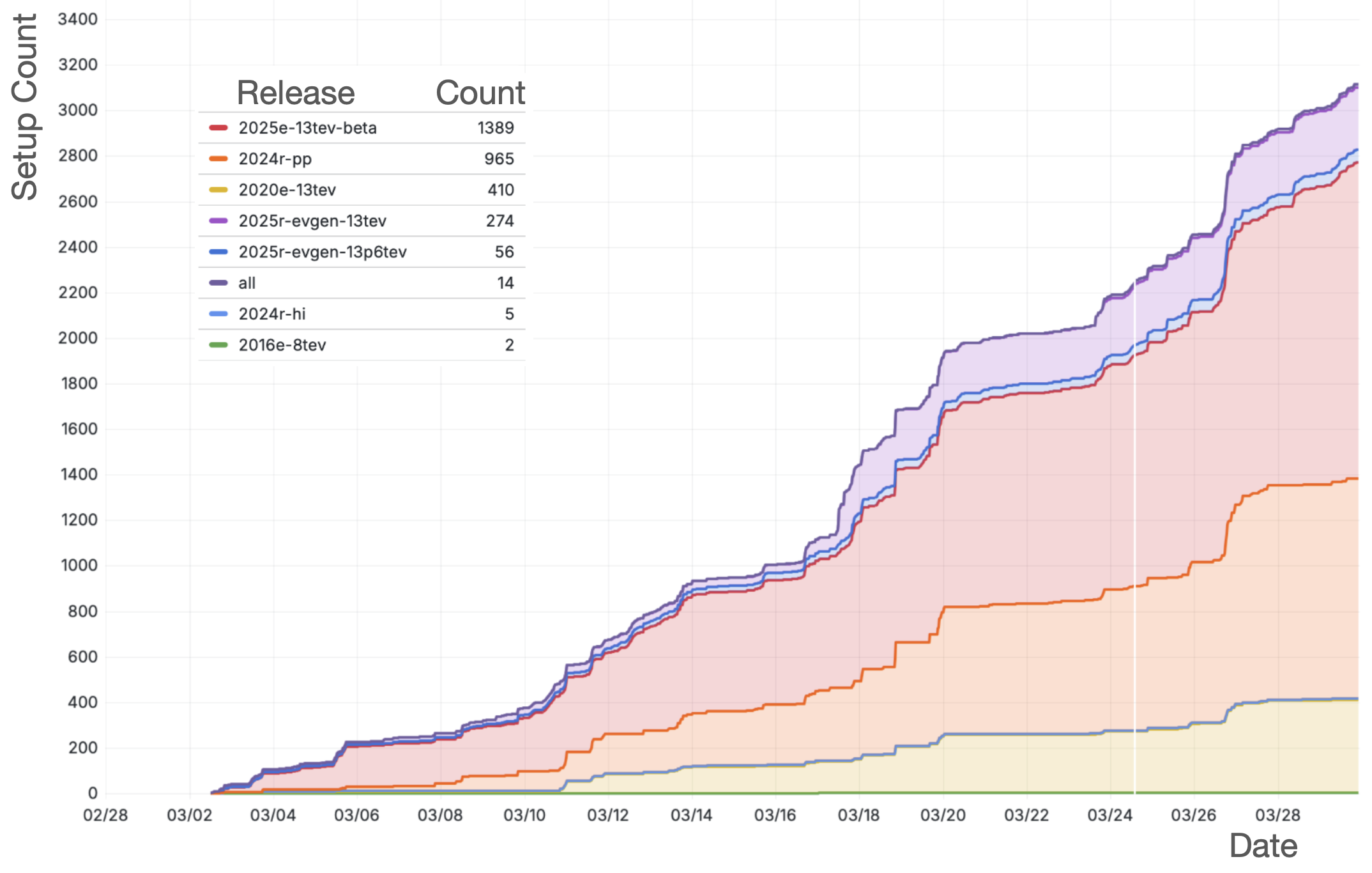}
    \caption{Popularity of the ATLAS Open Data releases based on monitoring from the \texttt{atlasopenmagic} metadata API during March 2026.}
    \label{fig:section_2.9:ATOMReleasePopularity}
\end{figure}

The collaboration has also expanded its monitoring and outreach to identify use of the Open Data around the world. A variety of courses have been developed in the US, Sweden, UK, Australia, Switzerland, and Greece, which support students at various levels. With these courses centrally cataloged and their material made available, it is more straightforward for other universities to adapt courses to their needs, rather than having to develop something new from scratch. Similarly, several projects have been gathered that form examples for others hoping to use the Open Data for research purposes. Monitoring of the ATLAS Open Data website, the Open Data portal access, and the metadata API provided by \texttt{atlasopenmagic} has provided considerable insights into the worldwide use of the Open Data. The Open Data are used on six continents, with large spikes in usage corresponding to both individual users' projects and tutorials that encourage interactive usage. Figure~\ref{fig:section_2.9:ATLASWebVisits} shows the visitors to the ATLAS Open Data website over the first eight months of monitoring. The metadata API monitoring, shown in Figure~\ref{fig:section_2.9:ATOMReleasePopularity}, makes clear that the education and outreach Open Data is the most popular offering, followed by the research Open Data. Feedback forms have been added to the Jupyter notebooks and web applications that indicate a very high level of satisfaction (4.9/5) and the appropriate level of the material (2.8/5, with 1 being ``too easy'' and 5 being ``too difficult'').

In November 2025, the first CERN-based ATLAS Open Data tutorial for the general public was held~\cite{ATLASODTutorial}. The tutorial provided an opportunity to further refresh and improve documentation. This first tutorial was divided into two ``education'' days and two ``research'' days, but made no attempt to target a specific audience based on particular learning goals. That resulted in a mixture of ATLAS members learning about the Open Data for use in the classroom, students learning about the Open Data to begin research projects, and experienced HEP physicists learning about the Open Data for their research work. This wide variety of goals made it particularly difficult to satisfy all the expectations of the attendees. Nevertheless, overall satisfaction with the tutorial was rated very high (4.7/5), and generally the difficulty of the material was rated appropriate, with the most complex sessions showing a wider variety of satisfaction and difficulty ratings. More targeted tutorials are anticipated as the next data offerings are released in the coming year.

Within the collaboration, several new policies were approved that have implications for software and data preservation. The new software policy~\cite{ATLASSoftwarePolicy} is more strongly open-source, recommending DOIs, mandating that all software be visible within the collaboration, and recommending that software be publicly available by default, with closed software being allowed with motivated exceptions (e.g.~for ongoing analyses). The collaboration documentation policy suggests that documentation be public by default as well, which will imply less duplication of documentation effort to serve the Open Data community's needs. Finally, the data preservation policy~\cite{ATLASDataPolicy} was extended and clarified significantly. The collaboration remains committed to preserving all raw data (level 4 data). However, there is no active support for the processing of Run 1 data. Further, support for the Run 2 and Run 3 data will be dropped in the coming HL-LHC era of Run 4, owing to the immense effort required to re-calibrate the data following any major reprocessing campaign. For physics analyses, this implies that statistical combinations will be possible, but the only data formats retained will be the same PHYSLITE formats released for use as Open Data. The production of new Monte Carlo simulation for Run 2 and Run 3 analyses will be possible through containerized workflows. This will allow considerable simplification of the experiment software, removing large amounts of legacy code required to support older detector configurations.

The ATLAS Collaboration remains committed to providing Level 1 data preservation information via HepData~\cite{HEPData} and Rivet~\cite{RivetAnalysis} analyses, for example, and has continued to provide data for publications. These data remain a valuable resource for event generator tuning efforts and phenomenology studies. The uptake of true ``data preservation'' via Reana and Recast remains limited. However, recently several groups have demonstrated re-implementations of analyses with the Open Data using agentic AI~\cite{ATLASAIOD}. This offers new hope in the analysis preservation space: if an agent can produce a preserved workflow by copying a paper, validating its results against the publication and perhaps against internal documentation, and the analysis team needs only cross-check the results and code, the threshold for preservation will be significantly lowered. This could also provide a path towards significantly more rapid updates of preserved analyses to be used on new datasets, which today requires substantial effort, often entirely from scratch.

Overall, the ATLAS Collaboration has made great strides on the Open Data, data preservation, and analysis preservation frontiers. New policies support the Open Science goals, and several new Open Data releases with expansive documentation serve a wide variety of audiences and of knowledge and skill levels. These Open Data releases have been very well received by audiences both in person at tutorials and asynchronously.

\subsection{LHCb Data Preservation and Open Data}
{\small
 \it Authors: Dillon S. Fitzgerald (University of Michigan (US)); Piet Nogga (University of Bonn (DE)); Tibor \v{S}imko (CERN); Cameron Duncan McClymont (CERN); Michael Buchar (CERN); Chris Burr (CERN); Ryunosuke O'Neil (CERN)
}

\begin{figure}[H]
    \centering
    \includegraphics[width=0.8\linewidth]{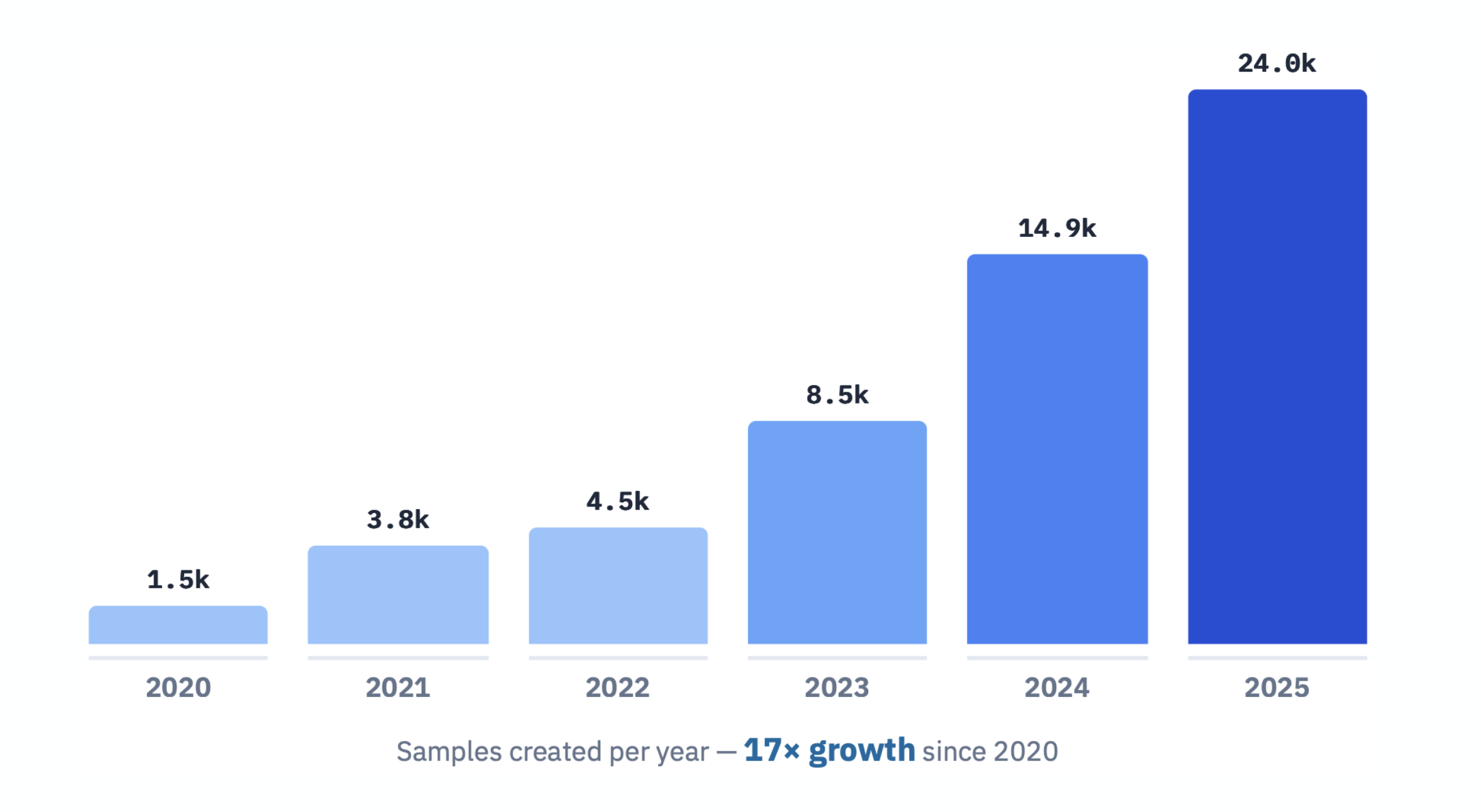}
    \caption{Number of samples per year produced with the Analysis Productions batch processing system, showing a 17 times growth since 2020.}
    \label{fig:section_2.10:AP_growth}
\end{figure}

The LHCb collaboration has accumulated over 800 scientific publications, making it increasingly important to preserve analysis workflows to facilitate both reusability and reinterpretation of the results and collected data. Data and analysis preservation are important steps towards producing publications at LHCb, and significant progress has been made in this area since the last DPHEP report~\cite{DPHEP:2023blx}, with Run 3 ntuple production for user analysis centralized in a way that preserves the data provenance, extensive use of workflow management systems like Snakemake, and requirements to upload published results to HEPData. The centralized batch processing system for producing ntuples is known as Analysis Productions~\cite{analysis_productions}, which has experienced a substantial uptake in usage following collection of the Run 3 data, with 0.5 exabyte per year throughput, and the number of samples produced increasing by a factor of 17 since 2020, as shown in Figure~\ref{fig:section_2.10:AP_growth}. The Analysis Productions system is used for ntuple production of both data and MC. The system also includes a tool known as \texttt{apd} (analysis productions data) useful for tagging datasets and facilitating queries of the data based on meaningful tags~\cite{lhcb_apd}.

Such valuable and complex data merits careful thought on how to preserve and provide open access to a broader community of researchers. Following the previous DPHEP report~\cite{DPHEP:2023blx}, LHCb has significantly improved the user experience with LHCb open data, with substantial improvements to tools for data access and exploration, as well as new and improved documentation~\cite{lhcb_opendata_guide}.

\begin{figure}[H]
    \centering
    \includegraphics[width=0.8\linewidth]{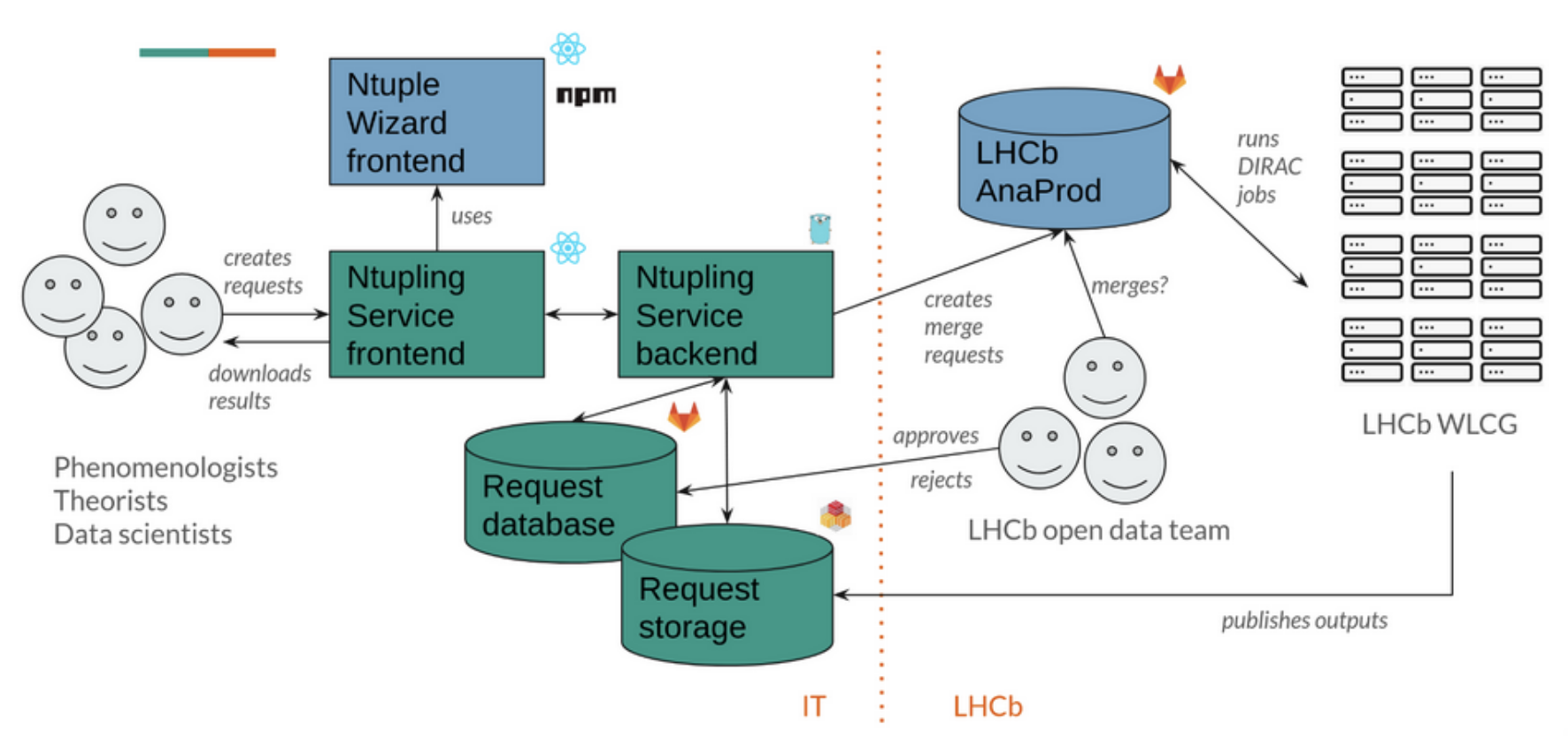}
    \caption{Schematic diagram of the LHCb Ntupling Service architecture.}
    \label{fig:section_2.10:ntupling_service_architecture}
\end{figure}

 In accordance with the CERN Open Data Policy, LHCb announced the release of the full Run 1 dataset gathered from proton-proton collisions, amounting to approximately 800 terabytes made public on the CERN Open Data portal (ODP)~\cite{OpenDataPortal} in 2023. However, the sheer volume of the Run 2 data and beyond makes the direct release approach used for Run 1 unfeasible, necessitating a fundementally different mechanism for providing open access. To address this, LHCb developed the LHCb Ntupling Service~\cite{LHCb-ntupleService, ntupling_service_proceedings} in collaboration with the CERN Department of Information Technology, the flagship open data platform from LHCb for on-demand production and publishing of LHCb open data. The service consists of the web interface frontend allowing users to create and review ntuple production requests, the backend application processing the user requests and storing them in the GitLab repositories (offering vetting capabilities to the LHCb Open Data team), and automatic dispatch of user requests to the LHCb Analysis Productions system after approval. The produced ntuples are then collected and delivered back to the users in the frontend web interface. Details of the LHCb Ntupling Service architecture can be seen in Figure~\ref{fig:section_2.10:ntupling_service_architecture}. A key feature is that the procedure of requesting and subsequently analyzing an ntuple requires no specific knowledge of the LHCb software stack.

The official release of the LHCb Ntupling Service was announced in February 2026. It is accessible directly through the CERN Open Data Portal, and provides public access to both Run 1, and for the first time, Run 2 $pp$ data collected at LHCb, amounting to over 4 PB of data to explore. At the time of writing, about 20 requests have come from a collection of theorists / phenomenologists looking into topics like exotic hadron searches and CP asymmetry, users testing fitting algorithms on mass spectra, both LHCb and non-LHCb members working with high school students, and others looking to learn more about high energy physics.

The Ntupling Service promotes flexibility in terms of storage requirements on \texttt{eospublic}, as we do not promise permanent storage of the produced ntuples, and provide the user a mechanism for batch downloading of the produced dataset. Continued usage of the service will inform policies on how long ntuples will be stored before deletion, but right now the data volume of the produced ntuples is small ($O$(1 TB)). Additionally, a mechanism for promoting certain produced ntuples to permenant records on the CERN Open Data portal is under development. In theory, as more records accumulate, users will have the ability to work with pre-produced ntuples, potentially trading off future CPU usage for the relatively small footprint of storing ntuples. This would also provide the user with a DOI to cite in the case of ntuples promoted to permanent records on the CERN ODP. A solution is under development to provide a general DOI to cite in the case that the open data used to produce a paper was not promoted to a permanent record on the CERN ODP.

Since the previous DPHEP report~\cite{DPHEP:2023blx}, LHCb held its first open data workshop for external participants in October 2024~\cite{LHCb_opendata_workshop}. This was during the development of the LHCb Ntupling Service, and served as a milestone to release a beta version of the application to be tested by workshop attendees. This allowed for the collection of valuable feedback from users to help steer development for the official release in February 2026. Additionally, a second workshop was held at the University of Bonn in 2025, attended mostly by scientists from the Belle collaboration. Feedback from attendees was very positive at both workshops, inspiring engagement between LHCb and the open science community.   

In summary, LHCb has made significant progress regarding data preservation and open data. A massive uptake in the Analysis Productions centralized batch processing system has been observed since Run 3 data taking, which maintains a provenance trace for the production of ntuples from DST files. Substantial milestones have been made in the open data space as well, from holding the first workshop for working with LHCb open data, to providing a new and unique web service for requesting custom ntuples of LHCb open data, the LHCb Ntupling Service. The barrier for entry to work with these data has been lowered such that knowledge of experiment-specific software is no longer a requirement. The official release of the LHCb Ntupling Service was very recent, but initial feedback and community engagement is very positive and encouraging.

\subsection{CMS Data Preservation and Open Access: status and plans}
{\small
 \it Authors: Matthew Bellis (Siena University (US)); Thomas McCauley (University of Notre Dame (US))
}

\begin{figure}[H]
    \centering
    \includegraphics[width=0.8\linewidth]{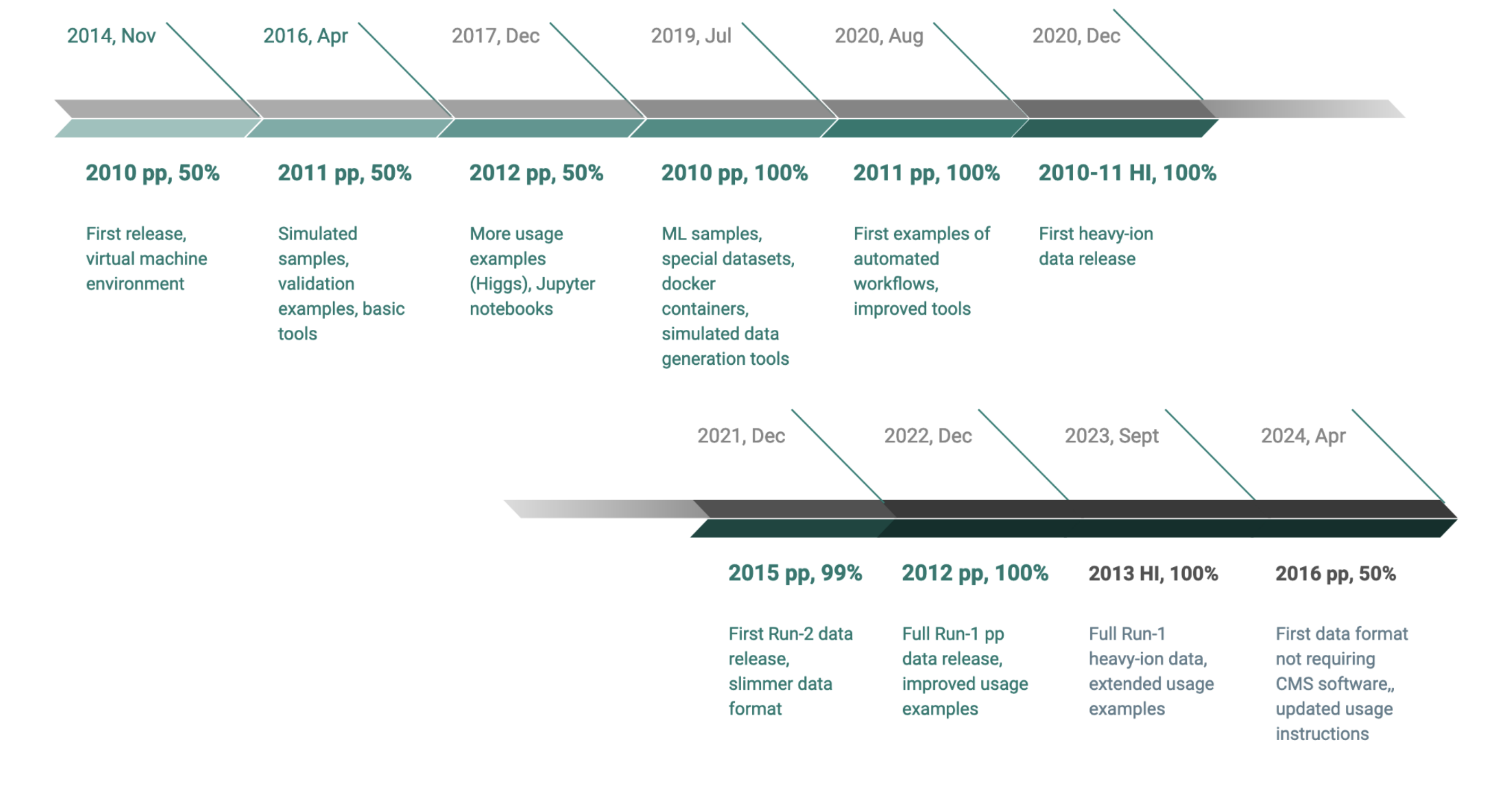}
    \caption{Timeline of CMS Open Data releases}
    \label{fig:section_2.11:cms_release_timeline}
\end{figure}

Since the last DPHEP report~\cite{DPHEP:2023blx} CMS has continued to implement its data preservation, re-use and open access policy~\cite{CMSOpenDataPolicy} (adopted in 2012 and revised in 2020). One aspect of the CMS policy is to release half of the data to the public after a six-year embargo period, with the remaining released after ten years. The first release (the first from an LHC experiment) was in 2014 and since then the CMS Open Data program has matured into a sustained pipeline of data to the public. Since 2012 there have regular releases up to the latest in 2024: proton-proton collision data and simulation from Run 2 of the LHC in 2016. The next release, half of collision data from 2017 (and all of the corresponding simulation), is under preparation. A timeline of the CMS open data releases may be seen in Figure~\ref{fig:section_2.11:cms_release_timeline}. There are currently over 300 collision datasets and over 20k simulated datasets available on the CERN Open Data Portal, a total of over 4 PB (a mixture of direct access on disk and requested access on tape). 

A dataset release does not just contain the datasets themselves but includes metadata, software, containers, and conditions data, without which the data itself would not be usable. These elements are either available from or linked to each dataset record on the CERN Open Data Portal (an example of which may be found here~\cite{CMSOpenDataRecord}).

Analysis of the complex open data from an LHC experiment is not a trivial task, often requiring extensive physics, data analysis, and software 
knowledge. To help provide this knowledge CMS created the CMS Open Data Guide~\cite{CMSOpenDataGuide}. The main goal of this guide is to facilitate the usage of CMS open/legacy data. The sections guide one through the main topics necessary to learn in order to conduct an analysis using CMS Open Data.

In order to help further lower this knowledge barrier CMS has held open data workshops~\cite{CMSOpenDataWorkshops} that provide training. The first, held in 2020, was focused on theorists unfamiliar with data analysis at a large experiment. Since then, CMS has organized five more hybrid and in-person workshops, the latest in 2024 at the CERN IdeaSquare. The next workshop will be held in 2026 at the University of Notre Dame from 28-30 July~\cite{CMSOpenDataWorkshopJuly} and will focus on teachers and university instructors interested in pedagogical applications of CMS Open Data.

\begin{figure}[htbp]
  \centering
  \includegraphics[width=0.45\linewidth]{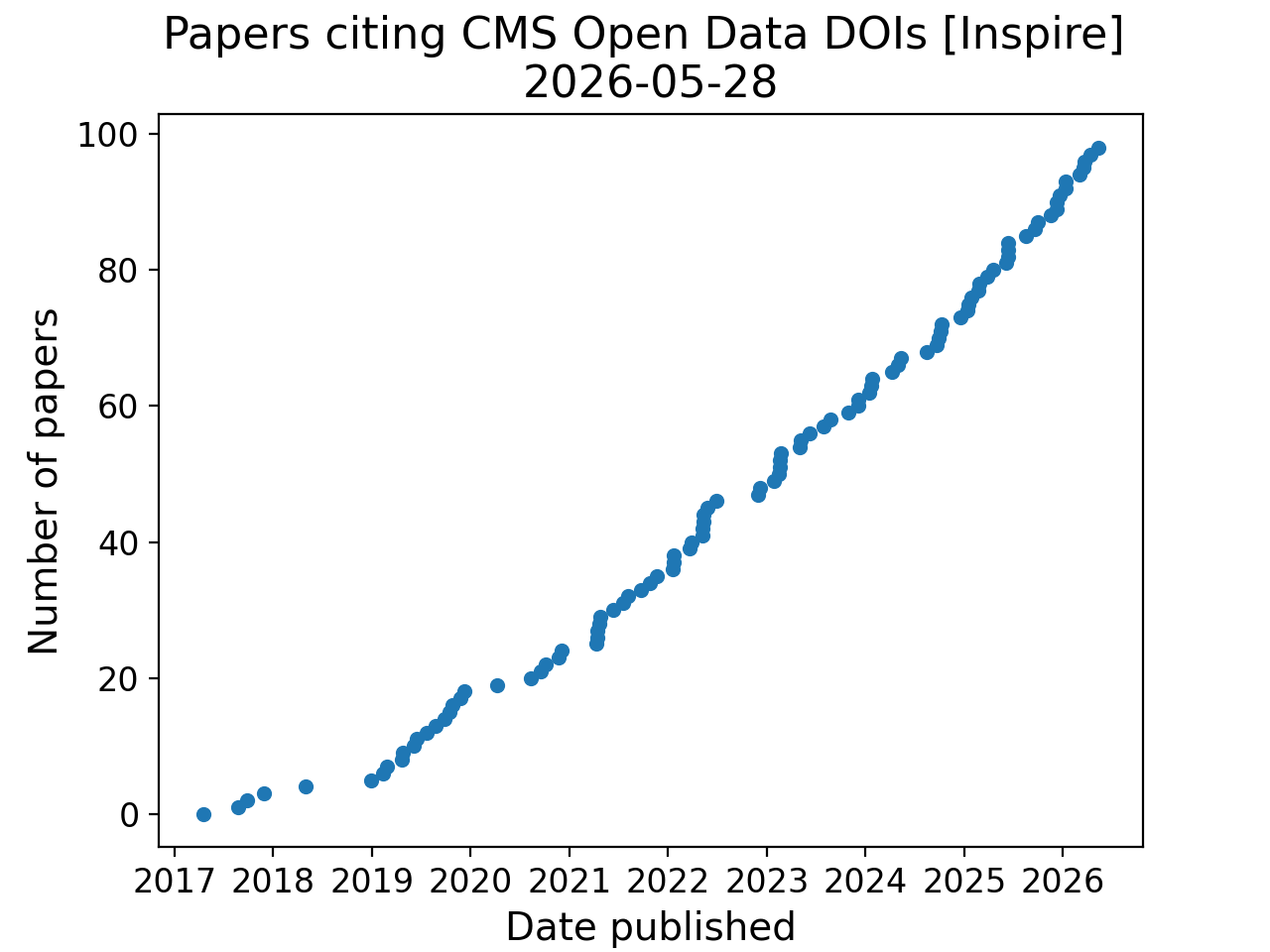}
  \hfill
  \includegraphics[width=0.45\linewidth]{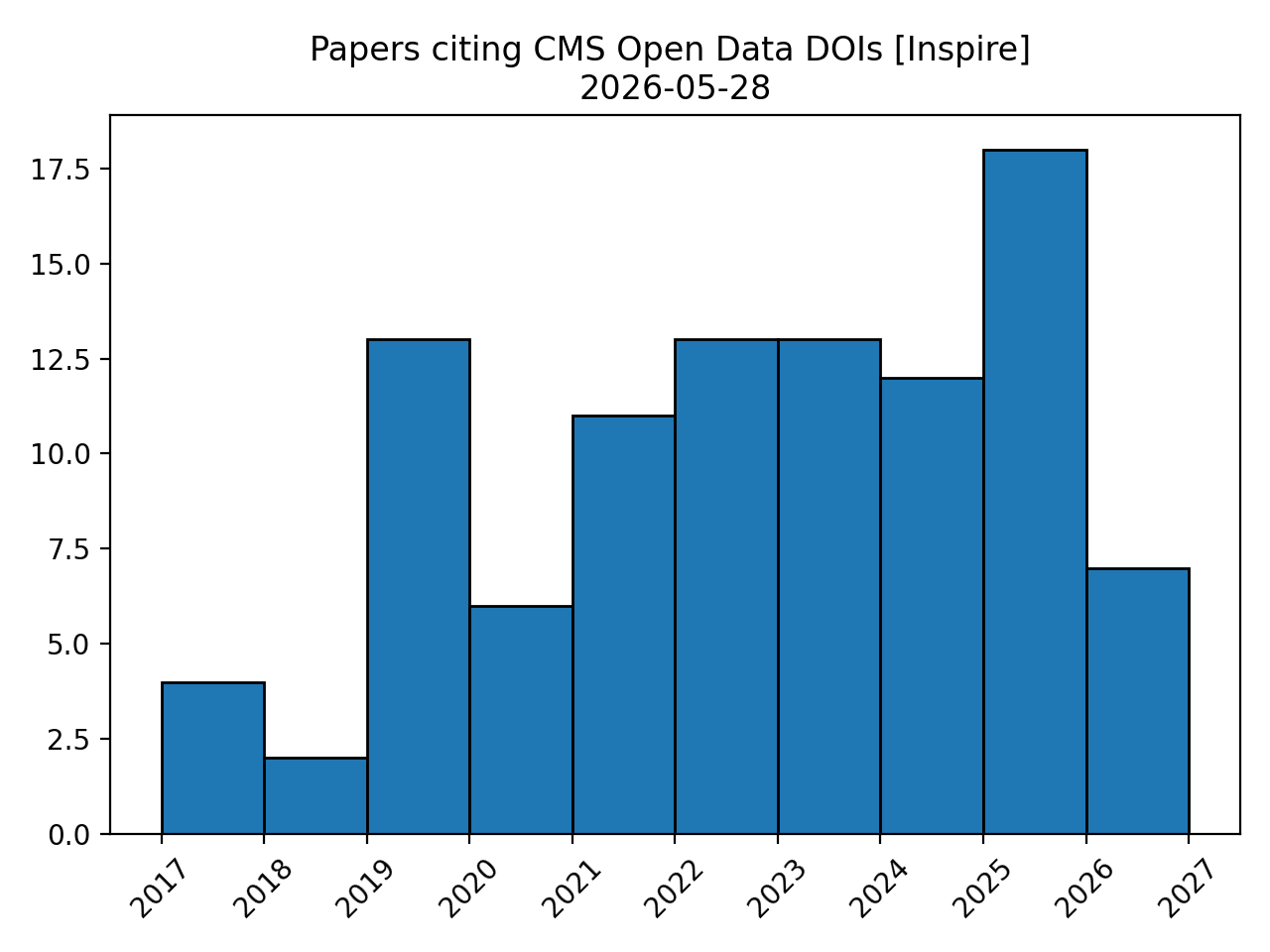}
  \caption{Research publications citing CMS Open Data DOIs over time~\cite{CMSDataUsage}. Papers by the CMS and CERN open data teams and by the CMS Collaboration are excluded.}
  \label{fig:section_2.11:cms_research_pubs}
\end{figure}

Beyond educational uses of CMS Open Data, such as in the International Masterclasses organized by the International Particle Outreach Group (IPPOG) (where the CMS masterclass was developed in collaboration with QuarkNet), CMS Open Data has found use in the larger HEP community which has produced published research results, involving topics such as studies of jets and machine learning applications. The steady production of research results can be found from analysis of papers that cite CMS Open Data DOIs and can be seen in Figure~\ref{fig:section_2.11:cms_research_pubs}.

As of this writing CMS has released all of Run 1 data as open data and is preparing its next release of Run 2 data: half of the proton-proton collision data collected in 2017 along with all of the corresponding Monte Carlo and relevant metadata. A regular program of open data workshops will continue in July of this year with a workshop focusing on pedagogical applications.

\subsection{Long-term data preservation in ALICE: status and plans}
{\small
 \it Authors: David Dobrigkeit Chinellato (Austrian Academy of Sciences (AT)); Stefano Piano (INFN Trieste)
}

\subsubsection{Strategy overview}

A central pillar of the ALICE data preservation philosophy is the ability to fully reproduce published physics results, not merely to store raw data. In compliance with the CERN Open Access Policy, all ALICE publications are available with open access, and data points together with additional information including the analysis code are made public at the time of publication through the HEPData portal and the Rivet toolkit. 

The preservation of analysis reproducibility is tightly coupled with ALICE' s long-standing tradition of organised, centralised analysis on the Grid. ALICE analyses mostly deal with large datasets using the distributed Grid infrastructure, and in Run 1 and Run 2 the collaboration developed a system of analysis trains, the so-called LEGO trains, that allowed users to configure analysis tasks (called wagons) running on the same data, building on the ALICE analysis framework and the Grid submission and monitoring infrastructure. For Run 3, this approach has been continued and extended through the new Hyperloop framework. Hyperloop is a platform to run and manage analysis trains on the Grid, building on the O2 analysis framework. Critically for data preservation purposes, Hyperloop archives all the information necessary to fully repeat a train run, including the software version, configuration parameters, input datasets, and wagon settings, ensuring that any analysis step can be reproduced exactly at a later time. The final step of each analysis is then archived on the ALICE Analysis Note portal, ensuring that the full chain from raw data to publication remains documented and traceable.

A key component of this strategy is the migration of Run 1 and Run 2
datasets to the new AO2D format developed within the ALICE O2 
framework for Runs 3 and 4~\cite{Buncic:2011297}. This 
format significantly reduces 
storage requirements while improving analysis performance, 
providing a sustainable solution for long-term preservation. 
For example, the 2015 Pb–Pb dataset was reduced from approximately 
1.2 PB in the original Run 2 format to about 52 TB in AO2D 
while retaining its general-purpose analysis capabilities.

\subsubsection{Replication and long-term storage}

To guarantee the long-term bit-level integrity of the preserved data, ALICE foresees that one copy of each AO2D dataset will be archived to tape. This tape archive constitutes the backbone of the experiment's long-term preservation guarantee, ensuring that the data remain recoverable over timescales extending well beyond the active lifetime of the experiment. This policy balances storage sustainability with the requirement that the definitive, publication-quality version of each dataset is never lost, complementing the open-data releases on the CERN Open Data Portal with a robust and auditable archival layer.

\subsubsection{Publication of open data}

An important component of the long-term data management
strategy of the ALICE experiment is
the publication of open data. 
In that regard, the open-data programme of ALICE has reached 
a key milestone: following earlier releases of a fraction 
of the 2010 Run 1 datasets, the complete 2015 Pb–Pb dataset (about 
55 TB, corresponding to roughly 5\% of Run 2 data) was 
successfully uploaded to the CERN Open Data Portal in October 
2025. Validation of the new format and analysis framework is 
largely complete, and benchmark analyses have demonstrated 
good agreement with published ALICE results. In continuation, 
ALICE plans to release approximately 105 TB of Run 2 data per 
year between 2026 and 2029, reaching about 50\% of the Run 2 
data volume. Open-data publication of Run 3 data is expected 
to begin around 2030. Given the much larger data volumes 
anticipated from Runs 3 and 4, ALICE is also studying the 
publication of derived and skimmed datasets to ensure the 
long-term sustainability of its preservation programme.

\subsection{Digital preservation and reanalysis of raw photographic data from the CERN 2m bubble chamber}
{\small
 \it Author: Andrew Chisholm (University of Birmingham (GB))
}

\subsubsection{Introduction}

The particle physics group of the University of Birmingham have a rich history of involvement in bubble chamber experiments, dating back to the 1950s. As part of this legacy, the group hold an extensive collection of photographic film, amounting to tens of thousands of frames, recorded by experiments at the CERN $2\,\text{m}$ Hydrogen Bubble Chamber (HBC), which operated between 1965 and 1976~\cite{bubble_2m}. These photographic records of particle interactions in the chamber volume represented the primary raw data format of such experiments, from which particle trajectories and momenta were then reconstructed from careful measurements of the film. The film in the Birmingham collection is remarkably well preserved and generally composed of complete sets of three reels in which individual beam exposures are recorded by all three of the chamber’s cameras. This aspect of the collection is particularly notable in that it facilitates, in principle, the extraction of quantitative, three-dimensional information on charged particle trajectories recorded within the chamber.
Motivated by the clear historical significance and scientific value of the collection, an effort was initiated to digitally preserve the collection of film for posterity and investigate the possibility of reopening its scientific exploitation~\cite{bubble_website}. In the following contribution, the digitisation procedure will be briefly outlined, together with a summary of a recent assessment of the feasibility of extracting quantitative scientific information from the digital images, based on archived technical documents associated with the CERN $2\,\text{m}$ HBC. In the context of data preservation in high energy physics, this study serves as a case study to assess the potential of recovering scientifically useful information in a rather extreme situation, involving an inherently analogue raw data format being re-analysed over 50 years since it was recorded, without input from the original scientific or technical personnel, performed on the basis of archived information alone.   

\subsubsection{Film Digitisation System}

\begin{figure}[h]
    \centering
    \includegraphics[height=0.2\textheight]{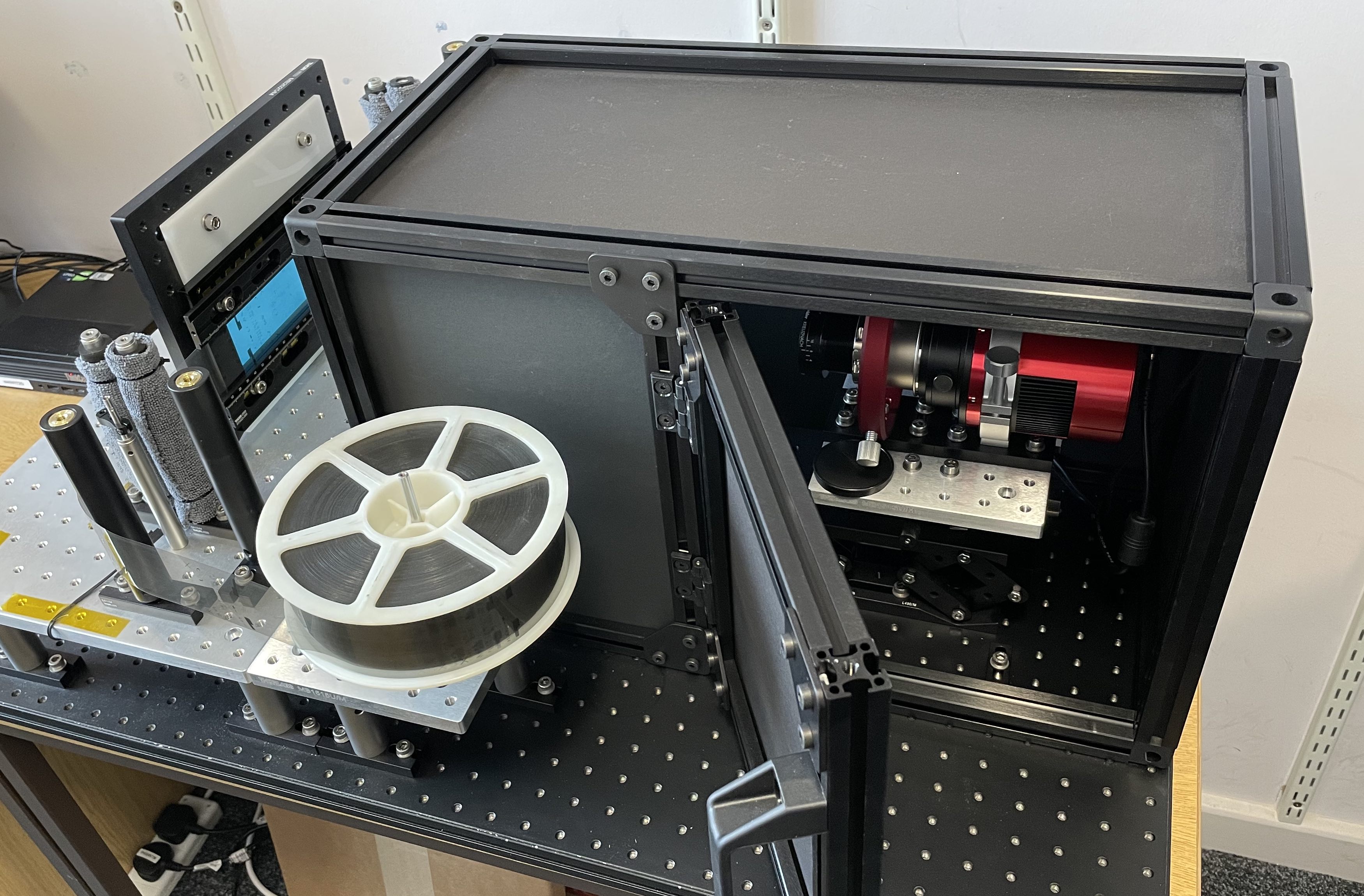}
    \includegraphics[height=0.2\textheight]{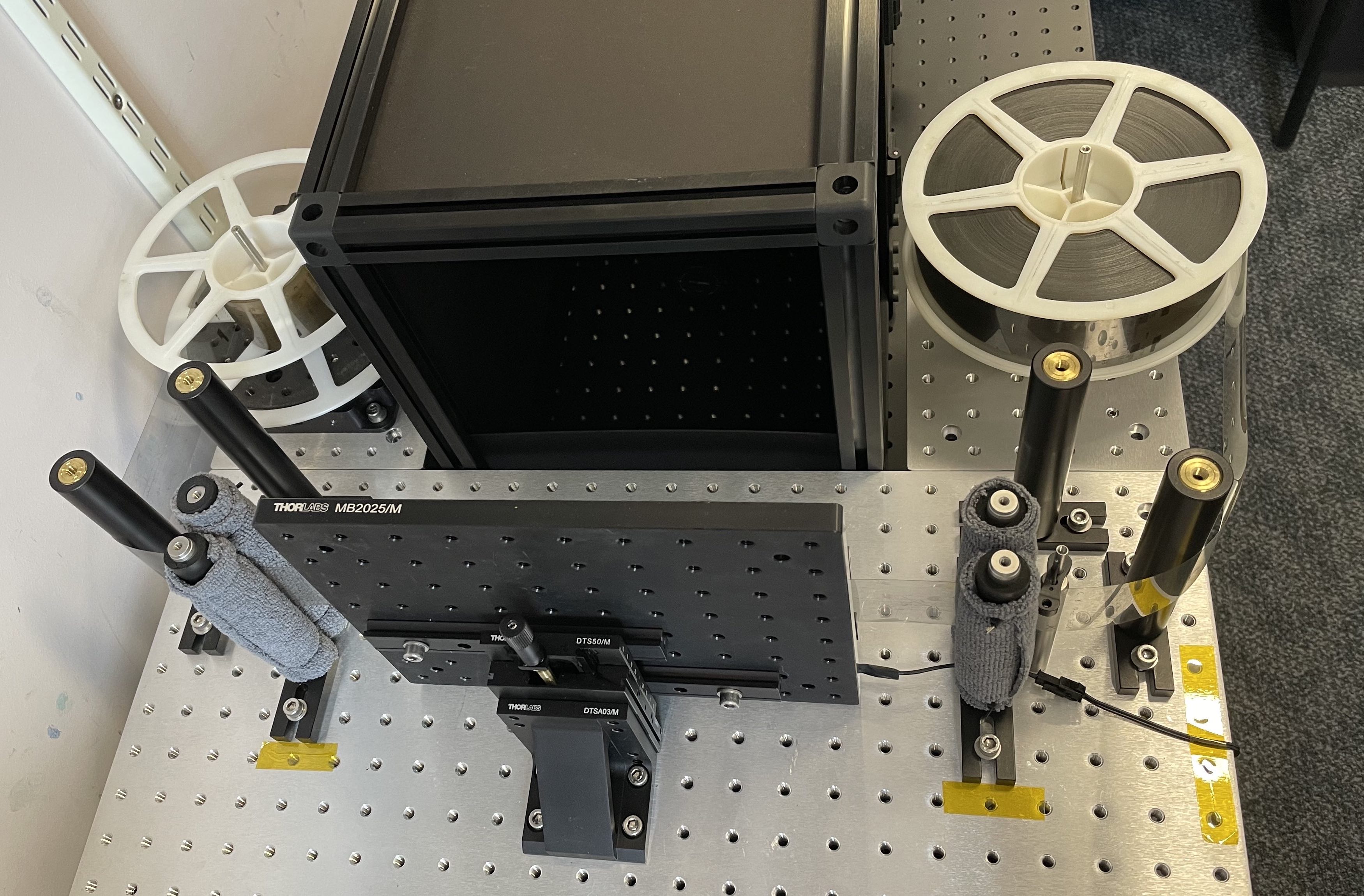}
    
    \caption{The film digitisation system, showing the key components of the digital camera (red device on left image), film illumination system (blue region on left image) and film feed mechanism (the path of which is shown on the right image). When in operation, the entire system, including the film illumination region, is fully is shielded from ambient light. }
    
    \label{fig:section_2.13:system}
\end{figure}

The digitisation system was developed to record digital images of individual film frames with a fidelity and resolution sufficient not compromise the intrinsic precision of the charged particle trajectories encoded on the film images. Such is the size of the Birmingham film collection, that automated operation, at a speed appropriate to process the entire collection on a practical timescale, was also a key design consideration. The film format used at the CERN $2\,\text{m}$ HBC has a width of 50 mm and the camera system produced image frames around 170 mm wide. The digitisation system, shown in Figure~\ref{fig:section_2.13:system}, is composed of three main components, namely a digital camera, film illumination system and film feed mechanism. The basic operation of the system amounts to images of individual film frames being recorded by the digital camera, while illuminated from behind, with an automated advance of the film feed to the next frame by a stepper motor. The key technical details of the system include:

\begin{itemize}
    \item \textbf{Digital Camera} - A monochrome CMOS imaging sensor (Sony IMX455), featuring a $9576\times6388$ pixel array and 16-bit intensity sensitivity, is coupled to a high quality, low distortion lens designed for machine vision applications, with a focal length of $80\,\text{mm}$, operated at a focal ratio of f/8. The objective element of the lens is placed around $350\,\text{mm}$ from the film plane.
    \item \textbf{Illumination System} - An electroluminescent panel ($200\,\text{mm}\times150\,\text{mm}$), diffused by a translucent plastic layer, illuminates the film from behind with high spatial uniformity.
    \item \textbf{Film Feed Mechanism} - The film bobbin is placed on a freely rotating stage, with the accessible end of the film roll attached to a second, receiving film bobbin, mounted on a computer-controlled, motorised rotational stage. These two stages straddle the camera’s field of view, such that the film roll passes in front of it, illuminated by the illumination system. In the region of the frame of interest, the film feed is maintained flat, parallel to the digital camera’s image plane, with a system of aluminium support structures. Either side of this region, a system of rollers maintain modest tension in the film, by friction, to further promote film flatness. The film feed is controlled by the motorised rotational stage, which rotates the receiving film bobbin to advance the frame.
\end{itemize}

\begin{figure}[h]
    \centering
    \includegraphics[width=\textwidth]{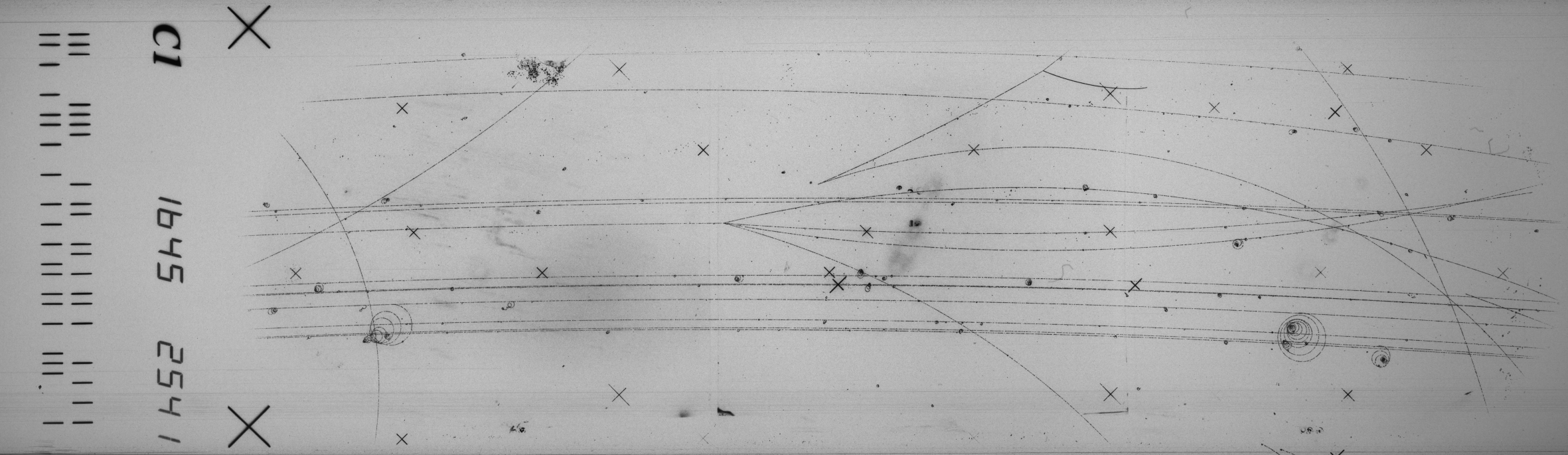}

    \caption{An example of a digitised film frame from the T209 experiment ($8.25\,\text{GeV}/c$ $K^{-}$ beam) at the CERN 2m HBC, showing a candiadate for $K^{-}p\rightarrow \Lambda^{0}\pi^{+}\pi^{-}\pi^{+}\pi^{-}$ with $\Lambda^{0}\rightarrow p \pi^{-}$. This image has been down-sampled by a factor of 4 relative to the original. }
    
    \label{fig:section_2.13:film}
\end{figure}

The system is controlled by dedicated, purpose-written control software, which allows for fully automated operation of the system at a speed of around 250 frames per hour (each film reel contains around 800 frames). The choices of the optical elements of the system were motivated by the spatial intensity profile of charged particle tracks within the film images, which exhibit a FWHM of around $40\,\mu\text{m}$. The system was designed to achieve a digital resolution sufficient to sample track images with multiple pixels, such that the centroid of their position on the film can be determined with sub-pixel precision from an intensity-weighted average of the pixel positions (the so-called centre-of-gravity approach). In practice, the system achieves a digital resolution of around $17\,\mu\text{m}$ per pixel, with tracks typically being sampled by $3-5$ pixels, depending on their trajectory. Figure~\ref{fig:section_2.13:film} shows an example of a film frame digitised by the system. To date, several thousands of frames have been digitised, a large fraction of which are hosted on the project’s website~\cite{bubble_website}.

\subsubsection{Data Analysis Feasibility Study}

The bulk of the film in the Birmingham collection originate from a single experiment, T209, which involved an $8.25\,\text{GeV}/c$ $K^{-}$ beam incident on the CERN $2\,\text{m}$ HBC~\cite{bubble_T209}. Owing to the beam setup, the experiment involves the copious production of strange hadrons, including $K_{s}^{0}$ and $\Lambda^{0}$. The experimentally attractive decays $K_{s}^{0}\rightarrow\pi^{+}\pi^{-}$ and $\Lambda^{0}\to p\pi^{-}$ (often collectively denoted $V^{0}$ decays), which are associated with the distinctive V-shaped topology (as shown in Figure~\ref{fig:section_2.13:film}), represent an ideal ``standard candle'' with which to assess the feasibility of reconstructing charged particle trajectories and momenta from measurements of the digitised film images. The basic technical information required to reconstruct charged particle trajectories and momenta from measurements of the film is documented in the CERN $2\,\text{m}$ HBC user's handbook, which was digitised and made publicly available by CERN~\cite{bubble_manual}.

As was typical of most bubble chambers of the era, beam exposures of the CERN $2\,\text{m}$ HBC were simultaneously imaged from three, distinct perspectives, in order to exploit the principles of stereoscopic image reconstruction. This approach facilitated the reconstruction of a three-dimensional space point in the frame of the chamber from multiple, two-dimensional, measurements of the position of the same space point as it appears on the film associated with different camera views. In this way, measurements of the path a charged particle track traces on the film of multiple camera views could be used to determine the physical trajectory of the particle, also allowing its momentum to be determined from an understanding of the magnetic field within the chamber.

The first step towards realising this approach to reconstruct particle trajectories from the digitised film images involved the development of an optical model which relates a single three-dimensional space point in the frame of the chamber to a two-dimensional position on the digital image of each of the three camera views. Such a model, based on the concept of a simple pinhole camera system, was implemented as three matrix equations, parameterised in homogenous coordinates. The parameters of these matrices can be determined through a calibration process facilitated by a set of corresponding chamber space points and image positions for each of the camera views. Corrections to this linear model were implemented to account for the refraction of light rays as they pass through the various media of the chamber (liquid hydrogen, glass and air) to the objective lens of the cameras. Such corrections can become sizable for the largest angles of incidence, corresponding to chamber positions near the beam entrance and exit (left and right edges of the images, respectively). Additional, much smaller corrections were also implemented to account for distortions in the original chamber cameras, based on data in the user’s manual. Effects associated with optical distortions in the digitisation system and potential, residual non-coplanarity of the film and digital image planes caused by film bowing are not yet considered. The data required to fulfil the calibration of the optical model are provided by fiducial marks (the crosses visible in Figure~\ref{fig:section_2.13:film}) which were etched into the glass of the chamber’s windows before its assembly, exactly for this purpose. The user’s manual contains a detailed account of the measured relative physical positions of these fiducial marks in the frame of the chamber, providing the means with which to calibrate the optical model. The position of up to 27 of these marks in the digital images of all three corresponding views of any beam exposure under consideration are automatically identified with a pattern recognition algorithm based on template matching. These fiducial mark positions on the digital images, in pixel coordinates, together with their corresponding physical positions from the user’s manual are used to determine the parameters of the optical model through a least-squares fit. The residuals distribution of the predicted and measured fiducial mark positions is consistently well-centered on zero, with an RMS of 3 pixels at most, corresponding to around $50\,\mu\text{m}$ on the film. In summary, the model can recover the expected fiducial mark positions with a precision roughly corresponding to the width of a charged particle track on the film.

The optical model may then be deployed to reconstruct the physical trajectory of a charged particle track from a set of digital film images. Beam exposures which include $V^{0}$ candidates are identified by a visual inspection of the digital images. Three sets, one for each camera view, of two-dimensional positions on the digitised film images corresponding to the path of the $V^{0}$ tracks are manually identified with the aid of a purpose-made computer software with a graphical interface. For the purposes of this limited feasibility study, this robust manual approach was preferred over algorithmic track finding, though there is considerable scope to extend this study to deploy modern algorithmic or machine learning-based approaches to find tracks within the digital images. In order to relate the momentum of the $V^{0}$ decay products at the decay vertex to trajectories within the chamber, a simple model of charged particle propagation in the chamber volume was developed, which simulates the effect of curvature due to the magnetic field and continuous energy loss in liquid hydrogen. The user’s manual includes detailed information on the magnetic field within the $2\,\text{m}$ HBC, including a field map. Since non-uniformities in the field are limited to around 2\%, a uniform field strength of $1.74\,\text{T}$ is assumed. Exploiting the aforementioned propagation model, sets of two-dimensional space points on three digital image views, are simultaneously fit for the two $V^{0}$ tracks to determine the momentum of each charged particle and the $V^{0}$ decay vertex. The track reconstruction procedure described above was developed and validated with the aid of a basic Geant4~\cite{GEANT4:2002zbu} simulation of the $2\,\text{m}$ HBC, interfaced to a simple optical model, which allowed the generation of simulated film frames which accurately mimic the features of the digitised film images.

\begin{figure}[h!]
    \centering
    \includegraphics[width=0.45\textwidth]{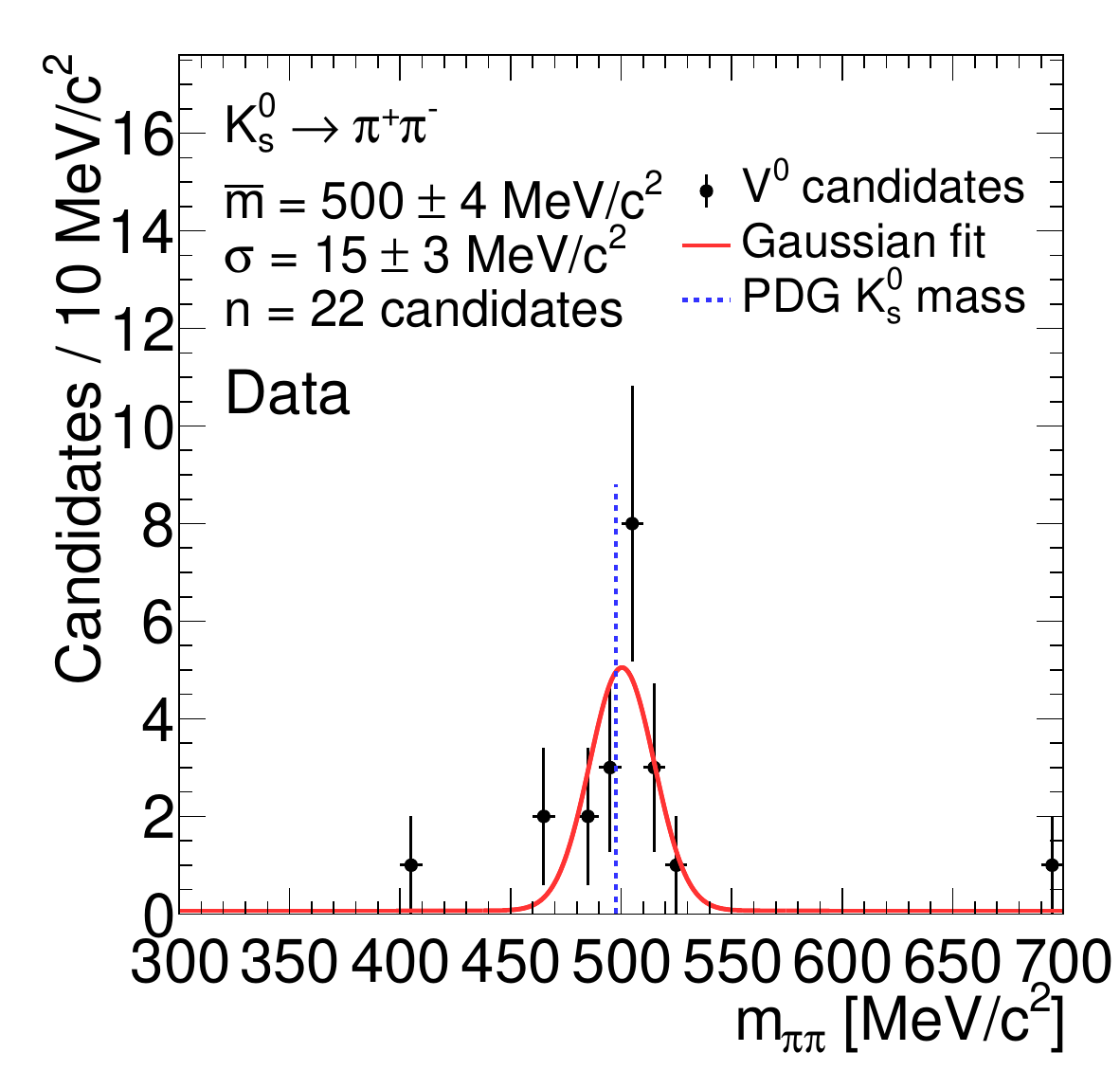}
    \includegraphics[width=0.45\textwidth]{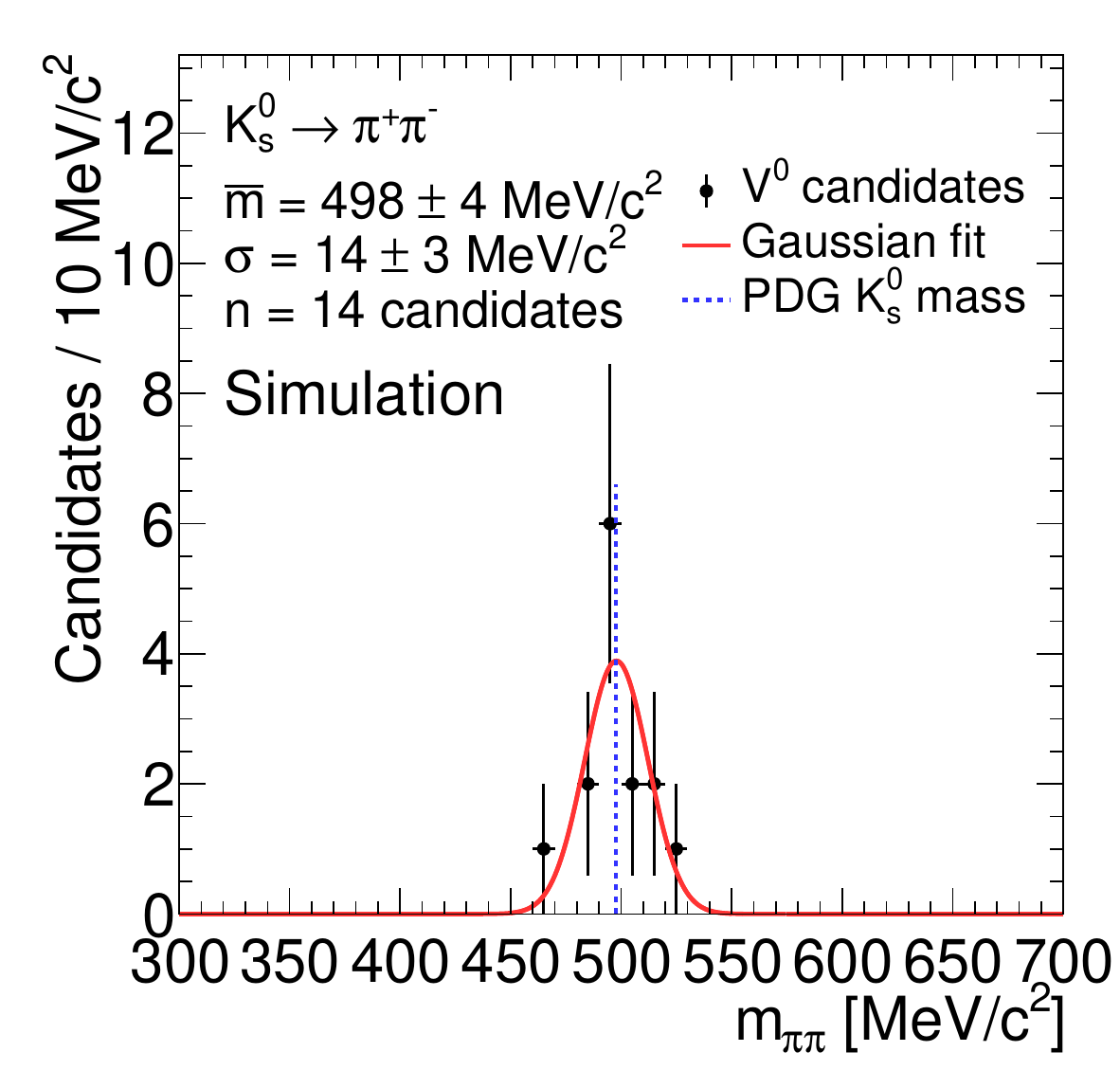}

    \caption{Invariant mass distributions of $K_{s}^{0}\rightarrow\pi^{+}\pi^{-}$ candidates reconstructed from digitised film images (left) and simulated film images (right) of the T209 experiment at the CERN $2\,\text{m}$ HBC. }
    
    \label{fig:section_2.13:mass}
\end{figure} 

For this limited feasibility study, the digital film images of around 300 beam exposures were studied, leading to the identification of 50 $V^{0}$ candidates. The current manual approach of image inspection and track finding limits the size of dataset which can be studied on a practical timescale. Clearly, the implementation of algorithmic approaches to these tasks could remove this limitation, though this was considered beyond the scope of the initial feasibility study. The identified $V^{0}$ candidates were classified as $K_{s}^{0}$, $\Lambda^{0}$ or photon conversions on the basis of their position in the Armenteros-Podolanski plane. Figure~\ref{fig:section_2.13:mass} shows the invariant mass distribution of $K_{s}^{0}$ candidates reconstructed from both the digitised original and simulated film frames. The distribution in data peaks close to the known $K_{s}^{0}$ mass and the resolution exhibited in data and simulation are in good agreement. The same conclusion is also drawn for $\Lambda^{0}$ candidates, with a lower level of statistical confidence, given only 10 candidates are identified in the dataset studied.

While the resolution achieved is likely significantly worse than that possible with contemporary reconstruction techniques, the feasibility study shows that in principle, digitised film images from bubble chamber experiments can be used to reconstruct charged particle vertices, trajectories and momenta at a useful level of precision, opening a wide range of possibilities to exploit the digitised film images for educational purposes, or with a more developed understanding, even renewed scientific exploitation.

\subsubsection{Conclusion}

Reanalysis of raw data from the CERN $2\,\text{m}$ HBC, starting from digitised film images, appears to be feasible at a limited but useful level of precision. While the achievements of this study are perhaps not too much of a surprise, the relative ease with which they were accomplished, without any interaction with or input from original collaborators, is an encouraging sign for HEP data preservation. However, the success of this endeavor is owed to the CERN archivists responsible for making the necessary technical information publicly available, long before this study was attempted. Without this information, it is unlikely that any comparable quantitative achievements could have been made.


\section{Transverse projects and technologies}

\subsection{k4GeneratorsConfig: A Unified Approach to MC Generator Benchmarking}
{\small
 \it Authors: Alan Price(Jagiellonian University (PL)); Dirk Zerwas (DMLab, DESY, CNRS/IN2P3 (DE))
}

The next generation of electron-positron colliders will require unprecedented precision in both theory and experiment. Sophisticated software frameworks are essential to evaluate detector concepts, optimize designs, and simulate physical processes. In this context, Monte Carlo (MC) event generators~\cite{Campbell:2022qmc} play a central role, enabling realistic simulations of Standard Model processes and providing the basis for physics studies. However, technical consistency across different generators remains critical, particularly in domains where agreement is expected. To address this need, we present k4GeneratorsConfig~\cite{Price:2025fzg}, a Python-based package that automates the benchmarking process for MC generators. The tool translates universal physics inputs into generator-specific configurations, ensuring consistency, reproducibility, and reduced human error. Its modular design allows for straightforward integration of additional generators and provides compatibility with the Key4hep software stack. While so far the focus has focused on the configuration of the generator, we are now aiming to integrate the production aspect of running generators with an emphasis on reproducibility and referencing for well defined MC generator versions.

In its current implementation k4GeneratorsConfig supports many modern MC generators~\cite{Alwall:2014hca,Kilian:2007gr,Sherpa:2024mfk,Bierlich:2022pfr,Jadach:2022mbe,Bellm:2015jjp} for $e^+e^-$ simulations. Physical input parameters are specified through a unified
YAML formatted configuration file, from which generator-specific runcards and
execution scripts are automatically derived. Common parameters such as the
centre-of-mass energy $\sqrt{s}$, beam polarisation, initial-state radiation (ISR)
settings, particle masses and widths , and phase-space
selectors are defined once and translated consistently across all requested generators.
This abstraction insulates users from the idiosyncrasies of individual generator
interfaces and substantially reduces the risk of human error in large-scale production
campaigns.

We use this framework to ensure that reproducibility of the generators is a first-class requirement.
Generator-specific output directories are created for each unique combination of
process, generator, and centre-of-mass energy, and random seed management is handled
centrally, with per-process overrides available. A Continuous Integration (CI)
pipeline, implemented via GitHub Actions and based on the Key4hep software stack,
automatically validates the generation of the runcards, the translation of physics inputs,
and the consistency of results across generators with each update to the codebase.

Post-generation analysis is tightly integrated into the workflow. k4GeneratorsConfig
supports two complementary analysis tools: the Rivet framework~\cite{Bierlich:2019rhm},
which provides a broad library of validated MC analyses, and the Key4hep-native analysis
chain, in which generated events are converted to the standardised EDM4hep event data
model~\cite{Gaede:2021izq} for direct use with downstream reconstruction and analysis
software. The Analysis key in the YAML configuration triggers the automatic appending
of the relevant analysis execution to the generated shell script, enabling a seamless
transition from event generation to physics observables within a single workflow
invocation.

Looking forward, the development of k4GeneratorsConfig will be focused in two areas.
We will expand the physics coverage by including more generators
as new MC tools reach sufficient maturity for $e^+e^-$ physics, leveraging the modular
architecture in which each generator is encapsulated as a self-contained module with
its own template and translation rules. We will tackle the challenge of extending our modular
approach to include more complicated physics modeling, such as higher-order corrections, 
which will be required for future experiments. While developed for e+e- colliders, the
code should work directly for muon colliders, where then only specific
process dependent settings will have to be added. For collisions
involving at least one hadron, further developments for PDF settings are
needed.  Secondly, the package
we aim to evolve from a pure configuration tool into a full production framework
for MC event generation, with a long-term view toward the preservation of MC generators
over the coming decades. Future $e^+e^-$ collider experiments will span timescales
of twenty years or more, during which generator codebases, dependencies, and software
environments will inevitably evolve. Ensuring that events generated today remain
reproducible and interpretable throughout the full lifetime of an experiment (and
beyond) requires not only capturing the generator executable and version tag, but also
preserving the precise software environment, parameter cards, and random seed state
associated with any given production.  k4GeneratorsConfig, through its tight integration
with the Key4hep software stack and the versioning infrastructure provided by git,
is being developed as a vehicle for this long-term stewardship: a given production run
will be uniquely referenceable, allowing future analysts to reconstruct the exact
conditions under which a sample was generated, irrespective of how the broader
software ecosystem has changed in the intervening years.
\begin{figure}
    \centering
    \includegraphics[width=\textwidth]{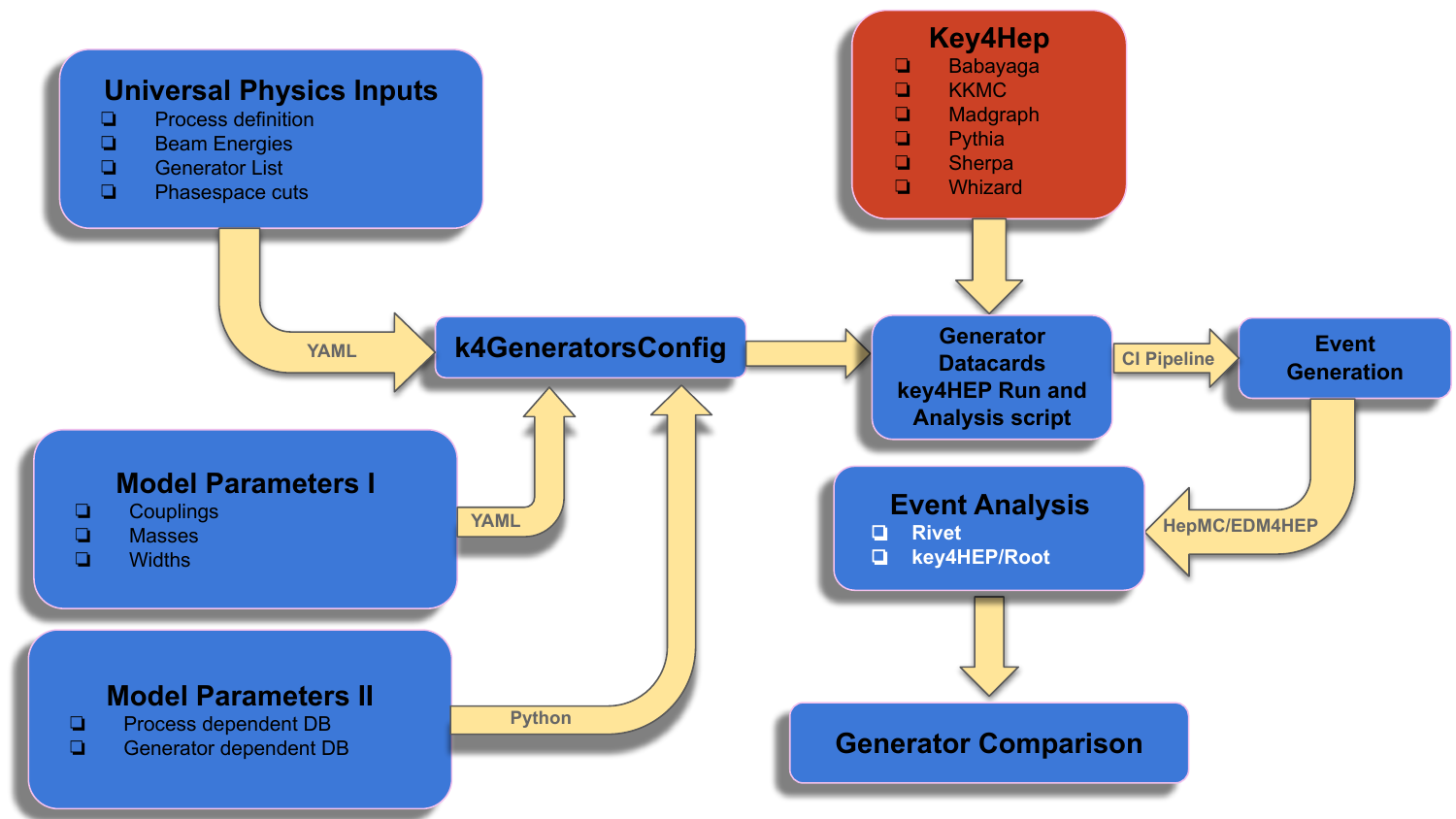}
    \caption{Workflow of the k4GeneratorsConfig~\cite{Price:2025fzg} framework.}
    \label{fig:k4gensworkflow}
\end{figure}

\subsection{Providing Cold Storage in the CERN Open Data portal}

{\small
 \it Authors: David Horvat (Deggendorf Institute of Technology (DE)); Diana Rand (CERN); Jose Benito Gonzalez Lopez (CERN); Pablo Saiz (CERN); Tibor \v{S}imko (CERN); Zacharias Zacharodimos (CERN)
}

The CERN Open Data portal \cite{CERN:OpenDataPortal} is the primary gateway for open access to high-energy physics data, serving the needs of research, education, and outreach communities worldwide. As of early 2026, the portal hosts over 5 petabytes of data, encompassing more than 80,000 entries and 3 million files from eight high-energy physics experiments. A ninth one, JADE, became available shortly after the workshop.

As the volume of hosted data continues to grow, a sustainable storage management strategy becomes essential. The challenge lies in balancing high-performance access for popular datasets, which need to remain readily available on disk, with cost-effective storage for rarely accessed data that must nonetheless be preserved for the long term. Simply expanding disk capacity is not a scalable solution given the pace at which new data releases are being made.

To address this, the CERN Open Data team developed and deployed a cold storage system that leverages tape archives to preserve massive volumes of data while freeing up primary disk resources. A central design principle was to maintain the portal’s open-access philosophy, meaning that any user, without the need for authentication, should be able to request access to archived datasets directly through the web interface.

\begin{figure}[hhh]
    \centering
    \includegraphics[width=0.5\linewidth]{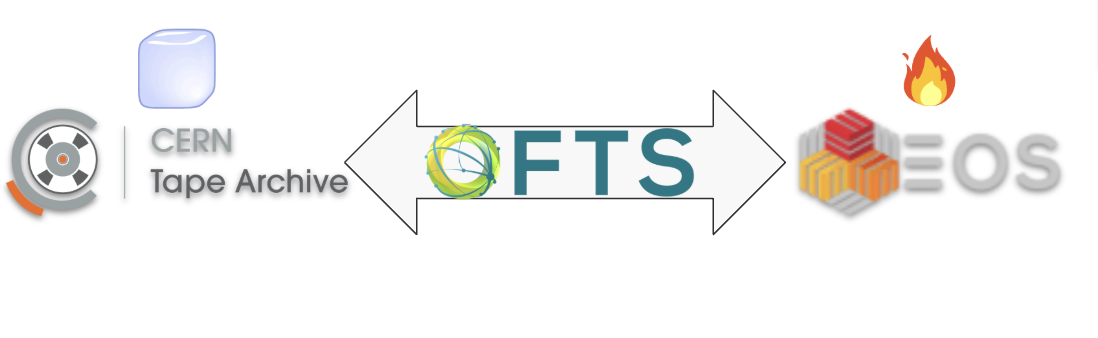}
    \caption{Cold storage infrastructure}
\end{figure}

The cold storage infrastructure relies on three pillars: 
\begin{itemize}
    \item EOS continues to serve as the primary hot storage for online data, as it did before the introduction of the cold storage layer. \cite{EOSOpenStorage}
    \item The CERN Tape Archive (CTA) serves as the cold storage layer, providing cost-effective long-term preservation on tape media. \cite{CERNTapeArchive}
    \item The File Transfer Service (FTS) manages the movement of files in both directions. Firstly, archiving data from hot to cold storage and then staging it back when requested by users. \cite{FileTransferSystem}
\end{itemize}

The transition to production was carried out in two phases. In May 2025, selected records were copied from EOS to CTA, and the user interface functionalities of the cold storage system were deployed in production. This initial phase allowed the team to monitor the system under real conditions while all data remained available on disk. In June 2025, the system moved fully into production. The copies on EOS were deleted for the migrated records, and automated transfers from CTA back to EOS were enabled. From this point on, the cold storage system was handling real staging requests from the public.

A key design goal was to make archived data accessible to any user without requiring authentication. The portal provides a self-service staging interface where users can request access to offline files directly through the web. By design, at least one file from every dataset is kept on hot storage to ensure that users can always access a sample immediately.

Once a request is submitted, an automated queuing system takes over. It manages the entire tape-to-disk transfer process without any manual intervention from the Open Data team. To ensure system stability and avoid overwhelming the File Transfer Service, a limit is enforced on the number of concurrent transfers. Users can view real-time progress of the transfer, including the number of successfully transferred files and the total size processed. This transparency helps manage expectations, particularly for large-scale data transfers. Importantly, this status is visible to all users, meaning that if another user is interested in the same record, they do not need to submit a separate request.

The usage and health of the cold storage system are tracked through CERN’s IT monitoring framework, including Grafana dashboards \cite{open-data-monitoring} that provide views of data access patterns, storage distribution, and EOS quota usage for both curators and developers.

\begin{figure}[hhh]
    \centering
    \includegraphics[width=0.9\linewidth]{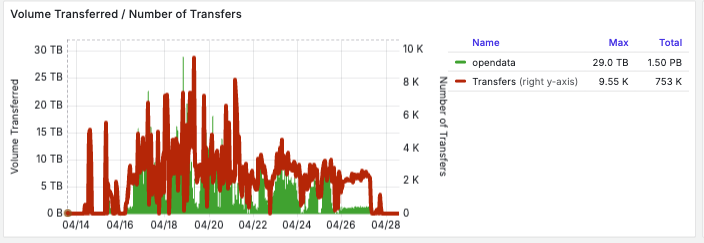}
    \caption{Files volume transferred and number of transfers in April 2025 \cite{fts-monitoring}}
    \label{fig:section3_2:files_volume}
\end{figure}

As of early 2026, over 1 million files, totaling approximately 1.8 petabytes, have been moved to cold storage. Figure \ref{fig:section3_2:files_volume} shows the files volume transferred and the number of transfers in April 2025, when the files were first moved to cold storage. Since the system entered production, over 140 staging requests have been processed from the public. This activity confirms that there is genuine demand for archived datasets and that the automated staging process is functioning effectively.

Several directions are planned for the further development of the cold storage system.

First, the team will continue to monitor usage patterns and refine automated procedures based on operational experience. One planned improvement is the addition of automated cleanup mechanisms to reclaim disk space by moving data that has been staged back to tape after a period of inactivity. Cold storage options will also be integrated into the Open Data client, giving users more ways to interact with archived data.

\begin{figure}[hhh]
    \centering
    \includegraphics[width=0.5\linewidth]{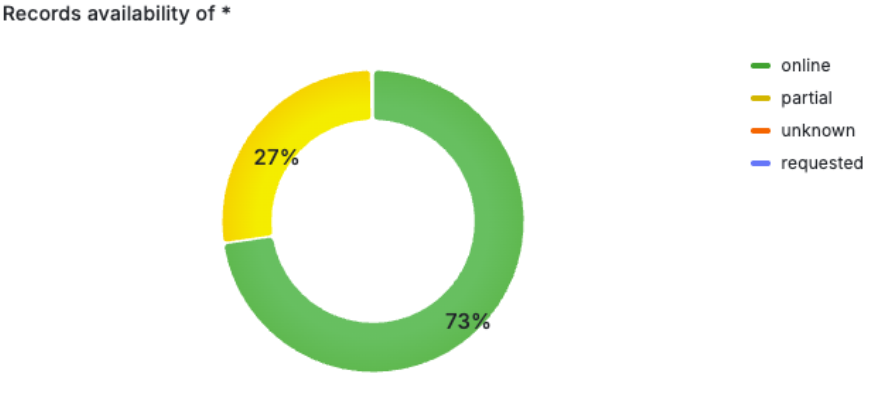}
    \caption{Records availability \cite{open-data-monitoring}}
    \label{fig:section3_2:records_availability}
\end{figure}

Second, the team aims to empower curators by allowing them to trigger the archiving of records to cold storage directly through the portal. Currently, these moves are performed manually by the Open Data team upon request. Enabling curators to initiate these transfers independently would help scale the system and increase the proportion of records in cold storage beyond the current 27\% (as seen on figure \ref{fig:section3_2:records_availability}).

Third, the incorporation of additional cold storage media is under consideration. One approach would be to consider the data hosted by the experiments as the cold storage, and use the experiment data management frameworks (like Rucio \cite{Rucio}), to stage the data directly. However, this would represent not only a technical change but also a governance shift, as it would move the long-term responsibility for data preservation and access to the experiments themselves. This transition will require extensive discussion with the relevant collaborations.

The deployment of a cold storage system has allowed the CERN Open Data portal to remain sustainable as it surpasses the 5 PB mark. By moving data to tape-based cold storage, the portal has freed disk resources while maintaining full accessibility for all records. The operational experience from the first year confirms both the viability of the approach and the genuine demand from the community for archived datasets.

\subsection{EOSC EDEN}
{\small
 \it Authors: Diana Rand (CERN); Jean-Yves Le Meur (CERN); Jose Benito Gonzalez Lopez (CERN); Pablo Saiz (CERN); Panna Liptak (CERN); Wesley Middelbos (CERN)
}

EOSC EDEN\cite{eosc-eden} is a European Union funded initiative aimed at improving long-term digital preservation across research infrastructures at both European and National level. The project is coordinated by CSC (IT Center for Science)\cite{eosc-eden-csc} and brings together 15 organizations across Europe.

Key drivers motivating the projects:
\begin{itemize}
    \item Digital preservation remains a fundamental challenge for a majority of research infrastructures.
    \item Data volumes are growing and technologies, file formats and software stack continue to evolve.
    \item Research data must remain usable, understandable and trustworthy well beyond the operations lifetime of the experiments that produced it.
\end{itemize}
    
The project consortium comprises 15 partners spanning research computing centers, data archives, universities and preservation-specialist organizations:

\begin{figure}[hhh]
    \centering
    \includegraphics[width=0.65\linewidth]{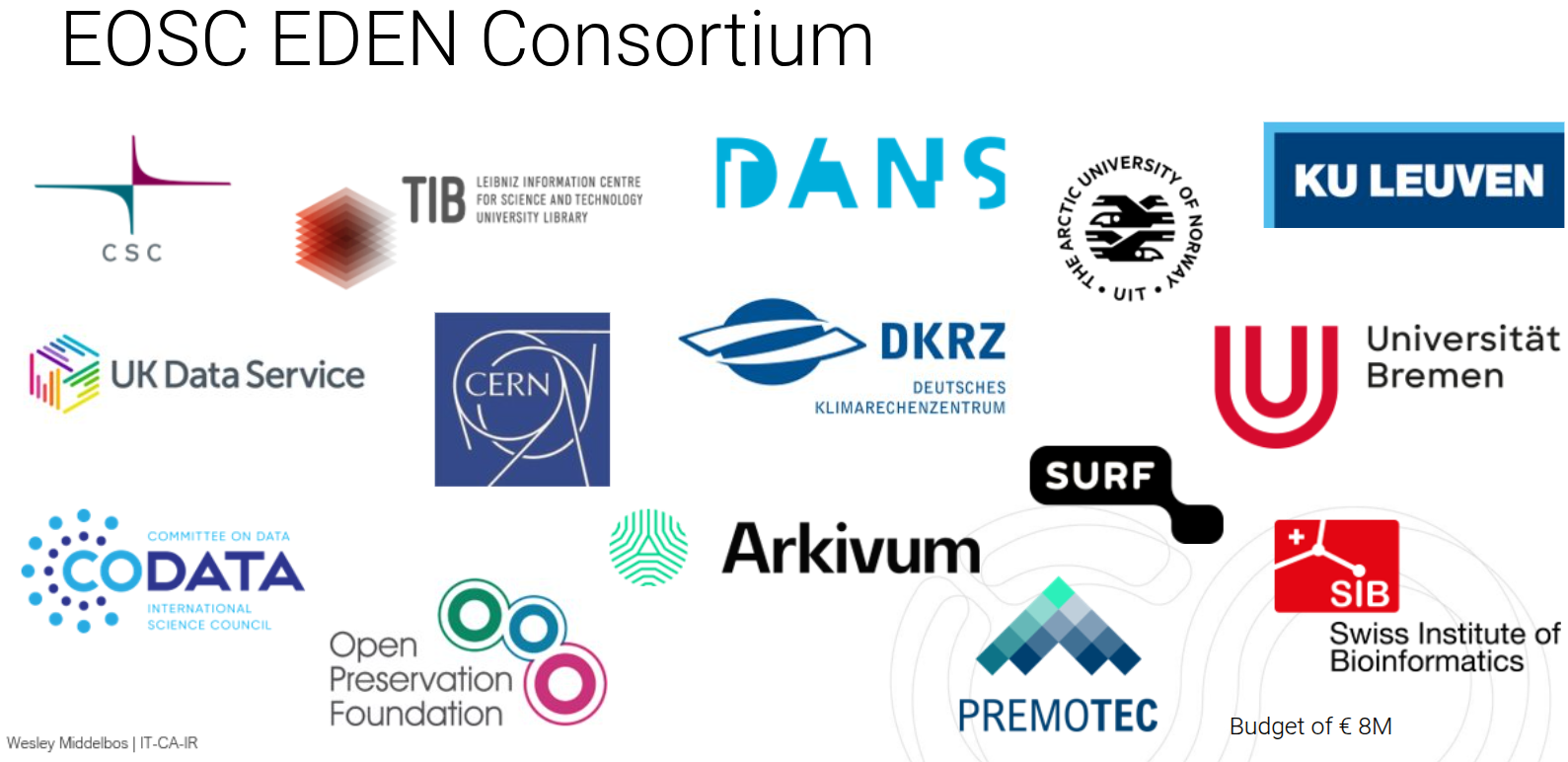}
    \caption{EOSC EDEN - Consortium}
    \label{fig:placeholder}
\end{figure}

The approach of EOSC EDEN is layered, to build shared digital preservation capabilities, based on gathered information and requirements from six scientific disciplines: Climate Simulations; Earth and Environmental Sciences; Food Sciences; High-Energy Physics; Life Sciences and Bioinformatics; Linguistics; and Social Sciences.

For these disciplines, the project gathered requirements by conducting desk research and interviews with stakeholders. Based on the results, the project sees a repetition of preservation challenges across the represented disciplines such as:

\begin{itemize}
    \item Need for standardized documentation and metadata practices
    \item Better management of sensitive data
    \item Improved support for open and interoperable file formats
    \item Lack of formal policies for data retention
\end{itemize}

One of the outputs of the project are the Core Preservation Processes (CPPs)\cite{eosc-eden-cpp-repository}. A Core Preservation Process is a specific action that every Trustworthy Digital Archive (TDA) must undertake adequately, either directly or through associated parties or services, in order to fulfill its digital preservation mission as documented in its preservation policy.

\begin{figure}[hhh]
    \centering
    \includegraphics[width=0.75\linewidth]{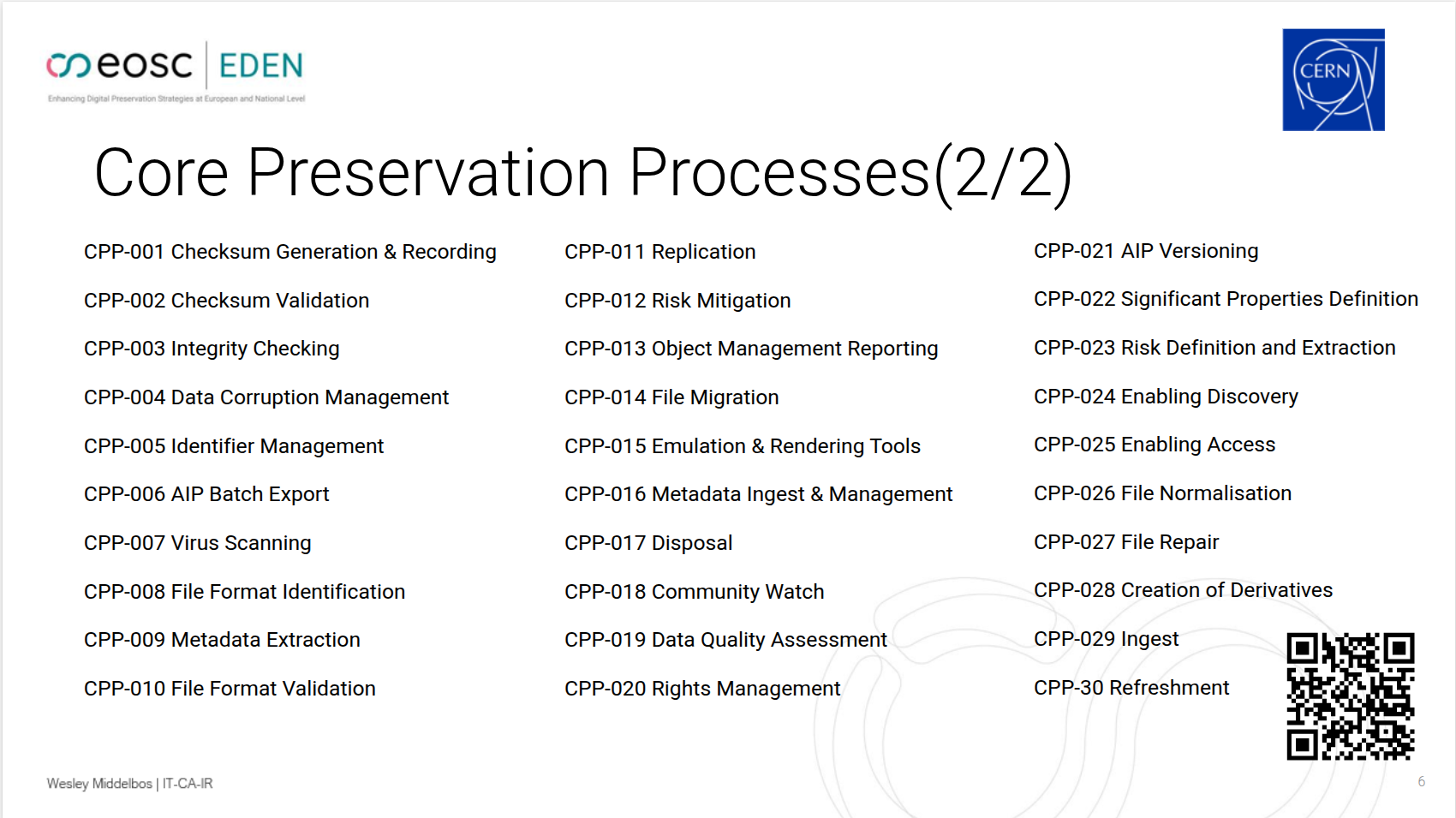}
    \caption{EOSC EDEN - Core Preservation Processes}
    \label{fig:placeholder}
\end{figure}

Building on top of the CPP framework and the discipline specific requirements \cite{eosc-eden-3.1}, EOSC EDEN is specifying and developing shared digital preservation and automated curation services that can be reused across research domains. Aside from the services being aligned with the CPPs and discipline specific requirements, also applicable principles such as FAIR, TRUST and standards like the OAIS (Open Archival Information System) are taken in consideration. 

The current status of the project is that there are over 30 services identified as candidates for further specification. The services are being specified in a way that they are domain-agnostic and are interoperable.

EOSC EDEN delivers value by:
\begin{itemize}
    \item Aligning preservation practices across disciplines without forcing an uniform solution.
    \item Improving interoperability and clarity in digital preservation practices.
    \item Reducing duplicated efforts by specifying domain-agnostic services.
    \item Developing reference and use case implementations of preservation and automated curation services.
\end{itemize}

One of the key priorities of the EOSC EDEN is sustainability of its output beyond its own lifetime. This is being addressed through the creation of expert networks, creating a project wide sustainability plan, collaboration with other horizon projects and search for possible transfer of ownership for the outputs of the project.

Key takeaways:
\begin{itemize}
    \item \textbf{Purpose}: Improve long-term digital preservation at European and National level.
    \item \textbf{Approach}: Define Core Preservation Processes for trustworthy digital archives, develop a registry and shared services to support interoperable preservation practices
    \item \textbf{Impact}: Align preservation practices across disciplines, reduce duplicated efforts, and strengthen interoperability of research data preservation practices.
\end{itemize}

\subsection{From tcl to awkward: analyzing old data with new tools\label{sec:babartcl}}
{\small
 \it Author: Matthew Bellis (Siena University (US))
}

The \babar experiment collected electron-positron collision data at the SLAC PEP-II collider from 1999 to 2008, amassing approximately 530~fb$^{-1}$
at and near the $\Upsilon(4S)$ resonance. Although data-taking ended over fifteen years ago, the collaboration continues to pursue new physics analyses through the Long Term Data Access (LTDA) system~\cite{Ebert:CHEP2024,Ebert:CHEP2025}. The LTDA infrastructure preserves the full \babar software stack — reconstruction code, databases, Monte Carlo generation tools, and analysis frameworks — in a frozen state running inside virtual machines. Originally hosted at SLAC, the system was migrated in 2022 to cloud computing resources at the University of Victoria (Canada), with approximately 1.5~PB of data stored at the GridKa Computing Centre (Germany) and accessed remotely via XRootD~\cite{Ebert:CHEP2024,Ebert:CHEP2025}. Analysts interact with the system by submitting batch jobs through HTCondor, which provisions VMs on demand, and retrieve output ntuples for further processing outside the frozen environment. This architecture ensures that the legacy software environment remains intact and reproducible while remaining accessible to collaborators worldwide.

In 2024, a new search for baryon-number violation in $B$-meson decays
was proposed. Rather than searching for processes in which baryon number
changes by $\Delta B =1$, we would search for the decay
$B^+ \rightarrow p \Lambda^0$, a processes in which $\Delta B = 2$.
This would be the first search for this process and the goal was to 
go from proposal to unblinding in a year, making use of newer computational tools.

The boundary between the legacy infrastructure and the modern analysis layer represents both the primary bottleneck in preserved-data analysis. Extracting data from the LTDA system requires using the same tools that existed when \babar was active: Tcl-based configuration scripts that configure the C++-based reconstruction framework, Perl scripts to query dataset databases, and the standard ROOT~\cite{Brun:1997pa} utilities to manage and merge output files. This is unavoidable — the reconstruction and event selection must be performed within the frozen environment to ensure reproducibility. However, once the relevant candidates have been selected and written to flat ntuples, the output is simply ROOT files containing standard TTrees, and the analysis can proceed with any tools the analyst chooses.

\begin{figure}[hhh]
    \centering
    \includegraphics[width=0.5\linewidth]{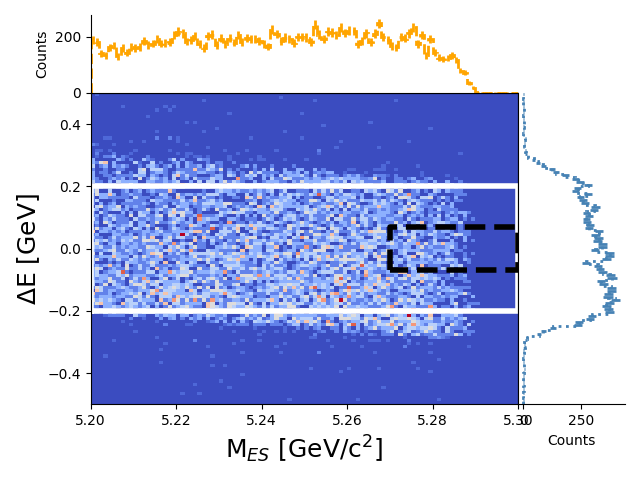}
    \caption{Plot from an intermediate stage in the \babar search for 
    $B^+\rightarrow p \Lambda^0$, showing two kinematic variables used
    for signal and background discrimination. Monte Carlo is shown here.
    The plot is made with {\it matplotlib}~\cite{Hunter:2007}, a modern python visualization
    library that was not widely used during \babar's data taking, but which
    is taught to many undergraduate students.}
    \label{fig:babarmesde}
\end{figure}

This is where the landscape has changed dramatically relative to what was available when \babar was running. 
For this search, we converted the ROOT output to the {\tt parquet} columnar format using uproot~\cite{uproot} and awkward-array~\cite{awkward}, enabling the full dataset (~3.7~GB of Monte Carlo and ~1.1~GB of collision data) to be loaded into memory and manipulated in Jupyter notebooks 
(see example in Fig.~\ref{fig:babarmesde}), much of which was done by an undergraduate physics major at Siena, Josie Swann. Statistical analysis and multivariate selection optimization were performed with scikit-learn, a tool that did not exist during \babar's operational lifetime. Tables and figures were auto-generated and piped directly into the analysis note LaTeX source. The experiment-agnostic nature of this tooling — the same packages used routinely in CMS and ATLAS analyses — means that student analysts trained on modern workflows can contribute productively to legacy-data analyses without needing to learn the full \babar software stack. The
analysis was unblinded in summer 2025 leading to the first upper-limits
on this process. The analysis took about 1 1/2 years, not including
writing the paper (in process at the time of this workshop). 

Several lessons emerge that are relevant to experiments currently planning data preservation strategies. First, the threshold for producing flat ntuples from the legacy system should be made as low as possible: this is the single step that requires legacy expertise, and once it is complete, the subsequent analysis is fully accessible to modern tools and modern analysts. Second, human knowledge is not fully captured by software preservation — having colleagues who remember the quirks of the original reconstruction and Monte Carlo production proved essential during this analysis. Third, the frozen-environment approach, while conservative, provides a high degree of confidence in reproducibility, which is central to the scientific value of preserved data. As more experiments in high-energy physics approach end-of-operation and plan for long-term data access, the \babar experience provides a concrete existence proof that nearly two-decade-old data can be brought to bear on new physics questions with modern computational methods.

{\it Portions of the text in this section (Section~\ref{sec:babartcl}) was generated using Claude. The authors of this section (listed below) are entirely responsible for the content of this text.}

\subsection{CERN Preserve Platform}
{\small
 \it Authors: Diana Rand (CERN); Jean-Yves Le Meur (CERN); Jose Benito Gonzalez Lopez (CERN); Pablo Saiz (CERN); Panna Liptak (CERN); Wesley Middelbos (CERN)
}

CERN has set up a preservation service to ensure that any data under its intellectual property can be treated for long-term preservation according to world-wide standards. The Preserve Platform is designed to communicate with live information repositories operated within the institute. Service level agreements establish how the data and metadata from these source systems are harvested in order to convert them into Archival Information Packages (AIP).

The Preserve Platform is acting as a hub, supporting pipelines and core processes, enabling a consistent approach to digital archive throughout the CERN laboratory. Data originated from various sources can be captured via the hub and treated for long-term preservation creating OAIS-compliant packages. OAIS~\cite{oais} stands for Open Archival Information System, this ISO standard provides a reference model to design a digital preservation system.

\begin{figure}[hhh]
    \centering
    \includegraphics[width=0.65\linewidth]{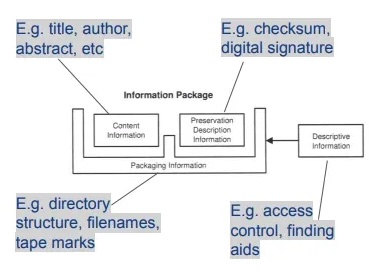}
    \caption{Information Package structure}
    \label{fig:placeholder}
\end{figure}

The major steps of the Preserve pipelines are covering the following actions:

\begin{itemize}
\item Harvesting from the source repository
\item Creation of standardized Submission Information Packages (SIP)
\item Validation of the SIPs
\item Creation of AIPs
\item Notifying the source about the successful preservation
\item Push to CERN Tape Archives~\cite{CERNTapeArchive} (cold storage)
\item Registry update
\end{itemize}

\begin{figure}[hhh]
    \centering
    \includegraphics[width=0.9\linewidth]{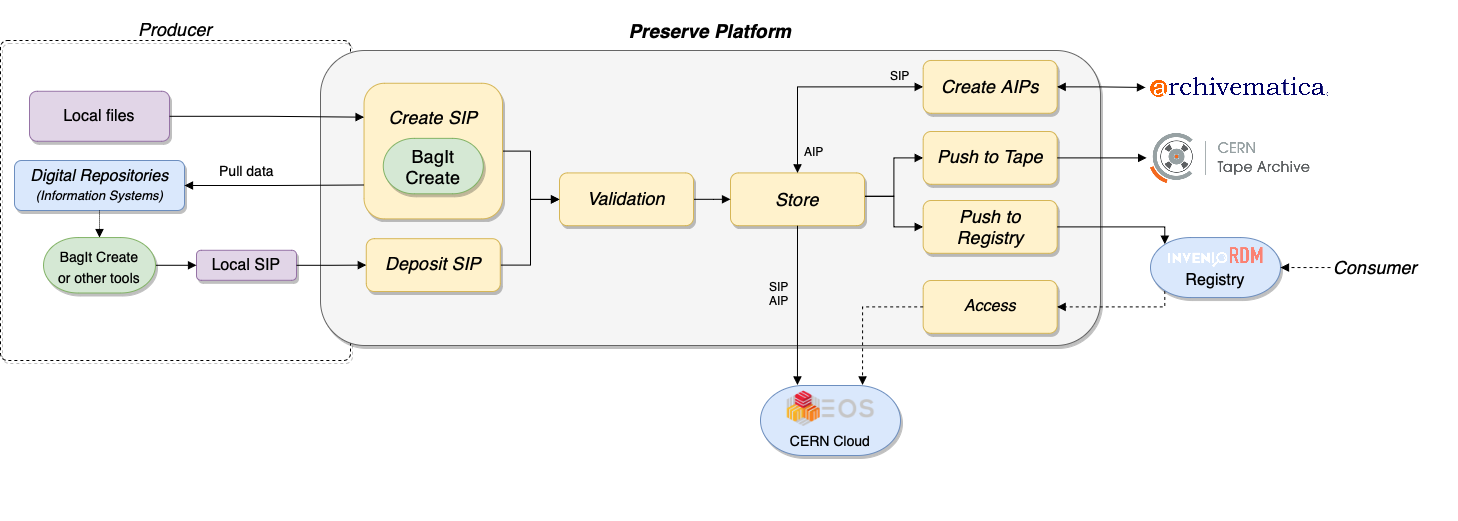}
    \caption{CERN Preserve Platform Architecture}
    \label{fig:placeholder}
\end{figure}

The transformation of SIP into AIP includes the activation of multiple micro-services. The current pipelines are delegating this task to the Archivematica open-source software, mostly running key components like:

\begin{itemize}
\item file identification
\item virus scanning
\item file normalization
\item standardized METS representation
\end{itemize}

At CERN, a strategy has been established to set up a Preservation infrastructure to handle digital content belonging to a predefined scope of ”must-be-preserved data”. This scope, mostly based on the criteria of the historical value of the data, is described in a lab-wide operational circular. Similarly to the paper archives, digital content must be collected within departments and transferred to the Digital Archives. Live information systems storing highly-valued records have to decide how the core preservation processes are activated, according to one of the following scenarios:
\begin{itemize}
    \item Permission is given to the Preserve Platform to harvest records in order to create, store and maintain AIPs on tape.
    \item SIPs are created and submitted to the Platform directly, which will ensure the creation of the AIPs, and their storage.
    \item AIPs are created and managed by the Information System; their description is pushed to the central registry.
\end{itemize}

The strategy is planning that the access to the registry is closed by default with controlled access only. This prevents the challenge of propagating the access controls of each enrolled information system, and it also reduces the risks of exposing restricted or sensitive data.

The platform provides a streamlined way to integrate InvenioRDM-based~\cite{inveniordm} repositories. It currently supports harvesting from Zenodo~\cite{Zenodo} and the new CERN Document Server (CDS)~\cite{cerndocumentserver}. CDS was also the first repository to be onboarded under a Service Level Agreement for the automated harvesting of all its records (after a grace period). Other supported sources include CodiMD~\cite{codimd}, Indico~\cite{indico}, GitLab, and, for metadata only, CERN Open Data~\cite{OpenDataPortal}.

In addition to onboarding different repositories, the platform also handles various types of data. In 2025, a pipeline was introduced to process proprietary mailboxes of the CERN Directorate. Support for handling restricted websites is currently under development.

The main challenge for the Preserve Platform is scalability and stability. As more repositories are onboarded in the future, the platform must be able to handle an increasing volume of records. Currently, Archivematica is the most resource-intensive component in the pipeline, creating a bottleneck.

Engaging with the community to collaborate and share experiences is expected to help build a more resilient and scalable digital preservation service.

\subsection{Facilitating Open Data Reuse with REANA: From Scalable Workflows to AI-Assisted Analysis Authoring}
{\small
 \it Authors: Cameron McClymont (CERN); Michael Buchar (CERN); Tibor \v{S}imko (CERN)
}

REANA is a platform for reusable and reproducible data analyses, allowing researchers to use declarative analysis workflows (CWL, Snakemake, Yadage) and run them on supported containerised compute clouds (Kubernetes, HTCondor, Slurm)~\cite{Simko:2018zzz}. One of its primary use cases is to facilitate analysis reinterpretation in light of new theories, such as ATLAS searches for new physics~\cite{Donadoni:2024nkq}.

\begin{figure}[hhh]
\begin{center}
\includegraphics[height=0.27\textheight]{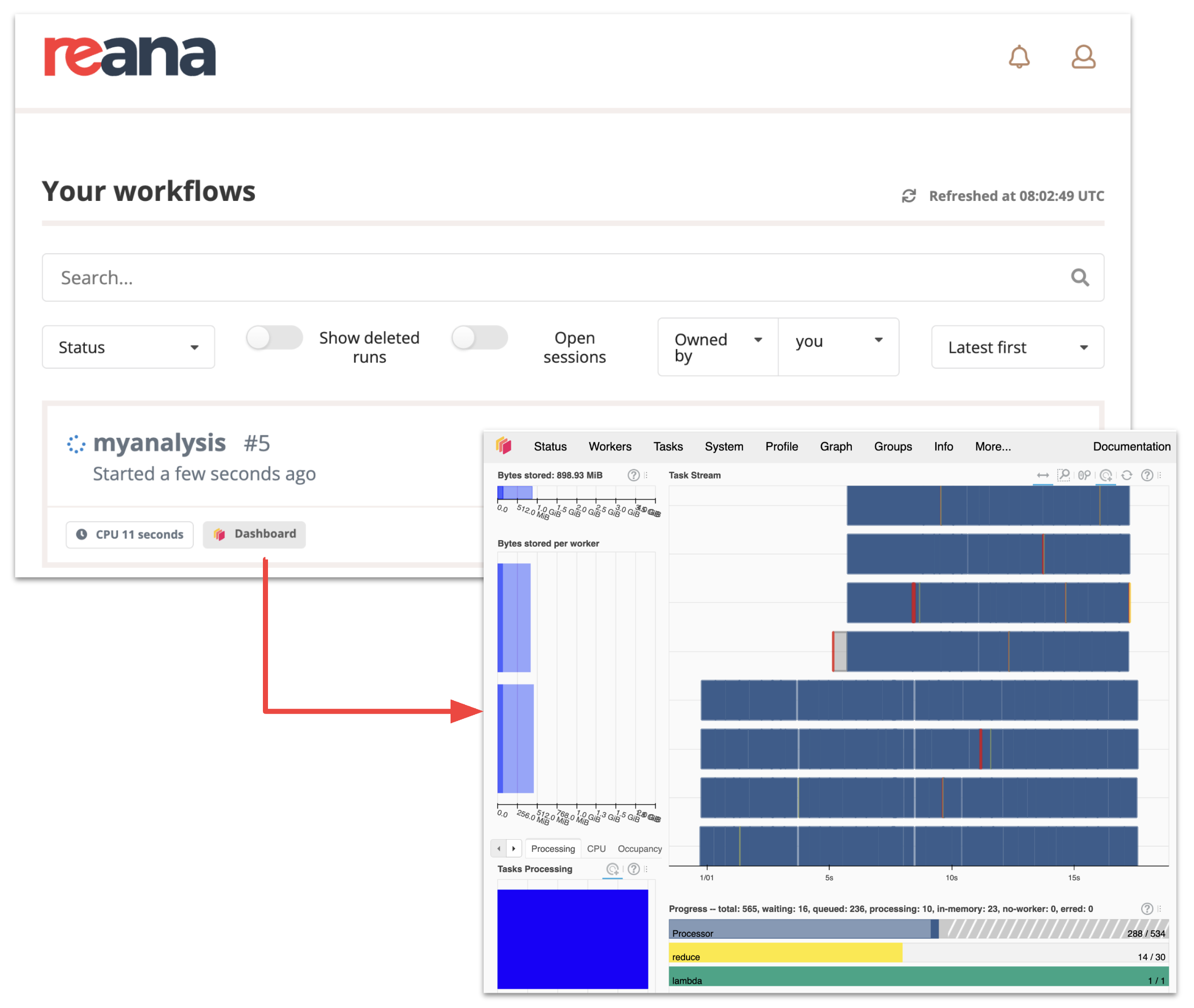} \hskip5mm
\includegraphics[height=0.27\textheight]{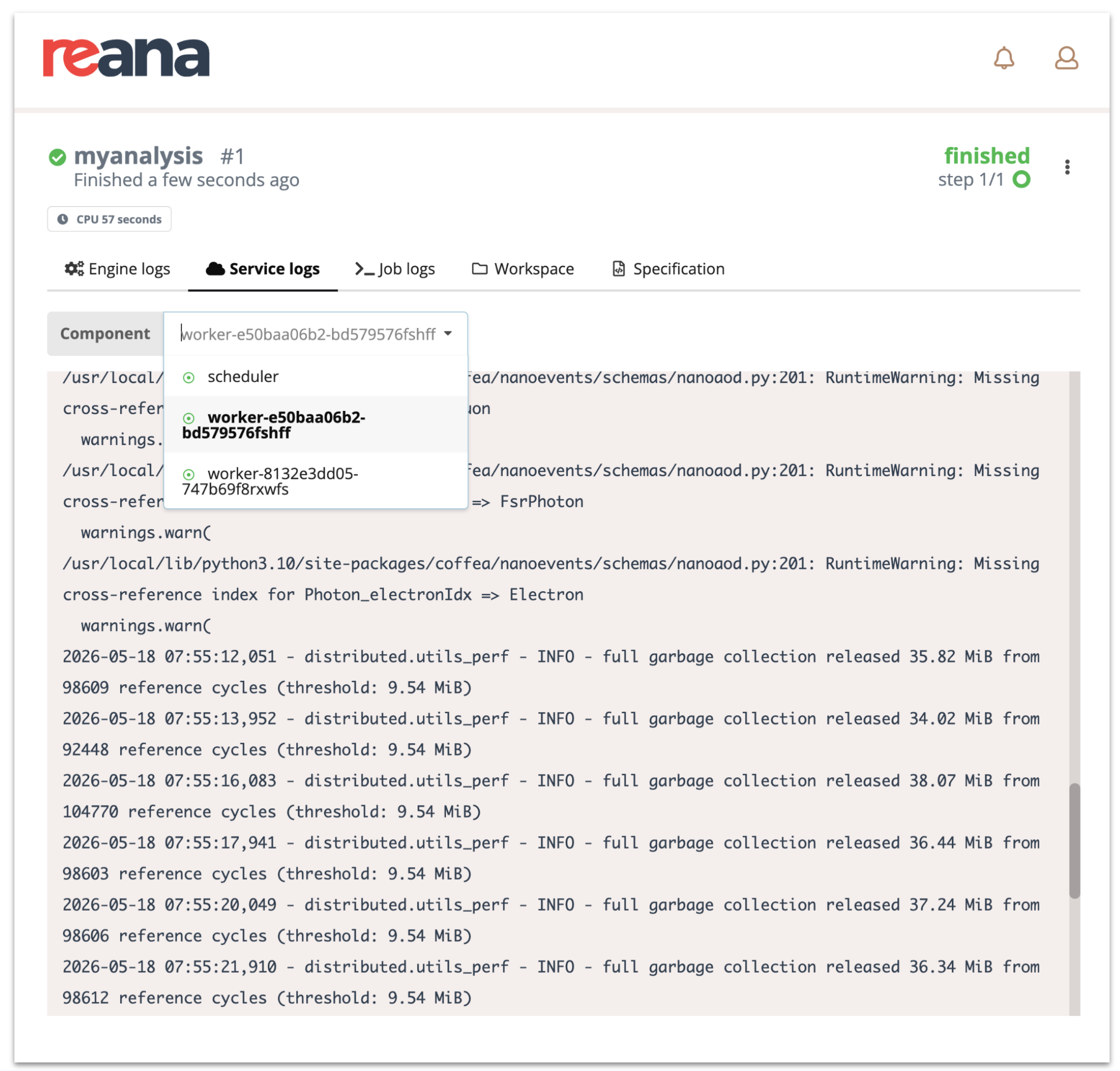}
\end{center}
\caption{REANA web interface with a running Dask-based analysis example. The Dask scheduler and worker logs are captured and exposed via a new Service tab on the usual web interface.
 \label{fig:reana_dask_ui}}
\end{figure}

Several recent developments in the REANA ecosystem illustrate on how to make the preserved data more easily reusable and reinterpretable in the future. One notable illustrating example is the recently-introduced support for Dask computational workflows (see Figure~\ref{fig:reana_dask_ui}). 

\begin{figure}[hhh]
\begin{center}
\includegraphics[height=0.18\textheight]{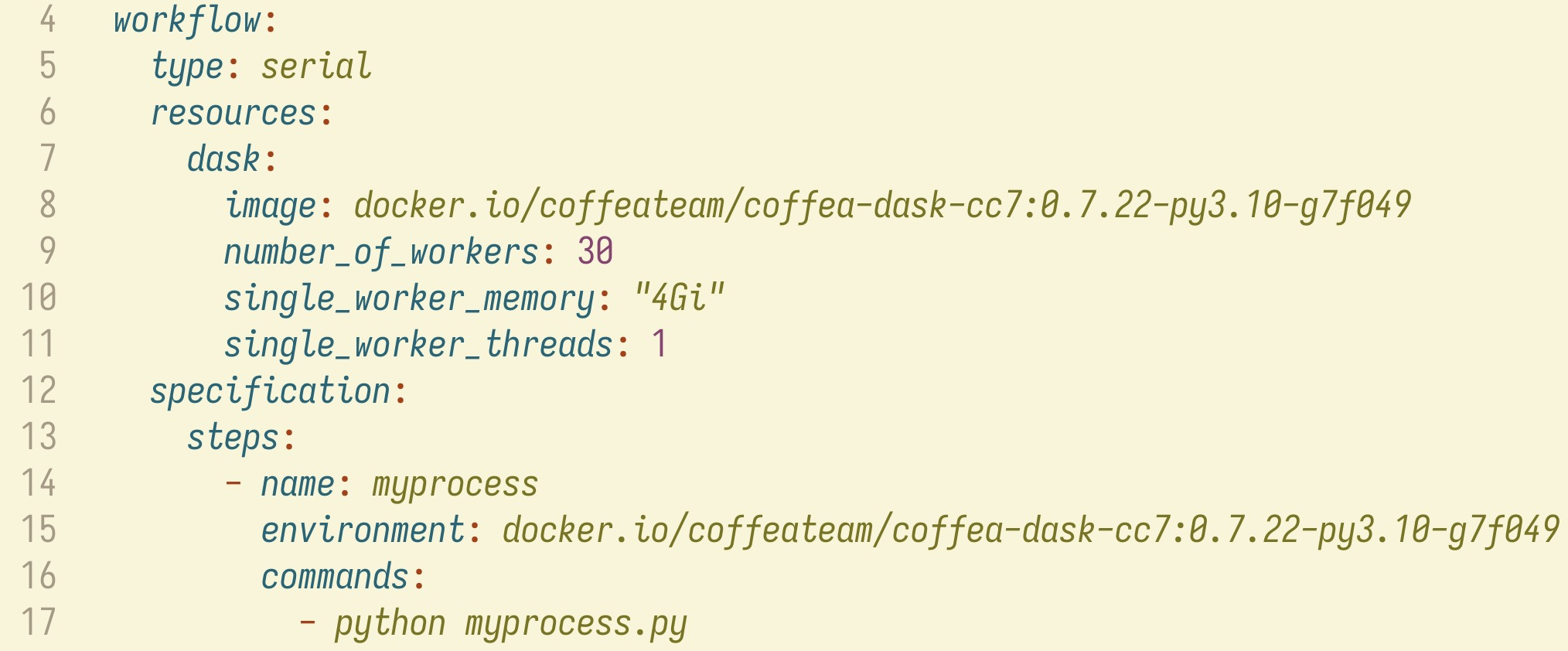}
\end{center}
\caption{An example of a REANA workflow specification using Dask resource hints to specify that this analysis needs notably Dask  Coffea version 0.7.22 with 30 workers of 4 GiB memory each and using single-thread computing. REANA will respect the hints specified by the  analyst and will spawn this concrete analysis environment for the execution. \label{fig:reana_dask_hints}}
\end{figure}

Dask is a popular Python library allowing analysts to scale data computations from local computers to large data centers, but it is not serialisable into a descriptive workflow language similar to CWL, Snakemake or Yadage. The support for Dask-based data analyses was therefore implemented by a new REANA component that is studying the workflow hints specified by analysts and then ensuring that the matching Dask cluster, the Dask scheduler and the Dask workers are spun up for the duration of the analysis and then tear down after its successful execution. It is noteworthy that each analyst can specify their custom desired Dask version (see Figure~\ref{fig:reana_dask_hints}) and the REANA platform automatically takes care of instantiating the specified original analysis environment for its future reuse. In this way, even if several researchers ask for mutually-incompatible Dask versions, REANA system spawns several Dask clusters to suit these needs. This helps to ensure that the original Dask analyses will be reinterpretable in the future even if the underlying data frame analysis stacks may change in incompatible ways.

Another notable recent development was prototyping the federated execution of scientific workflows across the nascent European Open Science Cloud Federation. We have amended REANA platform to allow dispatching of different part of analysis computations to different EOSC Nodes based on available resources or on data proximity. This ``near-data'' computing paradigm is well known from the experimental particle physics computing as well as other scientific disciplines; we have demonstrated how to bring it to the nascent EOSC Federation by means of allowing analysts to specify workflow hints similar to those of Figure~\ref{fig:reana_dask_hints} to instruct Snakemake workflow engine to dispatch various computation steps to various EOSC Nodes without the analyst having to worry about any implementation details~\cite{GuerrieriReanaEosc2025}. The nascent EOSC Federation efforts in distributed scientific workflows may lead to enabling theoretical physicists, machine-learning scientists, university teachers and students to study and reinterpret the preserved open data in a more accessible way.

The rising use of Artificial Intelligence and the success of Large Language Models (LLMs) in assisting researchers with reinterpreting past analyses is another topic of recent activity. Work has begun on enabling AI agents to communicate with the REANA system to verify LLM-generated workflows and validate them on a running platform in order to reduce hallucinations when assisting researchers in writing Snakemake-based analysis workflows. A nascent joint effort in the REANA community by the ATLAS collaboration and by the astronomy collaborators~\cite{ArmanPhysicsLLM2026} aims at consolidating and facilitating the use of LLMs in workflow authoring. 

These recent developments in the REANA ecosystem show how the data preservation efforts, focused on capturing data together with code and runnable usage examples, combined with workflow authoring through the Large Language Models, can lead to facilitating the open data reuse and reinterpretation in the light of testing new theories in the future.

\subsection{Using generative AI to extract dataset information from journal articles\label{sec:genai}}
{\small
 \it Authors: Emily Rensch (Siena University; Matthew Bellis (Siena University (US))
}

The CMS experiment has released approximately 5~petabytes of research-quality open data through the CERN Open Data Portal~\cite{Lassila-Perini:2021}, spanning collision data and simulated samples from the first two LHC runs. Roughly 100 publications to date have made use of these datasets, and the rate of new releases is set to accelerate significantly over the coming years. Storage planning for future releases requires trade-offs. One of these trade-offs includes the possibility of moving some datasets to tape-based cold storage. Retrieval could take days to weeks from this storage. Thus, it becomes increasingly valuable to understand in detail how the community actually uses the data. Existing tracking efforts can identify which specific datasets have been cited in publications. However, citation metadata alone does not reveal more granular information such as whether an analysis used an entire dataset or only a small fraction of it. It also doesn't reveal how much data was actually processed in terms of the number of events or bytes.

To address this gap, we have undertaken an exploratory study using large language models (LLMs) to extract structured dataset usage information from the text of journal articles. The core question is whether a generative AI tool can reliably identify, for each paper, which CMS open datasets were used, how many events were processed, and what fraction of the available dataset that represents. Initial prototyping used the LangExtract library paired with PyMuPDF for PDF text extraction and the Google Gemini model. However, this approach produced incomplete extractions with meaningful information loss during the PDF-to-text conversion. A subsequent attempt to use the OpenAI ChatGPT API~\cite{OpenAI} in five papers proved to be more structured. The results were returned in CSV format, but showed a tendency to hallucinate values not present in the source text and ran at a cost of approximately \$1.24 per paper.

The current approach uses the Anthropic Claude API~\cite{Anthropic:Claude}, specifically the Claude Sonnet~4 model. After prototyping prompt designs interactively on the Claude.ai web interface, a Python-based pipeline was developed. The python code extracts text from PDFs using PyMuPDF, chunks the text into manageable segments, and passes each chunk to the Claude API with a carefully constructed prompt specifying the fields to extract. Instructions are also provided for handling missing values. A key improvement over the earlier approach was the use of one of Claude's tools (structured output via the API's function-calling capability), which substantially reduced parsing errors in the returned data and enforced a consistent schema. The pipeline processes each paper and writes results to a CSV file containing fields for dataset name, number of events used, data size in bytes, and related metadata. Processing 15 papers took approximately 4 minutes and cost under \$10 in API costs. This result represented a substantial improvement in both speed and cost relative to the ChatGPT prototype.

\begin{figure}[hhh]
\includegraphics[height=0.18\textheight]{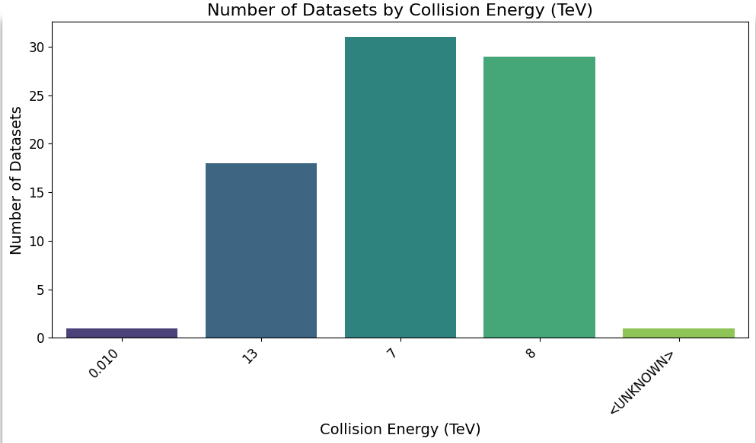}\hfill
\includegraphics[height=0.18\textheight]{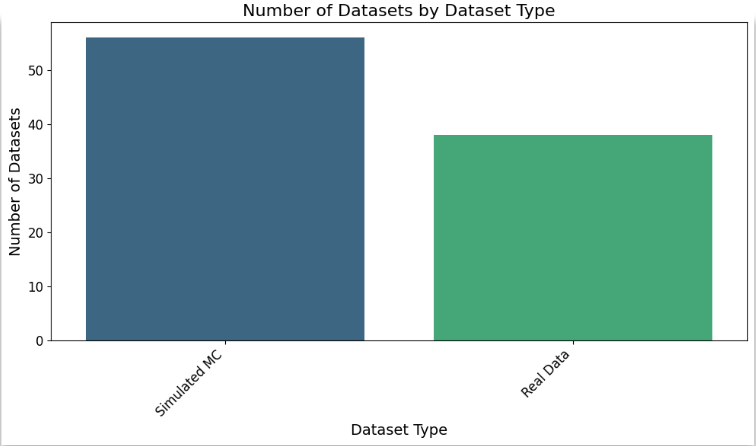}
\caption{Results from text analysis of 15 journal articles using Claude and the Anthropic python API. The left plot shows the collision energy of the open data datasets used in the articles and the right plots shows the dataset type. Multiple papers used multiple datasets, and so the number of entries in the plots is greater than 15. \label{fig:claude}}
\end{figure}

Early results are encouraging. The extracted information is broadly consistent with manual inspection of the papers, and the pipeline is well-positioned to be scaled up to the full scope of approximately 100 papers. Some examples showing the collision energy and dataset type are shown in Fig.~\ref{fig:claude}. Several challenges remain: not all papers report the granular dataset usage information we seek (e.g., many cite a dataset DOI but do not state the number of events analyzed), and the current workflow requires manual downloading and uploading of PDF files. Planned next steps include automating PDF retrieval directly from arXiv or publisher DOIs, expanding to the full paper scope, performing systematic spot-checks of extracted values against known dataset sizes from the CERN Open Data Portal, and eventually open-sourcing the pipeline code. This work demonstrates that LLM-based extraction is a viable and efficient approach to building a richer picture of how the particle physics open data community uses its available resources.

{\it Portions of the text in this section (Section~\ref{sec:genai}) was generated using Claude. The authors of this section (listed above) are entirely responsible for the content of this text.}


\subsection{Data ORchestration Agent (DORA) for AI-Ready Scientific Data in Large-Scale Facilities}

{\small \it\it Author:  Zhengde Zhang (IHEP) }

Scientific discovery increasingly relies on AI models, which require high-quality, AI-ready datasets. Yet, managing and preserving complex data from large-scale facilities remains a bottleneck. Large-scale research infrastructures such as the High-Energy Photon Source (HEPS) and the China Spallation Neutron Source (CSNS) generate staggering volumes of experimental data. However, this raw output remains largely inaccessible to AI-driven analysis due to pervasive issues of incompleteness, inconsistent curation, and poor interoperability. Bridging the gap between raw experimental output and AI-ready datasets is now one of the central challenges in scientific data management. FAIR principles~\cite{wilkinson2016} provide the guiding framework, but automated, scalable tools are required to enforce them in practice.

\paragraph{The DORA Framework}

\begin{figure}[h]
  \centering
  \includegraphics[width=0.75\textwidth]{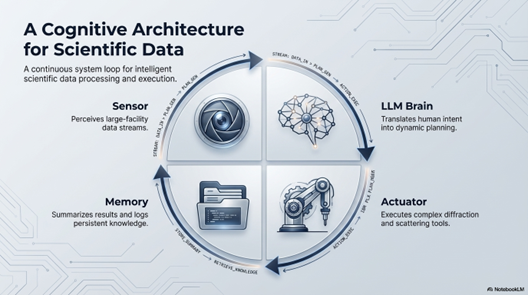}
  \caption{The cognitive architecture of DORA---a continuous system loop comprising Sensor, LLM Brain, Actuator, and Memory components.}
  \label{fig:dora_arch}
\end{figure}

We introduce the Data ORchestration Agent (DORA), an AI-driven framework that automates and optimizes the entire scientific data lifecycle---from ingestion and processing to preservation and provisioning---ensuring that all outputs are Findable, Accessible, Interoperable, and Reusable (FAIR) and ready for downstream machine learning applications. DORA is built on a cognitive architecture comprising four tightly coupled components (see Figure~\ref{fig:dora_arch}):

\begin{itemize}
  \item \textbf{Sensor:} Automatically ingests experimental parameters and detects facility outputs without manual intervention.
  \item \textbf{LLM Brain:} Parses natural language requests from scientists and decomposes complex analytical goals into actionable, multi-step workflows.
  \item \textbf{Actuator:} Seamlessly commands specialized domain software (e.g., GSAS-II) to perform advanced structural analysis.
  \item \textbf{Memory:} Generates automated LLM summaries, aligns multimodal data, and preserves results as FAIR-compliant records.
\end{itemize}

A key capability of DORA is its \textit{Engine of Completeness}---automated metadata completion that ensures missing experimental context never compromises downstream AI training. Given a user proposal in natural language, DORA can automatically infer and fill in missing experimental parameters, cross-reference related datasets, and produce richly annotated, standards-compliant records. This transforms sparse, heterogeneous facility outputs into dense, coherent training corpora.

\paragraph{Impact and Scale}

To demonstrate DORA's capabilities on a concrete scientific benchmark, we deployed a specialized DORA agent---named \textit{Dr.\ Sai-Rongzai}~\cite{rongzai2024}---at the CSNS Data Analysis Group, targeting autonomous neutron powder diffraction analysis. This agent achieved results comparable to domain experts in structural refinement tasks, validating DORA's ability to deliver expert-level scientific output without manual oversight.

\begin{figure}[h]
  \centering
  \includegraphics[width=0.75\textwidth]{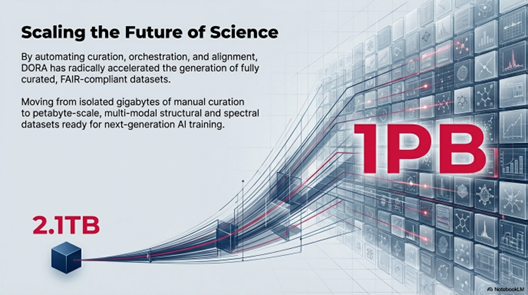}
  \caption{Data scale achieved by DORA at IHEP facilities---from 2.1\,TB of manually curated data to over 1\,PB of fully curated, FAIR-compliant, AI-ready multi-modal datasets.}
  \label{fig:data_scale}
\end{figure}

At the facility scale, DORA has radically accelerated the generation of fully curated, FAIR-compliant datasets. By automating curation, orchestration, and alignment, DORA has enabled the transition from terabyte-scale manually curated archives to petabyte-scale AI-ready multimodal data collections (see Figure~\ref{fig:data_scale}).

\paragraph{Outlook}

accelerates scientific discovery. It represents a shift toward autonomous, scalable, and sustainable data ecosystems for large-scale research infrastructures. Future work will extend DORA to additional beamlines and facility types, deepen integration with international data repositories, and incorporate federated learning workflows to enable cross-facility model training while respecting data sovereignty.

\section{The larger landscape}

\subsection{Progress and Prospects for Data Preservation at IHEP}
{\small
 \it Authors: Hao Hu, Gang Chen, Fazhi Qi (IHEP)
}

The Institute of High Energy Physics (IHEP) constructs and operates several large-scale scientific facilities, including BESIII, JUNO, Daya Bay (DYB), LHAASO and so on. The data generated from these experiments are crucial for driving discoveries and innovations in high-energy physics and related fields. The data volume and operation period of major experiments at IHEP are shown in the table 1. As summarized in the table, Daya Bay experiment has been retired in 2020 and BESIII experiment is scheduled for decommissioning in 2030. The data preservation of BESIII after its retirement requires immediate attention to its long-term data preservation strategy. In contrast, other major experiments such as JUNO and LHAASO, are poised for sustained operation over decades. For these experiments, we propose adopting DPHEP Level 4 framework to establish robust data policies, ensuring the enduring value and accessibility of these extensive scientific archives.

\begin{table}[htbp]
\centering
\begin{tabular}{|l|l|l|l|l|}
\hline
\textbf{Research Area} 
& \textbf{Experiments} & \textbf{Operation} & \textbf{Storage} & \textbf{Data produced} \\
& \textbf{/ Facilities} & \textbf{period} & & \textbf{(per year)} \\
\hline
\multirow{3}{*}{Particle Physics}
& BESIII (BEPCII) & 2009--2030 & 10\,PB & 750\,TB \\
\cline{2-5}
& JUNO & 2025--2055 & 11\,PB & 3\,PB \\
\cline{2-5}
& DAYA BAY & 2011--2020 (Retired) & 1.3\,PB & -- \\
\hline
\multirow{2}{*}{\begin{tabular}{@{}l@{}}Light source\\/ Neutron source\end{tabular}}
& HEPS & 2025--2055 & 400\,TB & $>$300\,PB \\
\cline{2-5}
& CSNS & 2018--2048 & 800\,TB & 1\,PB \\
\hline
\multirow{3}{*}{Astroparticle Physics}
& LHAASO & 2023--2043 & 45\,PB & 10\,PB \\
\cline{2-5}
& HXMT/GECAM/AliCPT & -- & 4.7\,PB & 500\,TB \\
\cline{2-5}
& HERD & 2027--2037 & 140\,TB & 3\,PB \\
\hline
\end{tabular}
\caption{Data Volume and Operation Period of Major Experiments at IHEP}
\label{tab:facilities_vline}
\end{table}

IHEP has been actively engaged in advancing the long-term preservation of data, software, and knowledge associated with these facilities.

A prime example of this commitment is the BESIII experiment. We have established a comprehensive data preservation framework encompassing raw and processed data, analysis software, documentation, and metadata. Moving beyond mere preservation, we are developing an AI-empowered Data Ecosystem for BESIII aimed at fully exploiting the scientific potential of the BESIII datasets. This initiative addresses user needs across short-, medium-, and long-term horizons. Specifically, our near- to mid-term focus lies in advanced data management and the extraction of structured knowledge from existing literature and documents. Looking further ahead, we aim to train a foundational AI model that serves as a compact, efficient representation of complex scientific information, including ditributions, detector effects and so on. With this model, researchers can do automated physics analysis. To steer these efforts, BESIII Data Ecosystem Committee~\cite{BESdataComittee}  was established in 2025, which is tasked with formulating strategies and mechanisms to preserve the capacity to extract new science from the BESIII data in the long term and coordinate efforts to align the BESIII data ecosystem with open data and FAIR practices.

Furthermore, IHEP promotes open data initiatives across other experiments. For the Daya Bay experiment, the full datasets~\cite{dybdatasets} was officially released by the collaboration in December 2025. The release comprises 5.55 million Inverse Beta Decay (IBD) events collected between 2011 and 2020 across eight detectors, providing sufficient statistical power to reproduce key oscillation parameters such as $\sin^2 2\theta_{13}$ and $\Delta m^2_{32}$. Committed to full transparency, the collaboration has made the dataset and corresponding analysis tools publicly accessible via the Zenodo repository and the NHEPSDC portal. Furthermore, streamlined analysis datasets are available on GitHub and PyPI, supported by open-source analysis code hosted on GitHub capable of reading data in multiple formats.

For the JUNO experiment, a preliminary data policy has been formulated, which implements a three-tier architecture. Raw data (RAW) are transferred to the Tier-0 center (IHEP EOS) for dual-copy tape archival and distributed to Tier-1 sites via the Distributed Computing Infrastructure (DCI). At Tier-0, data are converted to ROOT-based raw data (RTRAW) for preprocessing. Finally, the ESD reconstructed data are categorized into KUP and ReProd, and retained as single-copy disk storage at analysis sites without tape archive.

The LHAASO collaboration is committed to preserve and open its data, at different levels of complexity, and to allow their re-use by a wide community. Level-1 offers fully open access to publications, documentation, and code. Level-2 provides simplified datasets for education and outreach. Level-3 authorizes external researchers access calibrated reconstructed data for new analyses via partnership proposals. Finally, Level-4 strictly restricts access to the raw experimental data to collaboration members only, due to its complexity and volume, ensuring secure and expert-led handling.

In conclusion, the future of data preservation for large-scale scientific facilities relies on a multi-faceted approach encompassing long-term archival strategies, sustainable funding, and the integration of advanced AI technologies. While raw data from all experiments is securely archived on tape for long-term preservation, we are also prioritizing the curation of select open datasets to maximize scientific impact. The successful practices established by experiments like BESIII serve as a valuable model for extending robust data ecosystems to other major facilities such as JUNO. Securing resources for long-term data preservation requires sustained efforts to diversify funding streams. We are actively pursuing support from various channels, bolstered by the launch of several national-level projects specifically aimed at ensuring the longevity and accessibility of scientific data. Furthermore, the field is rapidly evolving towards intelligent data management, leveraging data agents to create AI-ready datasets and developing LLM-powered knowledge systems. Ultimately, our ambition is to cultivate foundation models capable of deeply understanding physics data, thereby automating analysis and unlocking new scientific insights from archived research.

\subsection{CERN Open Data: Policy to implementation}
{\small
 \it Author: Jamie Boyd (CERN)
}

In 2020 CERN released its Open Data policy document~\cite{CERN-OpenData-policy}, which was endorsed by the 4 large LHC experiments at CERN. The policy was written to balance the concerns from the experiments related to loss of ownership of their data, and resource issues with the desire to be as open as possible with the data. The policy had all of the experiments releasing a substantial part of their data with a latency of 5 years since the data was collected. Since then the smaller LHC experiments have also endorsed this policy, and discussions are ongoing with other experiments at CERN. 

As expected from the policy, all four large LHC experiments have now released substantial L3 datasets on the open data portal. The datasets released include proton-proton collisions from Run 1 ($\sqrt{s}$=7, 8 TeV) and the first part of Run 2 ($\sqrt{s}$=13 TeV) as well as heavy-ion collision data. In addition, the experiments all continue to release L1 data related to the published physics results, as well as tailored L2 datasets for outreach and education.

\begin{table}[thb]
  \centering
  \begin{tabular}{|l|c|c|c|c|}
  \hline
   & {\bf ALICE} &  {\bf ATLAS} & {\bf CMS}  & {\bf LHCb} \\
  \hline
    Expected   & 
    300 & 200 & 4600 & 4600 \\
    (end of 2025) (TB) & & & & \\
    \hline
    Current status  & 
    70 & 930 & 4600 & 800 \\
    (TB) &  &     & (36\% on tape) & \\
    \hline
    Comment & 
    Finalizing &
    New additional & 
    Started to  & 
    NtupleWizard \\
    & commissioning &
    generator level & 
    move significant & 
    model uses \\
    & of format &
    information &
    data to tape & 
    less resources \\
\hline
    \end{tabular}
    \caption{Summary of the storage resources expected and actually used at the end of 2025. 
    }
    \label{tab:OD-resources}
\end{table}

One of the key challenges with the implementation of the policy is the storage resource requirements. CERN agreed to provide storage for the first 5 years of the implementation phase, which has now come to an end, and discussions are ongoing with CERN for the next five year period. As can be seen in Table~\ref{tab:OD-resources}, currently there is more than 6 PB of open data available through the portal, and this is projected to grow substantially to around 20 PB by the end of 2030. In order to reduce the resource needs, CERN IT has implemented a tape backend to the CERN Open Data portal. Given that tape is around three times cheaper than disk, this will lead to significant savings. This feature has recently been put into production, and as can be seen from the table, already more than 1.5 PB has been moved to tape, with more data being moved. How best to use the tape backup will be a learning experience based on data use patterns,in order to balance the resources requirements with allowing efficient use of the data.

The LHCb experiment has implemented a new system (the LHCb Ntupling Service~\cite{LHCb-ntupleService}) to allow external users controlled access to the full data stored on internal experiment resources, and to output a small tailored dataset based on the user requirements. This is a good development which will save significant storage resources for the open data.

Progress has been made in the monitoring of the open data stored and accessed on the portal, with a grafana dashboard showing the relevant information~\cite{open-data-monitoring}, including several KPIs. Discussions are ongoing about adding additional monitoring metrics or KPIs, and this will likely evolve as the open data usage increases. 

Discussions are ongoing with the CERN non-LHC experiments about either endorsing the existing policy or writing their own policy document, targetted towards the specific scientific community and practices. In this context there are dedicated policy documents from the ISOLDE facility~\cite{ISOLDE-policy} and recently from the nTOF facility~\cite{nTOF-policy}. The hope is for CERN experiments based at the SPS will endorse the LHC policy, although there are concerns about the internal human resource needs for implementation of the policy.


\subsection{A Holistic Approach to Data, Analysis, and Knowledge Preservation at BNL}
{\small
 \it Author: Jerome Lauret (Brookhaven National Laboratory / BNL); Mohammad Atif (BNL); Zhihua Dong (BNL); ; Vincent Garonne (BNL); Eric Lancon (BNL); Alexander Prozorov (Czech Technical University) 
}

The RHIC Data and Analysis Preservation Program (DAPP) at Brookhaven National Laboratory safeguards 25 years of data, software, and institutional knowledge from the PHENIX, STAR, and sPHENIX experiments. As the community transitions to the Electron-Ion Collider (EIC) era, DAPP offers a complete model for preserving scientific value beyond the lifetime of active experiments, with direct relevance to the broader HEP/NP community preparing for HL-LHC, the EIC, and future facilities.

DAPP establishes three complementary pillars that together form a socio-technical preservation system. First, it establishes core preservation infrastructure for long-term data and software retention with enhanced metadata to enable result reanalysis and reproducibility. Second, it introduces SciBot, a locally deployed AI assistant using Retrieval-Augmented Generation (RAG) and locally hosted LLMs, which provides secure, natural-language access to preserved RHIC knowledge while critically guaranteeing data sovereignty for private documentation. Third, the program establishes institutional governance frameworks and formalized preservation workflows ensuring long-term accountability and stewardship. Together, these provide a practitioner's perspective on the full preservation stack—from raw data and analysis workflows to the organizational knowledge that contextualizes them—with cross-cutting lessons on metadata standards, access-control federation, AI integration, and institutional structures needed to sustain preservation throughout experiment life-cycles.

\vspace{0.5cm}
\textbf{The Preservation Challenge: Context and Constraints}

\begin{figure}[h]
  \centering
  \includegraphics[width=0.50\textwidth]{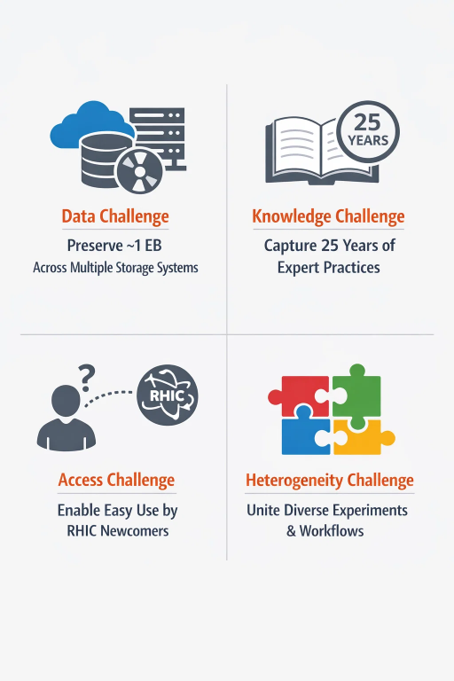}
  \caption{RHIC preservation requires a holistic approach addressing four complementary challenges: Data Challenge (safeguarding ~1 exabyte across multiple storage systems), Knowledge Challenge (capturing 25 years of expert practices before domain specialists retire), Access Challenge (enabling easy use by future RHIC newcomers), and Heterogeneity Challenge (unifying diverse experiment workflows and formats).}
  \label{fig:why_holistic}
\end{figure}

RHIC has generated an invaluable scientific legacy: 25 years of continuous operations across PHENIX (2000--2016), STAR (2000--2025), and sPHENIX (2023--2025), producing over 600 peer-reviewed publications and nearly one exabyte of irreplaceable datasets. The collider achieved unique physics signatures—notably, it remains the only machine to produce polarized proton-proton collisions at multiple energy scales, creating a rich experimental ecosystem unlikely to be replicated at future facilities. As shown in Figure~\ref{fig:why_holistic}, effective preservation requires addressing four complementary challenges.

However, preservation occurs under severe temporal and resource constraints. Detector expertise is retiring faster than active physics analysis ends; institutional knowledge accumulated over decades risks permanent loss as domain specialists depart. Simultaneously, computing infrastructure reaches end-of-life: CPU resources are projected to decline to approximately 50\% of current capacity by 2030, and disk storage to approximately 20\% of current levels. The RHIC operations program provides no expectation of hardware replacement beyond this window. These dual constraints, disappearing human expertise and finite computational resources, demand a preservation strategy that captures critical knowledge \textit{now} while infrastructure remains available for intensive reprocessing tasks.

This challenge extends beyond simple data storage. Following the DPHEP community framework~\cite{DPHEP:2023blx}, preservation must maintain the ability to perform reproducible analysis and verify published results, not merely archive files. For RHIC, this means preserving raw data alongside analysis workflows, reconstruction software, detailed metadata, and the tacit knowledge embedded in experiment-specific analysis techniques.

\vspace{0.5cm}
\textbf{Two-Phase Strategy: Exploiting the Temporal Window}

DAPP's core insight is that preservation must align with hardware and human life-cycles. The program divides into two distinct phases, each with different objectives and resource expectations.

Phase I (2025--2030) represents intensive preservation during the final years of operational computing capacity. This phase prioritizes urgent knowledge capture and AI integration, final global data reprocessing leveraging available computing infrastructure, and infrastructure stabilization and normalization. The 2025 Au+Au run, which may double RHIC's historical data volume, makes this window critical for completing comprehensive reprocessing before computing resources decline irreversibly. During Phase I, the team must also establish standardized metadata, document analysis workflows, containerize software environments, and begin knowledge transfer to AI systems.

Phase II (2030+) emphasizes sustainable operations with significantly reduced resources. Phase II deliberately abandons large-scale reprocessing; computing capacity will not permit it. Instead, emphasis shifts to functional preservation: ensuring analysis reproducibility from previously processed datasets, maintaining robust data discoverability through persistent metadata and AI-assisted search, and supporting continued scientific discovery. This phase succeeds only if Phase I completes its intensive preservation tasks successfully. This realistic phasing reflects hard resource constraints rather than idealized scenarios, enabling long-term sustainability.

\vspace{0.5cm}
\textbf{Technical Infrastructure: Functional Data and Workflow Preservation}

\begin{figure}[h]
  \centering
  \includegraphics[width=0.50\textwidth]{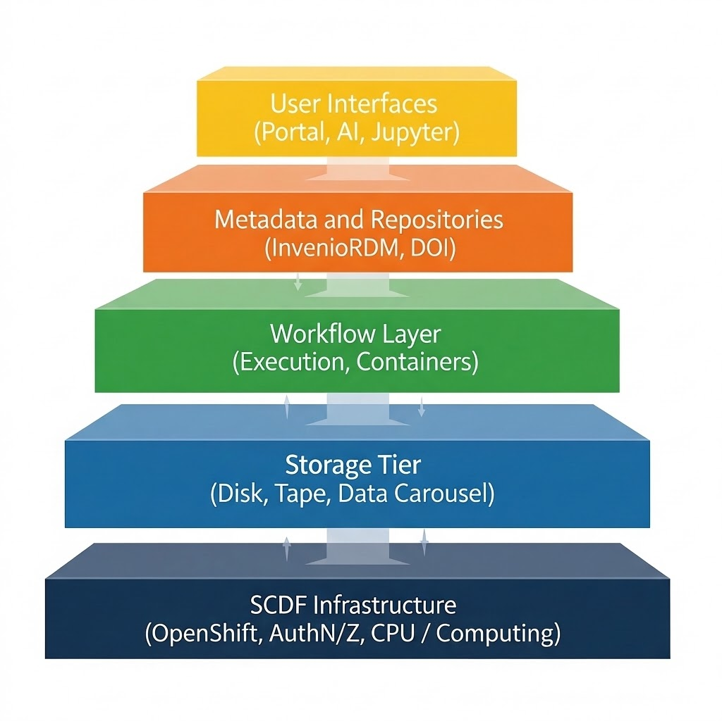}
  \caption{RHIC DAPP preservation infrastructure stack. Five integrated layers enable functional preservation prioritizing analysis-ready objects on disk for continued scientific usability over raw data archived on tape.}
  \label{fig:infrastructure}
\end{figure}

The preservation stack (Figure~\ref{fig:infrastructure}) comprises five integrated layers: foundational facility infrastructure (SCDF OpenShift, authentication/authorization systems), tiered storage (hot disk for frequently accessed analysis objects, cold tape for raw data archival with data carousel for intelligent staging), workflow execution environments (containerized analysis with tools such as the REANA platform), metadata repositories (InvenioRDM with DOI assignment), and user-facing interfaces (Portal, Jupyter environments, AI services).

The preservation strategy prioritizes \textit{functional} over \textit{bit} preservation. Raw data archives are maintained on tape for posterity, but large-scale reprocessing in Phase II is impractical. Instead, DAPP performs a final, comprehensive reprocessing in Phase I, producing consistent, well-documented analysis objects stored on accessible disk with complete provenance metadata. Containerized workflows preserve complete software environments (reconstruction code, analysis frameworks, dependencies), ensuring reproducibility long after original development.

Critically, metadata curation remains non-delegable human work. While AI tools assist in validation and discoverability enhancement, human expertise assessing data quality and scientific validity is irreplaceable. This reflects community consensus that automated preservation, while valuable, cannot substitute for domain-specific human judgment.

\vspace{0.5cm}
\textbf{SciBot: Knowledge Accessibility and Transfer}

\begin{figure}[h]
  \centering
  \includegraphics[width=0.90\textwidth]{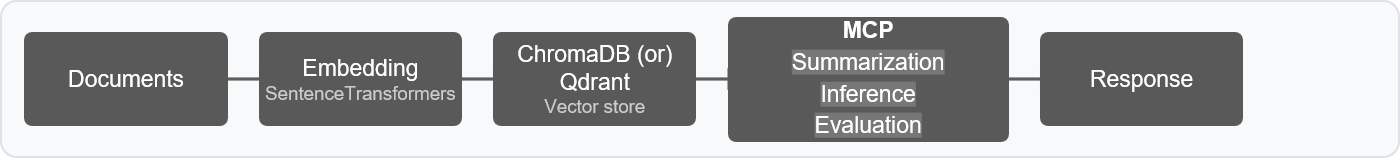}
  \caption{SciBot reasoning pipeline: Documents are embedded and indexed in vector database (ChromaDB or Qdrant), queried through modular reasoning stages (Retrieval → Summarization → Inference → Evaluation) with flexible LLM backends. Federated identity controls access to non-public materials.}
  \label{fig:scibot}
\end{figure}

RHIC accumulated 25 years of fragmented institutional knowledge: technical notes, detector logbooks, software repositories, wikis, mailing list FAQs, and informal collaboration documentation. Future researchers—lacking deep domain expertise—will struggle navigating this landscape. SciBot addresses this through Retrieval-Augmented Generation (RAG) architecture, aggregating institutional documentation into a unified semantic index queryable in natural language~\cite{NYSD2025:SciBot}. The system's architecture, shown in Figure~\ref{fig:scibot}, enables natural-language queries across heterogeneous sources while maintaining data sovereignty.

Critically, SciBot emphasizes \textit{local deployment with data sovereignty}. Sensitive RHIC collaboration data remains on BNL infrastructure rather than uploaded to external cloud services. The system supports flexible backend options (vLLM, Ollama, LlamaCPP) with optional API access to frontier models (GPT-4o, Gemini) limited to public information only. The RAG architecture is deliberately model-agnostic via the Model Context Protocol (MCP), allowing LLM substitution as technology evolves without requiring system redesign.

SciBot enables rapid on-boarding for newcomers to RHIC analysis, assists in translating legacy analysis scripts into modern formats, and dramatically improves knowledge discoverability. However, it explicitly does not replace curated metadata, experiment-level validation, or human scientific judgment. The system assists discovery and accelerates learning; rigorous scientific validation remains human responsibility.

\vspace{0.5cm}
\textbf{Institutional Framework and Governance}

Beyond infrastructure and knowledge access, long-term preservation requires explicit institutional commitment and formalized workflows. DAPP establishes clear governance structures (Steering Committee oversight, Technical Working Groups, Implementation Teams) with defined responsibilities, regular risk assessments, and documented procedures. Roles are formalized (Readers, Curators, Managers, Owners), metadata standards explicitly documented, and compliance with FAIR principles and DOE requirements systematically tracked.

This institutional framework becomes essential in Phase II, when declining resources demand that organizational structure compensate for reduced staffing. By embedding preservation requirements into formal institutional processes during Phase I, DAPP creates conditions for sustainable stewardship independent of individual personnel transitions.

\vspace{0.5cm}
\textbf{Broader Lessons for the HEP and Nuclear Physics Communities}

Several cross-cutting insights from RHIC DAPP apply broadly to future large-scale experiments. First, preservation planning must precede operations end. Designing preservation after experiments cease is too late to capture expertise and optimize infrastructure. Future facilities (EIC, HL-LHC collaborations) should integrate preservation planning from inception. Second, realistic resource constraints enable sustainability. Explicitly accepting that Phase II cannot match Phase I intensity allows pragmatic, survivable long-term strategies rather than aspirational plans that fail due to unfunded mandates. Third, metadata and human expertise remain irreplaceable. AI tools enhance discovery but cannot replace domain-specific human judgment in data validation and scientific interpretation. Finally, institutional control over infrastructure matters. Deploying preservation systems on institutional infrastructure, rather than outsourcing to commercial providers, ensures security, long-term independence, and alignment with compliance and sovereignty requirements.

The overarching principle guiding DAPP is straightforward: \textit{the goal is not to store data, but to preserve the ability to use it.}

\subsection{Analysis Preservation as a Pillar of Long-Term Knowledge Reuse in HEP}
{\small
 \it Authors: Andy Buckley (University of Glasgow (GB)); Christian Gutschow (UCL (UK)); Martin Habedank (University of Glasgow); Tomasz Procter (Jagiellonian University (PL))
}


Discussions of long-term preservation in high-energy physics often emphasise event data, software environments, and documentation. Equally important, and increasingly mature, is the preservation of analyses: the executable logic that connects experimental data to published measurements through event-selections, derived observables, statistical procedures, and -- more recently -- machine-learning models.

Over the past decades, community-driven efforts have demonstrated that analysis-level preservation is both technically feasible and scientifically impactful, enabling reinterpretation of legacy measurements, cross-experiment validation, and educational reuse. Analysis preservation can take different forms. At one extreme, the ATLAS RECAST/REANA workflow~\cite{Simko:2018zzz} in principle allows the exact reproduction of the procedure carried out by the analysis team. However, in many cases simpler and less computationally demanding approaches can be used to achieve the same end more efficiently.

Not every re-use of an experiment's data needs to start from raw detector hits and work through the entire experimental data-processing workflow; for example, re-using the results from unfolded SM measurements requires only the analysis logic (in the form of a Rivet~\cite{Bierlich:2019rhm,Bierlich:2024vqo} analysis), as well as the unfolded results of the measurement (typically stored as histograms on HEPData~\cite{Maguire:2017ypu}). Such a workflow is used by the CONTUR~\cite{CONTUR:2021qmv} tool for BSM studies, as well as by PROFESSOR~\cite{Buckley:2009bj} (and the wider tuning ecosystem) for tuning of hadronisation and parton-shower MC models.

In this vein, as a practical example, we consider the re-use of LEP analyses preserved in Rivet. As shown in Figure~TODO, more than two decades after the collider ceased collecting data, LEP data is still being actively re-used via Rivet. These are primarily measurements of distributions crucial to event-generator tuning (e.g. Reference~\cite{DELPHI:1996sen} at DELPHI and Reference~\cite{ALEPH:2003obs} at ALEPH), though there are many other potential uses, including Dark Matter~\cite{DiMauro:2025vxp} and Information theory~\cite{Assi:2025ibi} studies. The MC tunes produced from the re-use of this LEP data are essential to producing reliable MC predictions for the LHC and beyond. However, this should also come with a warning for the LHC programme: while most Rivet analyses for LEP experiments were only written many years after the fact, the increase in analysis complexity since the LEP-era means analysis preservation from the LHC is fraught with more complexity and pitfalls than was the case for LEP. Preserving the LHC's analyses so they can be as influential for future physics as the LEP results have been requires active planning and effort from the LHC experiments.

\begin{figure}[htb]
\begin{subfigure}[b]{0.45\textwidth}
    \centering
    \includegraphics[width=\textwidth]{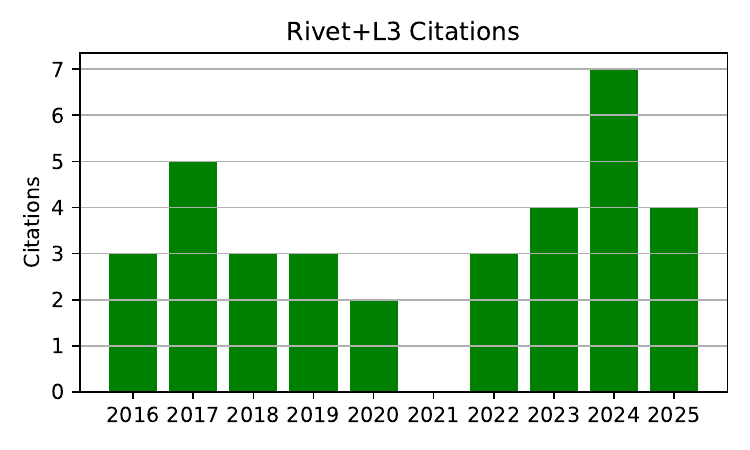}
    \caption{}
    \label{subfig:rivet_L3}
\end{subfigure}
\begin{subfigure}[b]{0.45\textwidth}
    \centering
    \includegraphics[width=\textwidth]{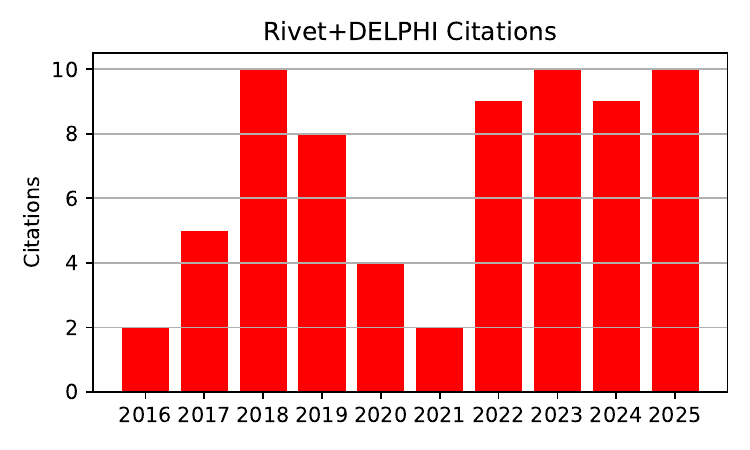}
    \caption{}
    \label{subfig:rivet_DELPHI}
\end{subfigure} \\
\begin{subfigure}[b]{0.45\textwidth}
    \centering
    \includegraphics[width=\textwidth]{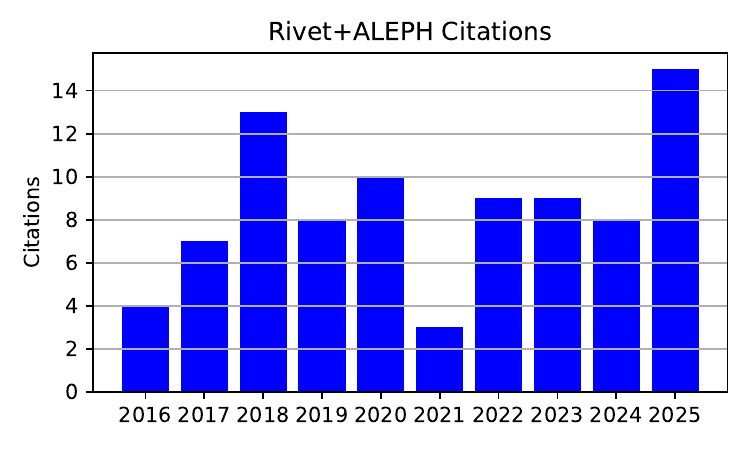}
    \caption{}
    \label{subfig:rivet_ALEPH}
\end{subfigure}
\begin{subfigure}[b]{0.45\textwidth}
    \centering
    \includegraphics[width=\textwidth]{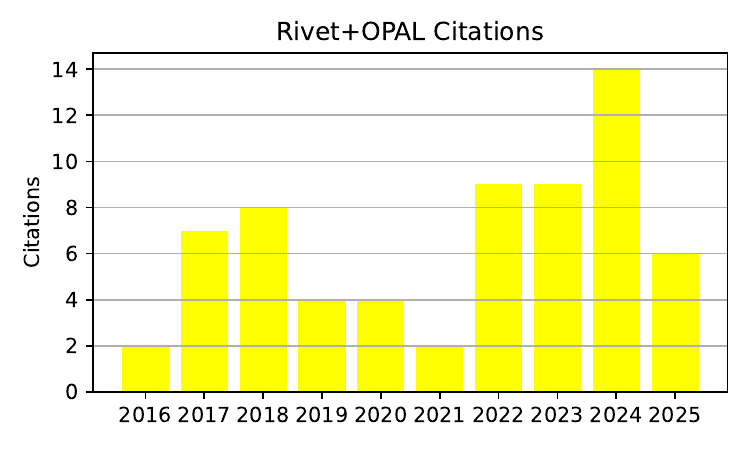}
    \caption{}
    \label{subfig:rivet_OPAL}
\end{subfigure}
\caption{Papers on the Inspire database citing both Rivet and a LEP analysis with a public Rivet implementation, split by the four LEP experiments. While this does not guarantee that Rivet was actually used, based on a random sample of 12 papers that where checked thoroughly, only one -- a review paper -- did not actually use Rivet. Note that the subfigures are not independent, i.e. it is possible for a single study to contribute a reference to multiple LEP experiments.}
\label{fig:rivet_lep_citations}
\end{figure}

It should be stressed that preservation of an analysis via Rivet is not in competition with the aim of fully preserving and re-using the data and software from e.g.~the LEP or B-factory experiments (as has been outlined elsewhere in this report). Indeed, the two can often be complementary: returning to the tuning example, re-analysis of LEP data with modern techniques by groups like the electron-positron alliance~\cite{} has the potential to even further improve the unfolded datasets available for tuning via Rivet.

A similar methodology is applied to detector-level ``search'' analyses, which can be preserved using tools such as Rivet, CheckMATE, MadAnalysis or ColliderBit. In this case, simplified detector simulations (such as DELPHES) or emulations (using e.g.~the Gambit BuckFast system~\cite{GAMBIT:2017qxg}, MadAnalysis' SFS~\cite{Araz:2020lnp}, or the Rivet smearing system~\cite{Buckley:2019stt}) are used instead of unfolding (estimating the true values leading to detector-reconstructed data) to bridge the gap between MC ``truth'' and detector-level results.

A growing challenge for analysis preservation is the ever-increasing reliance of modern particle-physics analyses on machine learning (ML). As emphasised by the Les Houches guidelines on reinterpretable ML~\cite{Araz:2023mda}, when dealing with black-box ML algorithms, it is essential that not only are the ML weights preserved and published, but also that the feature interface is thoroughly documented. Standardised formats such as ONNX are preferred, but tools such as petrifyML~\cite{Buckley:2025pqk} (which converts older proprietary formats into either ONNX or plain C++/Python code) can help when these are not available. As the trend in ML usage favours increasingly low-level inputs which may be difficult for reinterpreters to reproduce, the LH Guidelines suggest the use of surrogate ML models usable on detectorless MC-truth records, as were recently released for a recent ATLAS LLP search~\cite{}.

Analysis preservation provides a direct bridge between data-preservation and knowledge-preservation by encoding expert intent in forms that are both human-readable and machine-actionable. It also creates natural interfaces for AI-assisted documentation, automated validation workflows, and outreach-oriented reproducibility. While significant progress has been made—supported by broad collaboration-level engagement—coverage remains incomplete and sustainability often depends on limited dedicated resources. Recognising analysis preservation explicitly alongside data-, software-, and hardware-preservation will help consolidate existing successes into a durable, community-supported infrastructure for future research and public engagement.

\subsection{ICFA recommendations on data preservation and open science and their assessment}
{\small
 \it Author: Kati Lassila-Perini (Helsinki Institute of Physics (FI))
}

The Data Lifecycle panel was established by the International Committee for Future Accelerators (ICFA) in 2024 with a mandate to enhance global coordination on all aspects of the data lifecycle with the focus on FAIR practices and open science. As the first major undertaking, the panel called on scientists and experts involved in the preservation of high-energy physics data to join the effort of defining best-practice recommendations for data preservation and open science in HEP. The guiding principle of the editorial work was to keep recommendations actionable by the community by ensuring they remained concrete, specific and relevant to the indicated groups. The overall goal is the long-term scientific usability of experimental data within the community and beyond. The recommendations were released as a web application~\cite{icfa-best-pratices-app} and an ArXiv note~\cite{icfa-best-practices-note} in August 2025.

Data preservation initiatives in the high-energy physics domain, as indicated in this document, have made solid progress and attract increasing interest. However, the driving force behind the progress has often been individual efforts, and an important success factor is the presence of collaborators who maintain the knowledge of correct use of preserved data. The recommendations aim to provide a concrete to-do list for stakeholders at different levels, and clearly show how actions at one level can enable or facilitate actions at subsequent steps during the data and analysis lifecycle, ensuring that long-term scientific usability relies not on individual actions, but on systematic, step-wise work.

A special emphasis is given to knowledge preservation. Among the components of open science, the HEP community excels in open access publishing and is making good progress in releasing open data, but significant challenges remain in open methodology, i.e. preserving analysis knowledge: the workflows, software, and contextual documentation. The recommendations therefore cover infrastructure, working practices, and skills as the cornerstones for efficient knowledge sharing. They advocate for analysis software and workflow descriptions to become integral research outcomes and request that all supplementary information and contextual knowledge for data understanding and (re)use be preserved. Furthermore, the community must ensure researchers can acquire the necessary skills to work respecting FAIR practices.

To make these practices sustainable, the recommendations call for concrete planning, clear policies, and dedicated resources.

The panel will undertake regular evaluation rounds on how these recommendations are followed in the HEP domain. This will start with surveys specific to host laboratories and experiment management teams, in which each recommendation is assessed: whether it is applied (fully or partially), planned, not yet considered, or considered non-applicable. The input will be analysed and summarised in a public report with a focus on the overall picture, e.g. why certain actions were not taken, to identify common challenges and highlight successful actions. The goal of this work is to raise awareness and promote the actions needed to keep the full scientific value of unprecedented volumes of data with extraordinary complexity and richness that modern HEP experiments generate.

\subsection*{Conclusions and trends}
The workshop reemphasized (once more) that data preservation (in HEP, as elsewhere) is not merely archival but a scientific-dynamic and technological-rich process, enabling new discoveries, education, and interdisciplinary innovation. Several key trends and challenges deserve attention.

First, there is a noticeable shift toward AI and automation, with increasing reliance on artificial intelligence and machine learning for documentation, data curation, metadata extraction, and workflow storage and optimization. This reflects HEP's adaptation to modern computational tools, enabling more efficient and accurate data handling, but also documentation availability and collaboration in the long term.

Second, the momentum toward open science continues to grow, driven by the adoption of FAIR principles and open data policies. Major experiments, such as those at the LHC, are committing to public data releases, adhering to rules like the 5-year latency policy, which promotes transparency and scientific collaboration. This trend enables valuable opportunities for legacy data from former experiments, the transition being however jeopardized by the lack of resources and/or momentum. 

Another significant development is the revival of legacy data. Archival datasets from past experiments, such as OPAL and ALEPH, have been successfully reanalyzed using contemporary methods. This demonstrates the long-term value of data preservation, allowing new insights to emerge from historical resources, in particular towards future projects such as FCC and EIC. 

Additionally, collaborative frameworks are being strengthened, with international initiatives like EOSC EDEN and REANA working to standardize preservation practices across disciplines. These efforts enhance consistency and efficiency in global data management.

However, sustainability challenges remain. Long-term funding and institutional support are essential to maintain data preservation infrastructure, particularly for legacy experiments transitioning to archival modes. Without these resources, the accessibility and usability of valuable datasets could be at risk, limiting their scientific potential. Furthermore, the patrimonial value of HEP data has to be addressed for experiments such as \babar where, after a very successful exploitation of a long term potential; the data preservation may indeed come to an end due to the absence of resources, and in spite of a remarkable effort by specific groups within the collaboration.

These trends highlight the need for an integrated approach---combining technological innovation, international cooperation, and institutional commitment---to ensure the preservation and optimal use of HEP data. The communication forum provided by the DPHEP collaboration, operating as a component of the ICFA Data Lifecycle panel,  is essential in order to expose and promote in the longer term and at the global level the dedicated preservation activities at experiments and sites.  

\bibliographystyle{utphys} 
{
\bibliography{bib}}

@article{DPHEP:2023blx,
    author = "Basaglia, T. and others",
    collaboration = "DPHEP",
    title = "{Data preservation in high energy physics}",
    eprint = "2302.03583",
    archivePrefix = "arXiv",
    primaryClass = "hep-ex",
    reportNumber = "DPHEP-2023-01",
    doi = "10.1140/epjc/s10052-023-11885-1",
    journal = "Eur. Phys. J. C",
    volume = "83",
    number = "9",
    pages = "795",
    year = "2023"
}

@misc{defranchis2026modernjetflavourtagging,
      title={Modern jet flavour tagging in hadronic Z decays with archived ALEPH data}, 
      author={Matteo M. Defranchis and Jacopo Fanini and Apranik Fatehi and Gerardo Ganis and Taj Gillin and Loukas Gouskos and Luka Lambrecht and Michele Selvaggi and Birgit Stapf},
      year={2026},
      eprint={2603.06524},
      archivePrefix={arXiv},
      primaryClass={hep-ex},
      url={https://arxiv.org/abs/2603.06524}, 
}

@misc{fanini2026lepdataedm4hepmitigatingdata,
      title={LEP Data@EDM4hep: mitigating data loss risks by increasing data FAIRness, with a view on FCC-ee}, 
      author={Jacopo Fanini and Gerardo Ganis and Marcello Maggi},
      year={2026},
      eprint={2603.15493},
      archivePrefix={arXiv},
      primaryClass={hep-ex},
      url={https://arxiv.org/abs/2603.15493}, 
}

@article{Barate:321135,
      author        = {{ALEPH Collaboration}},
      title         = "{A Measurement of $R_b$ using Mutually Exclusive Tags}",
      reportNumber  = "CERN-PPE-97-018, FSU-SCRI-97-78",
      journal       = "Phys. Lett. B",
      volume        = "401",
      pages         = "163-175",
      year          = "1997",
      url           = "https://cds.cern.ch/record/321135",
      doi           = "10.1016/S0370-2693(97)00407-3",
}

@article{ALEPH:2001mdb,
    author = "Heister, A. and others",
    collaboration = "ALEPH",
    title = "{Measurement of $A_{\rm FB}^b$ using inclusive $b$-hadron decays}",
    eprint = "hep-ex/0107033",
    archivePrefix = "arXiv",
    reportNumber = "CERN-EP-2001-047",
    doi = "10.1007/s100520100812",
    journal = "Eur. Phys. J. C",
    volume = "22",
    pages = "201--215",
    year = "2001"
}

@misc{qu2024particletransformerjettagging,
      title={Particle Transformer for Jet Tagging}, 
      author={Huilin Qu and Congqiao Li and Sitian Qian},
      year={2024},
      eprint={2202.03772},
      archivePrefix={arXiv},
      primaryClass={hep-ph},
      url={https://arxiv.org/abs/2202.03772}, 
}

@misc{CC0,
      title = "{CC0 1.0 Universal License}",
      url = "https://creativecommons.org/publicdomain/zero/1.0/",
      accessed = "2024-12-02",
      author = "{Creative Commons}",
}

@misc{OpenDataPortal,
    title          = "{{CERN Open Data Portal}}",
    url            = "http://opendata.cern.ch",
    year           = "2024"
}

@misc{Zenodo,
  doi = {10.25495/7GXK-RD71},
  url = {https://www.zenodo.org/},
  author = {{European Organization For Nuclear Research} and {OpenAIRE}},
  language = {en},
  title = {Zenodo},
  publisher = {CERN},
  year = {2013}
}

@misc{codimd,
  author       = {{CodiMD}},
  title        = {{CodiMD}},
  url          = {https://github.com/hackmdio/codimd},
}

@misc{inveniordm,
  author       = {{CERN}},
  title        = {{InvenioRDM}},
  url          = {https://inveniosoftware.org/},
}

@misc{cerndocumentserver,
  author       = {{CERN}},
  title        = {{CERN Document Server}},
  url          = {https://repository.cern/},
}

@misc{indico,
  author       = {{CERN}},
  title        = {Indico},
  url          = {https://indico.cern.ch/},
}

@techreport{oais,
  author       = {{Consultative Committee for Space Data Systems (CCSDS)}},
  title        = {{Reference Model for an Open Archival Information System (OAIS)}},
  institution  = {CCSDS},
  year         = {2024},
  number       = {CCSDS 650.0-M-3},
  url          = {https://ccsds.org/Pubs/650x0m3.pdf}
}

@misc{CERNTapeArchive,
    author         = {{CERN}},
    title          = "{{CERN Tape Archive}}",
    url            = "https://cta.web.cern.ch/cta/",
}

@misc{atlasopenmagic,
  author = "{ATLAS Collaboration}",
  title = "{ATLAS Open Magic}",
  url = "https://pypi.org/project/atlasopenmagic/",
  year = "2026"
}

@misc{Phoenix,
      title = "Phoenix Event Display",
      url = "https://github.com/HSF/phoenix",
      accessed = "2024-12-02",
}

@booklet{ATLASSoftwarePolicy,
  author = "{ATLAS Collaboration}",
  title = "{ATLAS Software Policy}",
  howpublished = "{ATL-CBPOLICY-PUB-2026-003}",
  year = "2026",
  url = "https://cds.cern.ch/record/2939069"
}

@booklet{ATLASDataPolicy,
  author = "{ATLAS Collaboration}",
  title = "{ATLAS Data Preservation Policy}",
  howpublished = "{ATL-CBPOLICY-PUB-2026-001}",
  year = "2026",
  url = "https://cds.cern.ch/record/2954457"
}

@misc{ATLASODTutorial,
  title = "{ATLAS Open Data Tutorial}",
  year = "2025",
  url = "https://indico.cern.ch/event/1564767",
  author = "{ATLAS Collaboration}"
}

@misc{RivetAnalysis,
    author         = "{Rivet Collaboration}",
    title          = "{Rivet analysis coverage}",
    url            = "https://rivet.hepforge.org/rivet-coverage",
    year           = "2023"
}

@misc{HEPData,
  url = "https://www.hepdata.net",
  year = "2026",
  title = "HEPData"
}

@article{ATLASAIOD,
  author = "E. Gendreau-Distler and others",
  title = "{Automating High Energy Physics Data Analysis with LLM-Powered Agents}",
  eprint = "2512.07785",
  archivePrefix = "arXiv",
  primaryClass = "physics.data-an",
  doi = "10.48550/arXiv.2512.07785",
  year = "2025"
}

@misc{ATLASOpenData,
    author         = "{ATLAS Collaboration}",
    title          = "{ATLAS Open Data}",
    url            = "http://opendata.atlas.cern",
    year           = "2026"
}

@misc{ATLASClassroomApp,
  author = "{ATLAS Collaboration}",
  url = "https://opendata.atlas.cern/docs/webapps/teachersapp",
  year = "2026",
  title = "Classroom Application"
}

@misc{CERN-OpenData-policy,
    title = "{CERN Open Data Policy for the LHC Experiments}",
      url = "https://cds.cern.ch/record/2745133",
      accessed = "2020-11-20",
}

@misc{LHCb-ntupleService,
    title = "{LHCb Ntupling Service}",
      url = "https://opendata-lhcb-ntupling-service.app.cern.ch/"
}

@misc{open-data-monitoring,
    author = {{CERN}},
    title = "{CERN Open Data dashboard}",
    url = "https://monit-grafana.cern.ch/d/da06d76c-24f0-4d23-b51e-da08d36c4ece/welcome?orgId=93",
}

@misc{ISOLDE-policy,
      title = "{ISOLDE Collaboration Open Data Policy and Advice}",
      url = "https://isolde.cern/sites/default/files/ISOLDE-Open-Data-Policy-and-Advice-Approved.pdf",
      accessed = "2023-06-15",
}

@misc{nTOF-policy,
      title = "{nTOF Collaboration Open Data Policy}",
      url = "https://cds.cern.ch/record/2955617",
      accessed = "2025-12-12",
}

@article{analysis_productions,
    author = "Abdelmotteleb, Ahmed and others",
    title = "{The LHCb Sprucing and Analysis Productions}",
    eprint = "2506.20309",
    archivePrefix = "arXiv",
    primaryClass = "hep-ex",
    doi = "10.1007/s41781-025-00144-5",
    journal = "Comput. Softw. Big Sci.",
    volume = "9",
    number = "1",
    pages = "15",
    year = "2025"
}

@article{ntupling_service_proceedings,
    author = "Aidala, Christine and others",
    title = "{LHCb Open Data Ntupling Service: On-demand production and publishing of custom LHCb Open Data}",
    eprint = "2504.00610",
    archivePrefix = "arXiv",
    primaryClass = "hep-ex",
    doi = "10.1051/epjconf/202533701231",
    journal = "EPJ Web Conf.",
    volume = "337",
    pages = "01231",
    year = "2025"
}

@misc{lhcb_opendata_guide,
    title = "{LHCb Open Data Guide}",
      url = "https://lhcb-opendata-guide.web.cern.ch/"
}

@misc{lhcb_opendata_workshop,
    title = "{First LHCb Open Data and Ntuple Wizard Workshop}",
      url = "https://indico.cern.ch/event/1429526/"
}

@article{lhcb_apd,
      author        = "Burr, Chris and Couturier, Ben and O’Neil, Ryunosuke",
      title         = "{Facilitating the preservation of LHCb Analyses with APD}",
      journal       = "EPJ Web Conf.",
      volume        = "295",
      pages         = "08008",
      year          = "2024",
      url           = "https://cds.cern.ch/record/2919288",
      doi           = "10.1051/epjconf/202429508008",
}

@misc{bubble_manual,
      title         = "{CERN 2m HBC user's handbook}",
      url           = "https://cds.cern.ch/record/2227678",
}

@techreport{bubble_T209,
      author        = "Carney, J N and Cox, G F and Kinson, J B and Votruba, F
                       and Bossen, Gerrit Jan and Quercigh, Emanuele and Tallini,
                       B",
      title         = "{Proposal for a large statistics $K^{-}p$ exposure at 8.25
                       GeV/c in the CERN 2 meter HBC}",
      institution   = "CERN",
      reportNumber  = "CERN-TCC-74-14",
      address       = "Geneva",
      year          = "1974",
      url           = "https://cds.cern.ch/record/732701",
}

@misc{bubble_website,
      title         = "{Bubble Chamber Data Preservation Initiative}",
      url           = "https://bubblechamber.web.cern.ch",
}

@article{bubble_2m,
      author        = "Weiss, L",
      title         = "{The construction of CERN's first hydrogen bubble
                       chambers}",
      reportNumber  = "CERN-CHS-26",
      year          = "1988",
      url           = "https://cds.cern.ch/record/194093",
}

@article{GEANT4:2002zbu,
    author = "Agostinelli, S. and others",
    collaboration = "GEANT4",
    title = "{GEANT4 - A Simulation Toolkit}",
    reportNumber = "SLAC-PUB-9350, FERMILAB-PUB-03-339, CERN-IT-2002-003",
    doi = "10.1016/S0168-9002(03)01368-8",
    journal = "Nucl. Instrum. Meth. A",
    volume = "506",
    pages = "250--303",
    year = "2003"
}

@article{OPAL:2000puu,
    author = "Abbiendi, G. and others",
    collaboration = "OPAL",
    title = "{Photonic events with missing energy in $\mathrm{e}^{+} \mathrm{e}^{-}$ 
    collisions at $\sqrt{s}=$ 189 GeV}",
    eprint = "hep-ex/0005002",
    archivePrefix = "arXiv",
    reportNumber = "CERN-EP-2000-050",
    doi = "10.1007/s100520000522",
    journal = "Eur. Phys. J. C",
    volume = "18",
    pages = "253--272",
    year = "2000"
}

@article{ Lassila-Perini:2021,
    author = {{Lassila-Perini, Kati} and {Lange, Clemens} and {Carrera Jarrin, Edgar} and {Bellis, Matthew}},
    title = {Using CMS Open Data in research – challenges and directions},
    DOI= "10.1051/epjconf/202125101004",
    url= "https://doi.org/10.1051/epjconf/202125101004",
    journal = {EPJ Web Conf.},
    year = 2021,
    volume = 251,
    pages = "01004",
}

@misc{OpenAI,
  author        = {{OpenAI}},
  title         = {{ChatGPT: Language Models from OpenAI}},
  year          = {2023},
  howpublished  = {\url{https://openai.com/chatgpt}},
  note          = {API access via \url{https://platform.openai.com}}
}

@misc{Anthropic:Claude,
  author        = {{Anthropic}},
  title         = {{Claude: AI Assistant}},
  year          = {2024},
  howpublished  = {\url{https://www.anthropic.com/claude}},
  note          = {Model card available at
                   \url{https://www-cdn.anthropic.com/de8ba9b01c9ab7cbabf5c33b80b7bbc618857627/Model_Card_Claude_3.pdf}}
}

@misc{BESdataComittee,
  author        = {{BESIII collaboration}},
  title         = {{BESIII Data Ecosystem Committee}},
  year          = {2025},
  note          = {\url{https://english.ihep.cas.cn/bes/co/or/co/202109/t20210924_284102.html}}
}

@misc{dybdatasets,
  author        = {{Dayabay collaboration}},
  title         = {{Full Data Release of the Daya Bay Reactor Neutrino Experiment}},
  year          = {2025},
  note          = {v1.0.0. Zenodo, \url{DOI:10.5281/zenodo.17587229; 2025}}
}

@misc{CERN:OpenDataPortal,
  author        = {{CERN}},
  title         = {{CERN Open Data Portal}},
  year          = {2014},
  howpublished  = {\url{https://opendata.cern.ch}}
}

@misc{eosc-eden,
  author       = {{EOSC EDEN Project}},
  title        = {EDEN-FIDELIS},
  url          = {https://eden-fidelis.eu/about-us},
}

@misc(eosc-eden-csc,
author  =   "{{CSC - IT Center for Sciene}}",
title = "CSC",
url = "https://csc.fi/en/"
)

@misc{eosc-eden-cpp-repository,
author = "{{EOSC EDEN}}",
title = "Core Preservation Processes",
url = "https://github.com/EOSC-EDEN/wp1-cpp-descriptions"
}

@misc{EOSOpenStorage,
    author         = {{CERN}},
    title          = "{{EOS Open Storage}}",
    url            = "https://eos-web.web.cern.ch/eos-web/",
}

@misc{FileTransferSystem,
    author         = {{CERN}},
    title          = "{{File Transfer System}}",
    url            = "https://fts.web.cern.ch/fts/",
}

@misc{fts-monitoring,
    author         = {{CERN}},
    title          = "{FTS Servers Dashboard}",
    url            = "https://monit-grafana.cern.ch/d/veRQSWBGz/fts-servers-dashboard?orgId=25",
}

@misc{rucio,
    author         = {{CERN}},
    title          = "{Rucio}",
    url            = "https://rucio.cern.ch/",
}

@misc{eosc-eden-3.1,
author = "{{EOSC EDEN}}",
title = "EOSC EDEN D3.1 - Report on Discipline Requirements and Needs",
url = "https://doi.org/10.5281/zenodo.15789261"
}

@misc{icfa-best-pratices-app,
  author        = {{ICFA}},
  title         = {{Recommendations for best practices for data preservation and open science in HEP}},
  year          = {2025},
  howpublished  = {\url{https://icfa-data-best-practices.app.cern.ch/}}
}

@article{icfa-best-practices-note,
    author = "Campana, Simone and others",
    collaboration = "ICFA Data Lifecycle panel ",
    title = "{Recommendations for best practices for data preservation and open science in HEP}",
    year={2025},
    eprint={2508.18892},
    archivePrefix={arXiv},
    primaryClass={hep-ex},
    url={https://arxiv.org/abs/2508.18892}, 
}

@inproceedings{Ebert:CHEP2024,
  author       = {Ebert, Marcus and others},
  title        = {{BABAR's Experience with the Preservation of Data and Analysis Capabilities}},
  booktitle    = {EPJ Web of Conferences},
  year         = {2024},
  volume       = {295},
  pages        = {08006},
  doi          = {10.1051/epjconf/202429508006},
  url          = {https://www.epj-conferences.org/articles/epjconf/pdf/2024/05/epjconf_chep2024_08006.pdf}
}

@inproceedings{Ebert:CHEP2025,
  author       = {Ebert, Marcus and others},
  title        = {{The BABAR Long Term Data Preservation and Computing Infrastructure}},
  booktitle    = {EPJ Web of Conferences},
  year         = {2025},
  volume       = {317},
  pages        = {01054},
  doi          = {10.1051/epjconf/202533701054},
  url          = {https://www.epj-conferences.org/articles/epjconf/abs/2025/22/epjconf_chep2025_01054/epjconf_chep2025_01054.html}
}

@article{Brun:1997pa,
  author        = {Brun, R. and Rademakers, F.},
  title         = {{ROOT: An object oriented data analysis framework}},
  journal       = {Nucl. Instrum. Meth. A},
  volume        = {389},
  pages         = {81--86},
  year          = {1997},
  doi           = {10.1016/S0168-9002(97)00048-X}
}

@software{uproot,
  author        = {Pivarski, Jim and Schreiner, Henry and others},
  title         = {{uproot: ROOT I/O in pure Python and NumPy}},
  year          = {2017},
  publisher     = {Zenodo},
  doi           = {10.5281/zenodo.2552892},
  url           = {https://doi.org/10.5281/zenodo.2552892}
}

@software{awkward,
  author        = {Pivarski, Jim and Osborne, Ianna and Ifrim, Ioana and Schreiner, Henry and others},
  title         = {{Awkward Array}},
  year          = {2018},
  publisher     = {Zenodo},
  doi           = {10.5281/zenodo.4341376},
  url           = {https://doi.org/10.5281/zenodo.4341376}
}

@Article{Hunter:2007,
  Author    = {Hunter, J. D.},
  Title     = {Matplotlib: A 2D graphics environment},
  Journal   = {Computing in Science \& Engineering},
  Volume    = {9},
  Number    = {3},
  Pages     = {90--95},
  publisher = {IEEE COMPUTER SOC},
  doi       = {10.1109/MCSE.2007.55},
  year      = 2007
}

@misc{CMSOpenDataPolicy,
author = "{{CMS Collaboration}}",
title = "{CMS Data Preservation, Re-use, and Open Access Policy}",
url = "http://doi.org/10.7483/OPENDATA.CMS.1BNU.8V1W"
}

@misc{CMSOpenDataRecord,
author = "{{CMS Collaboration}}",
title = "{Tau primary dataset in NANOAOD format from RunH of 2016}",
url = "http://doi.org/10.7483/OPENDATA.CMS.TTK7.008J"
}

@misc{CMSDataUsage,
author = "{{McCauley, T.}}",
title = "cms-dpoa/data-usage: 28-05-2026 (28-05-2026). Zenodo",
url = "https://doi.org/10.5281/zenodo.20431719"
}

@article{Campbell:2022qmc,
    author = "Campbell, J. M. and others",
    title = "{Event generators for high-energy physics experiments}",
    eprint = "2203.11110",
    archivePrefix = "arXiv",
    primaryClass = "hep-ph",
    reportNumber = "CP3-22-12, DESY-22-042, FERMILAB-PUB-22-116-SCD-T, IPPP/21/51,
  JLAB-PHY-22-3576, KA-TP-04-2022, LA-UR-22-22126, LU-TP-22-12, MCNET-22-04,
  OUTP-22-03P, P3H-22-024, PITT-PACC 2207, UCI-TR-2022-02",
    doi = "10.21468/SciPostPhys.16.5.130",
    journal = "SciPost Phys.",
    volume = "16",
    number = "5",
    pages = "130",
    year = "2024"
}

@article{Price:2025fzg,
    author = "Price, Alan and Zerwas, Dirk",
    title = "{Configuration and Benchmarking of $\mathrm{e}^+\mathrm{e}^-$ Processes with K4GeneratorsConfig}",
    eprint = "2509.20116",
    archivePrefix = "arXiv",
    primaryClass = "hep-ph",
    month = "9",
    year = "2025"
}

@article{Alwall:2014hca,
    author = "Alwall, J. and Frederix, R. and Frixione, S. and Hirschi, V. and Maltoni, F. and Mattelaer, O. and Shao, H. -S. and Stelzer, T. and Torrielli, P. and Zaro, M.",
    title = "{The automated computation of tree-level and next-to-leading order differential cross sections, and their matching to parton shower simulations}",
    eprint = "1405.0301",
    archivePrefix = "arXiv",
    primaryClass = "hep-ph",
    reportNumber = "CERN-PH-TH-2014-064, CP3-14-18, LPN14-066, MCNET-14-09, ZU-TH-14-14",
    doi = "10.1007/JHEP07(2014)079",
    journal = "JHEP",
    volume = "07",
    pages = "079",
    year = "2014"
}

@article{Kilian:2007gr,
    author = "Kilian, Wolfgang and Ohl, Thorsten and Reuter, Jurgen",
    title = "{WHIZARD: Simulating Multi-Particle Processes at LHC and ILC}",
    eprint = "0708.4233",
    archivePrefix = "arXiv",
    primaryClass = "hep-ph",
    reportNumber = "DESY-11-126, EDINBURGH-2010-36, FR-PHENO-2010-037, SI-HEP-2010-18",
    doi = "10.1140/epjc/s10052-011-1742-y",
    journal = "Eur. Phys. J. C",
    volume = "71",
    pages = "1742",
    year = "2011"
}

@article{Sherpa:2024mfk,
    author = "Bothmann, Enrico and others",
    collaboration = "Sherpa",
    title = "{Event generation with Sherpa 3}",
    eprint = "2410.22148",
    archivePrefix = "arXiv",
    primaryClass = "hep-ph",
    reportNumber = "IPPP/24/67, LTH-1385, FERMILAB-PUB-24-0748-T, ZU-TH 51/24, MCNET-24-17, CERN-TH-2024-171",
    doi = "10.1007/JHEP12(2024)156",
    journal = "JHEP",
    volume = "12",
    pages = "156",
    year = "2024"
}

@article{Bierlich:2022pfr,
    author = "Bierlich, Christian and others",
    title = "{A comprehensive guide to the physics and usage of PYTHIA 8.3}",
    eprint = "2203.11601",
    archivePrefix = "arXiv",
    primaryClass = "hep-ph",
    reportNumber = "LU-TP 22-16, MCNET-22-04, FERMILAB-PUB-22-227-SCD",
    doi = "10.21468/SciPostPhysCodeb.8",
    journal = "SciPost Phys. Codeb.",
    volume = "2022",
    pages = "8",
    year = "2022"
}

@article{Jadach:2022mbe,
    author = "Jadach, S. and Ward, B. F. L. and Was, Z. and Yost, S. A. and Siodmok, A.",
    title = "{Multi-photon Monte Carlo event generator KKMCee for lepton and quark pair production in lepton colliders}",
    eprint = "2204.11949",
    archivePrefix = "arXiv",
    primaryClass = "hep-ph",
    reportNumber = "IFJPAN-IV-2022-6, BU-HEPP-22-02, MCnet-22",
    doi = "10.1016/j.cpc.2022.108556",
    journal = "Comput. Phys. Commun.",
    volume = "283",
    pages = "108556",
    year = "2023"
}

@article{Bellm:2015jjp,
    author = "Bellm, Johannes and others",
    title = "{Herwig 7.0/Herwig++ 3.0 release note}",
    eprint = "1512.01178",
    archivePrefix = "arXiv",
    primaryClass = "hep-ph",
    reportNumber = "CERN-PH-TH-2015-289, MAN-HEP-2015-15, IFJPAN-IV-2015-13, KA-TP-18-2015, DCPT-15-142, MCNET-15-28, IPPP-15-71, HERWIG-2015-01",
    doi = "10.1140/epjc/s10052-016-4018-8",
    journal = "Eur. Phys. J. C",
    volume = "76",
    number = "4",
    pages = "196",
    year = "2016"
}

@article{Bierlich:2019rhm,
    author = "Bierlich, Christian and others",
    title = "{Robust Independent Validation of Experiment and Theory: Rivet version 3}",
    eprint = "1912.05451",
    archivePrefix = "arXiv",
    primaryClass = "hep-ph",
    reportNumber = "MCnet-19-26",
    doi = "10.21468/SciPostPhys.8.2.026",
    journal = "SciPost Phys.",
    volume = "8",
    pages = "026",
    year = "2020"
}

@article{Bierlich:2024vqo,
    author = "Bierlich, Christian and Buckley, Andy and Butterworth, Jonathan Mark and Gutschow, Christian and Lonnblad, Leif and Procter, Tomasz and Richardson, Peter and Yeh, Yoran",
    title = "{Robust independent validation of experiment and theory: Rivet version 4 release note}",
    eprint = "2404.15984",
    archivePrefix = "arXiv",
    primaryClass = "hep-ph",
    reportNumber = "MCNET-24-05",
    doi = "10.21468/SciPostPhysCodeb.36",
    journal = "SciPost Phys. Codeb.",
    volume = "36",
    pages = "1",
    year = "2024"
}

@article{CONTUR:2021qmv,
    author = "Buckley, A. and others",
    collaboration = "CONTUR",
    title = "{Testing new physics models with global comparisons to collider measurements: the Contur toolkit}",
    eprint = "2102.04377",
    archivePrefix = "arXiv",
    primaryClass = "hep-ph",
    reportNumber = "MCnet-21",
    doi = "10.21468/SciPostPhysCore.4.2.013",
    journal = "SciPost Phys. Core",
    volume = "4",
    pages = "013",
    year = "2021"
}

@article{Buckley:2009bj,
    author = "Buckley, Andy and Hoeth, Hendrik and Lacker, Heiko and Schulz, Holger and von Seggern, Jan Eike",
    title = "{Systematic event generator tuning for the LHC}",
    eprint = "0907.2973",
    archivePrefix = "arXiv",
    primaryClass = "hep-ph",
    reportNumber = "IPPP-09-52, DCPT-104-22, LU-TP-09-18, HU-EP-09-33, MCNET-09-14",
    doi = "10.1140/epjc/s10052-009-1196-7",
    journal = "Eur. Phys. J. C",
    volume = "65",
    pages = "331--357",
    year = "2010"
}

@article{GAMBIT:2017qxg,
    author = "Bal{\'a}zs, Csaba and others",
    collaboration = "GAMBIT",
    title = "{ColliderBit: a GAMBIT module for the calculation of high-energy collider observables and likelihoods}",
    eprint = "1705.07919",
    archivePrefix = "arXiv",
    primaryClass = "hep-ph",
    reportNumber = "gambit-code-2017",
    doi = "10.1140/epjc/s10052-017-5285-8",
    journal = "Eur. Phys. J. C",
    volume = "77",
    number = "11",
    pages = "795",
    year = "2017"
}

@article{Araz:2020lnp,
    author = "Araz, Jack Y. and Fuks, Benjamin and Polykratis, Georgios",
    title = "{Simplified fast detector simulation in MADANALYSIS 5}",
    eprint = "2006.09387",
    archivePrefix = "arXiv",
    primaryClass = "hep-ph",
    doi = "10.1140/epjc/s10052-021-09052-5",
    journal = "Eur. Phys. J. C",
    volume = "81",
    number = "4",
    pages = "329",
    year = "2021"
}

@article{Buckley:2019stt,
    author = {Buckley, Andy and Kar, Deepak and Nordstr{\"o}m, Karl},
    title = "{Fast simulation of detector effects in Rivet}",
    eprint = "1910.01637",
    archivePrefix = "arXiv",
    primaryClass = "hep-ph",
    reportNumber = "MCNET-19-22",
    doi = "10.21468/SciPostPhys.8.2.025",
    journal = "SciPost Phys.",
    volume = "8",
    pages = "025",
    year = "2020"
}

@article{Araz:2023mda,
    author = "Araz, Jack Y. and others",
    title = "{Les Houches guide to reusable ML models in LHC analyses}",
    eprint = "2312.14575",
    archivePrefix = "arXiv",
    primaryClass = "hep-ph",
    doi = "10.21468/SciPostPhysCommRep.3",
    month = "12",
    year = "2023"
}

@article{Buckley:2025pqk,
    author = "Buckley, Andy and Corpe, Louie and Habedank, Martin and Procter, Tomasz",
    title = "{Enabling stable preservation of ML algorithms in high-energy physics with petrifyML}",
    eprint = "2509.11830",
    archivePrefix = "arXiv",
    primaryClass = "hep-ph",
    reportNumber = "MCNET-26-13; OPENMAPP-26-02",
    month = "9",
    year = "2025"
}

@article{Gaede:2021izq,
    author = "Gaede, Frank and Ganis, Gerardo and Hegner, Benedikt and Helsens, Clement and Madlener, Thomas and Sailer, Andre and Stewart, Graeme A. and Volkl, Valentin and Wang, Joseph",
    collaboration = "Key4hep",
    title = "{EDM4hep and podio - The event data model of the Key4hep project and its implementation}",
    doi = "10.1051/epjconf/202125103026",
    journal = "EPJ Web Conf.",
    volume = "251",
    pages = "03026",
    year = "2021"
}

@misc{CMSOpenDataGuide,
    title = "{CMS Open Data Guide}",
    url = "https://cms-opendata-guide.web.cern.ch/"
}

@misc{CMSOpenDataWorkshops,
    title = "{CMS Open Data Workshops}",
    url = "https://cms-opendata-guide.web.cern.ch/cmsOpenData/workshops/",
}

@misc{CMSOpenDataWorkshopJuly,
    title = "{CMS Open Data Workshop 28-30 July 2026}",
    url = "https://indico.cern.ch/event/1672496/"
}

@misc{DPHEP2025,
      title={Data Preservation in High Energy Physics}, 
      author={Alexandre Arbey and others},
      year={2025},
      eprint={2503.23619},
      archivePrefix={arXiv},
      primaryClass={hep-ex},
      url={https://arxiv.org/abs/2503.23619}
}

@misc{babar:openaccess,
title={{The \babar Associates Open-access Program}},
accessed={June-5th-2026},
url={https://babar.heprc.uvic.ca/www/join_BaBar.html},
author={\babar}
}

@article{ROOT_NIMA_1997,
    author = "Brun, R. and Rademakers, F.",
    editor = "Werlen, M. and Perret-Gallix, D.",
    title = "{ROOT: An object oriented data analysis framework}",
    doi = "10.1016/S0168-9002(97)00048-X",
    journal = "Nucl. Instrum. Meth. A",
    volume = "389",
    pages = "81--86",
    year = "1997"
}

@article{H1:1996jzy,
    author = "Abt, I. and others",
    collaboration = "H1",
    title = "{The Tracking, calorimeter and muon detectors of the H1 experiment at HERA}",
    reportNumber = "SLAC-REPRINT-1996-034",
    doi = "10.1016/S0168-9002(96)00894-7",
    journal = "Nucl. Instrum. Meth. A",
    volume = "386",
    pages = "348--396",
    year = "1997"
}

@article{H1:1996prr,
    author = "Abt, I. and others",
    collaboration = "H1",
    title = "{The H1 detector at HERA}",
    reportNumber = "SLAC-REPRINT-1996-078",
    doi = "10.1016/S0168-9002(96)00893-5",
    journal = "Nucl. Instrum. Meth. A",
    volume = "386",
    pages = "310--347",
    year = "1997"
}

@article{Steder:2011zz,
    author = "Steder, Michael",
    editor = "Lin, Simon C.",
    collaboration = "H1",
    title = "{H100: A centralised analysis framework for the H1 experiment}",
    doi = "10.1088/1742-6596/331/3/032051",
    journal = "J. Phys. Conf. Ser.",
    volume = "331",
    pages = "032051",
    year = "2011"
}

@article{H1:2016goa,
    author = "Andreev, V. and others",
    collaboration = "H1",
    title = "{Measurement of Jet Production Cross Sections in Deep-inelastic $ep$ Scattering at HERA}",
    eprint = "1611.03421",
    archivePrefix = "arXiv",
    primaryClass = "hep-ex",
    reportNumber = "DESY-16-200, DESY-16-200",
    doi = "10.1140/epjc/s10052-017-4717-9",
    journal = "Eur. Phys. J. C",
    volume = "77",
    number = "4",
    pages = "215",
    year = "2017",
    note = "[Erratum: Eur.Phys.J.C 81, 739 (2021)]"
}

@article{Britzger:2021xcx,
    author = "Britzger, Daniel and Levonian, Sergey and Schmitt, Stefan and South, David",
    collaboration = "H1",
    title = "{Preservation through modernisation: The software of the H1 experiment at HERA}",
    eprint = "2106.11058",
    archivePrefix = "arXiv",
    primaryClass = "hep-ex",
    reportNumber = "MPP-2021-87, DESY-21-097",
    doi = "10.1051/epjconf/202125103004",
    journal = "EPJ Web Conf.",
    volume = "251",
    pages = "03004",
    year = "2021"
}

@article{Arratia:2021tsq,
    author = "Arratia, Miguel and Britzger, Daniel and Long, Owen and Nachman, Benjamin",
    title = "{Reconstructing the kinematics of deep inelastic scattering with deep learning}",
    eprint = "2110.05505",
    archivePrefix = "arXiv",
    primaryClass = "hep-ex",
    reportNumber = "MPP-2021-174",
    doi = "10.1016/j.nima.2021.166164",
    journal = "Nucl. Instrum. Meth. A",
    volume = "1025",
    pages = "166164",
    year = "2022"
}

@article{H1:2021wkz,
    author = "Andreev, V. and others",
    collaboration = "H1",
    title = "{Measurement of Lepton-Jet Correlation in Deep-Inelastic Scattering with the H1 Detector Using Machine Learning for Unfolding}",
    eprint = "2108.12376",
    archivePrefix = "arXiv",
    primaryClass = "hep-ex",
    reportNumber = "DESY 21-130",
    doi = "10.1103/PhysRevLett.128.132002",
    journal = "Phys. Rev. Lett.",
    volume = "128",
    number = "13",
    pages = "132002",
    year = "2022"
}

@article{H1:2023fzk,
    author = "Andreev, V. and others",
    collaboration = "H1",
    title = "{Unbinned deep learning jet substructure measurement in high Q2ep collisions at HERA}",
    eprint = "2303.13620",
    archivePrefix = "arXiv",
    primaryClass = "hep-ex",
    reportNumber = "DESY-23-034",
    doi = "10.1016/j.physletb.2023.138101",
    journal = "Phys. Lett. B",
    volume = "844",
    pages = "138101",
    year = "2023"
}

@article{H1:2024mox,
    author = "Andreev, V. and others",
    collaboration = "H1",
    title = "{Machine Learning-Assisted Measurement of Lepton-Jet Azimuthal Angular Asymmetries in Deep-Inelastic Scattering at HERA}",
    eprint = "2412.14092",
    archivePrefix = "arXiv",
    primaryClass = "hep-ex",
    reportNumber = "DESY24-200",
    month = "12",
    year = "2024"
}

@article{H1:2024nde,
    author = "Andreev, V. and others",
    collaboration = "H1",
    title = "{Observation and differential cross section measurement of neutral current DIS events with an empty hemisphere in the Breit frame}",
    eprint = "2403.08982",
    archivePrefix = "arXiv",
    primaryClass = "hep-ex",
    reportNumber = "DESY-24-034",
    doi = "10.1140/epjc/s10052-024-13003-1",
    journal = "Eur. Phys. J. C",
    volume = "84",
    number = "7",
    pages = "720",
    year = "2024"
}

@article{H1:2024aze,
    author = "Andreev, V. and others",
    collaboration = "H1",
    title = "{Measurement of the 1-jettiness event shape observable in deep-inelastic electron-proton scattering at HERA}",
    eprint = "2403.10109",
    archivePrefix = "arXiv",
    primaryClass = "hep-ex",
    reportNumber = "DESY-24-035",
    doi = "10.1140/epjc/s10052-024-13115-8",
    journal = "Eur. Phys. J. C",
    volume = "84",
    number = "8",
    pages = "785",
    year = "2024"
}

@article{H1:2024pvu,
    author = "Andreev, V. and others",
    collaboration = "H1",
    title = "{Measurement of groomed event shape observables in deep-inelastic electron-proton scattering at HERA}",
    eprint = "2403.10134",
    archivePrefix = "arXiv",
    primaryClass = "hep-ex",
    reportNumber = "DESY-24-036",
    doi = "10.1140/epjc/s10052-024-12987-0",
    journal = "Eur. Phys. J. C",
    volume = "84",
    number = "7",
    pages = "718",
    year = "2024"
}

@article{Roiser_2010,
	doi = {10.1088/1742-6596/219/4/042022},
	year = 2010,
	month = {apr},
	publisher = {{IOP} Publishing},
	volume = {219},
	number = {4},
	pages = {042022},
	author = {Stefan Roiser and Ana Gaspar and Yves Perrin and Karol Kruzelecki},
	title = {Servicing {HEP} experiments with a complete set of ready integreated and configured common software components},
	journal = {Journal of Physics: Conference Series},
}

@article{Maguire:2017ypu,
    author = "Maguire, Eamonn and Heinrich, Lukas and Watt, Graeme",
    editor = "Mount, Richard and Tull, Craig",
    title = "{HEPData: a repository for high energy physics data}",
    eprint = "1704.05473",
    archivePrefix = "arXiv",
    primaryClass = "hep-ex",
    reportNumber = "IPPP-17-31",
    doi = "10.1088/1742-6596/898/10/102006",
    journal = "J. Phys. Conf. Ser.",
    volume = "898",
    number = "10",
    pages = "102006",
    year = "2017"
}

@article{DELPHI:1996sen,
    author = "Abreu, P. and others",
    collaboration = "DELPHI",
    title = "{Tuning and test of fragmentation models based on identified particles and precision event shape data}",
    reportNumber = "CERN-PPE-96-120",
    doi = "10.1007/s002880050295",
    journal = "Z. Phys. C",
    volume = "73",
    pages = "11--60",
    year = "1996"
}

@article{ALEPH:2003obs,
    author = "Heister, A. and others",
    collaboration = "ALEPH",
    title = "{Studies of QCD at e+ e- centre-of-mass energies between 91-GeV and 209-GeV}",
    reportNumber = "CERN-EP-2003-084",
    doi = "10.1140/epjc/s2004-01891-4",
    journal = "Eur. Phys. J. C",
    volume = "35",
    pages = "457--486",
    year = "2004"
}

@article{DiMauro:2025vxp,
    author = "Di Mauro, Mattia and Jueid, Adil and Koechler, Jordan and de Austri, Roberto Ruiz",
    title = "{Robust determination of antinuclei production from dark matter via weakly decaying beauty hadrons}",
    eprint = "2504.07172",
    archivePrefix = "arXiv",
    primaryClass = "hep-ph",
    doi = "10.1103/s6cm-45b4",
    journal = "Phys. Rev. D",
    volume = "112",
    number = "8",
    pages = "083017",
    year = "2025"
}

@article{Assi:2025ibi,
    author = {Assi, Beno{\^\i}t and H{\"o}che, Stefan and Lee, Kyle and Thaler, Jesse},
    title = "{QCD Theory Meets Information Theory}",
    eprint = "2501.17219",
    archivePrefix = "arXiv",
    primaryClass = "hep-ph",
    reportNumber = "FERMILAB-PUB-25-0029-T, MIT-CTP 5827, MCNET-25-01",
    doi = "10.1103/gf42-qzd9",
    journal = "Phys. Rev. Lett.",
    volume = "135",
    number = "13",
    pages = "131901",
    year = "2025"
}

@article{MINERvA:2011bff,
    author = "Tagg, N. and others",
    collaboration = "MINERvA",
    title = "{Arachne - A web-based event viewer for MINERvA}",
    eprint = "1111.5315",
    archivePrefix = "arXiv",
    primaryClass = "hep-ex",
    reportNumber = "FERMILAB-PUB-11-625-E",
    doi = "10.1016/j.nima.2012.01.059",
    journal = "Nucl. Instrum. Meth. A",
    volume = "676",
    pages = "44--49",
    year = "2012"
}

@article{MINERvA:2025cfj,
    collaboration = "MINERvA",
    title = "{MINERvA Experiment Open Data Release}",
    reportNumber = "FERMILAB-DATA-2025-07",
    doi = "10.15484/3022562",
    month = "10",
    year = "2025"
}

@article{MINERvA:2025lya,
    author = "Lin, Z. and others",
    collaboration = "MINERvA",
    title = "{Boosted decision tree reweighting of simulated neutrino interactions for $O(1)$ GeV neutrino cross section measurements}",
    eprint = "2510.07463",
    archivePrefix = "arXiv",
    primaryClass = "hep-ex",
    reportNumber = "FERMILAB-PUB-25-0766-ETD-PPD",
    month = "10",
    year = "2025"
}

@article{Simko:2018zzz,
    author = "{\v{S}}imko, Tibor and Heinrich, Lukas and Hirvonsalo, Harri and Kousidis, Dinos and Rodr{\'\i}guez, Diego",
    editor = "Forti, A. and Betev, L. and Litmaath, M. and Smirnova, O. and Hristov, P.",
    title = "{REANA: A System for Reusable Research Data Analyses}",
    reportNumber = "CERN-IT-2018-003",
    doi = "10.1051/epjconf/201921406034",
    journal = "EPJ Web Conf.",
    volume = "214",
    pages = "06034",
    year = "2019"
}

@article{Donadoni:2024nkq,
    author = "Donadoni, Marco and Feickert, Matthew and Heinrich, Lukas and Liu, Yang and Me{\v{c}}ionis, Audrius and Moisieienkov, Vladyslav and {\v{S}}imko, Tibor and Stark, Giordon and Garc{\'\i}a, Marco Vidal",
    title = "{Scalable ATLAS pMSSM computational workflows using containerised REANA reusable analysis platform}",
    eprint = "2403.03494",
    archivePrefix = "arXiv",
    primaryClass = "cs.DC",
    doi = "10.1051/epjconf/202429504035",
    journal = "EPJ Web Conf.",
    volume = "295",
    pages = "04035",
    year = "2024"
}

@misc{ArmanPhysicsLLM2026,
  title = "{Agentic Systems Status}",
  year = "2026",
  howpublished = "{Physics-LLM Kick-Off Meeting}",
  url = "https://indico.desy.de/event/51692/",
  author =  "{Khalatyan, Arman}"
}

@misc{GuerrieriReanaEosc2025,
  author = "Guerrieri, Giovanni",
  title = "{The REANA use case in the EOSC Federation: FAIR (re)analysis of the LHC data in a distributed environment}",
  howpublished = "{EOSC Symposium}",
  year = "2026",
  url = "https://indico.cern.ch/event/1543880/",
}

@article{NYSD2025:SciBot,
    author = "Atif, Mohammad and Garonne, Vincent and Lancon, Eric and Lauret, Jerome and Prozorov, Alexandr and Vranovsky, Michal",
    title = "{AI-Powered Assistant for Long-Term Access to RHIC Knowledge}",
    journal = "New York Scientific Data Summit 2025",
    pages = "21--24",
    doi = "10.1137/1.9781611978933.6",
    year = "2025",
    url = "https://epubs.siam.org/doi/10.1137/1.9781611978933.6",
}

@techreport{Buncic:2011297,
      author        = "Buncic, P and Krzewicki, M and Vande Vyvre, P",
      title         = "{Technical Design Report for the Upgrade of the
                       Online-Offline Computing System}",
      reportNumber  = "CERN-LHCC-2015-006, ALICE-TDR-019",
      year          = "2015",
      url           = "https://cds.cern.ch/record/2011297",
}

@article{wilkinson2016,
  author = {Wilkinson, M. D. and others},
  title = {The {FAIR} Guiding Principles for scientific data management and stewardship},
  journal = {Scientific Data},
  volume = {3},
  pages = {160018},
  year = {2016},
  doi = {10.1038/sdata.2016.18},
  url = {https://doi.org/10.1038/sdata.2016.18}
}

@article{rongzai2024,
  author = {Li, Q. and Wang, H. and Xiong, D. and Zhong, J. and Ji, W. and Hu, H. and Zhang, Y. and Zhang, B. and Wang, H. and Zhu, Y. and Du, R. and Zhang, Z. and Qi, F. and Zhang, J.},
  title = {{Rongzai} agent: A Large Language Model-Based Autonomous Assistant for {Rietveld} Refinement},
  journal = {arXiv},
  year = {2025},
  eprint = {2605.13911},
  url = {https://arxiv.org/abs/2605.13911}
}

\end{document}